\documentclass[twocolumn,final]{IEEEtran}
\IEEEoverridecommandlockouts
\pdfoutput=1
\usepackage[cmex10]{amsmath}
\usepackage{epsfig}
\usepackage{color}
\usepackage{latexsym}
\usepackage{graphics}
\usepackage{amssymb}
\usepackage{amsthm}
\usepackage{subfigure}
\usepackage{algorithm}
\usepackage{algorithmic}

\title{Base station cooperation on the downlink: Large system analysis}

\author
{Randa Zakhour and Stephen V. Hanly
\thanks {
    R. Zakhour is with the Electrical and Electronic Engineering Department,
    University of Melbourne, Australia: rzakhour@unimelb.edu.au, S. V. Hanly is with the Department of Electrical and Computer Engineering,
    National University of Singapore: elehsv@nus.edu.sg.  This work was supported by the Australian Research Council (ARC) under grant DP0984862, and NUS grant R-263-000-572-133.}
}

\begin{document}

\newtheorem{theorem}{Theorem}
\newtheorem{proposition}{Proposition}
\newtheorem{corollary}{Corollary}
\newtheorem{lemma}{Lemma}

\newcommand \bol {\mbox{\boldmath $\lambda$}}
\newcommand \bal {\bar{\lambda}}
\newcommand \bald {\bar{\lambda}(\delta)}
\newcommand \bobl {\mbox{\boldmath $\bar{\lambda}$}}
\newcommand \bld {\bar{\lambda}(\delta)}
\newcommand \bloned {\bar{\lambda}_1(\delta)}
\newcommand \bltwod {\bar{\lambda}_2(\delta)}
\newcommand \blonemd {\bar{\lambda}_1(-\delta)}
\newcommand \bltwomd {\bar{\lambda}_2(-\delta)}
\newcommand \blone {\bar{\lambda}_1}
\newcommand \bltwo {\bar{\lambda}_2}
\newcommand \bljd {\bar{\lambda}_j(\delta)}
\newcommand \blmd {\bar{\lambda}(-\delta)}
\newcommand \bobld {\mbox{\boldmath $\bar{{\lambda}}$}(\delta)}
\newcommand \boblmd {\mbox{\boldmath $\bar{\lambda}$}(-\delta)}
\newcommand \bap {\bar{p}}
\newcommand \baj {\bar{j}}
\newcommand \bak {\bar{k}}
\newcommand \bamu {\bar{\mu}}
\newcommand \bog {\mbox{\boldmath $\gamma$}}
\newcommand \bobloned {\mbox{\boldmath ${\bar{\lambda}_1(\delta)}$}}
\newcommand \bobltwod {\mbox{\boldmath ${\bar{\lambda}_2(\delta)}$}}
\newcommand \boblonemd {\mbox{\boldmath ${\bar{\lambda}_1(-\delta)}$}}
\newcommand \bobltwomd {\mbox{\boldmath ${\bar{\lambda}_2(-\delta)}$}}

\renewcommand{\theenumi}{\roman{enumi}}

%
\maketitle
\begin{abstract}
This paper considers maximizing the network-wide minimum supported
rate in the downlink of a two-cell system, where each base station
(BS) is endowed with multiple antennas. This is done for different
levels of cell cooperation. At one extreme, we consider single cell
processing where the BS is oblivious to the interference it is
creating at the other cell. At the other extreme, we consider full
cooperative macroscopic beamforming. In between, we consider
coordinated beamforming, which takes account of inter-cell
interference, but does not require full cooperation between the BSs.
We combine elements of Lagrangian duality and large system analysis
to obtain limiting SINRs and bit-rates, allowing comparison between
the considered schemes. The main contributions of the paper are
theorems which provide concise formulas for optimal transmit power,
beamforming vectors, and achieved signal to interference and noise
ratio (SINR) for the considered schemes. The formulas obtained are
valid for the limit in which the number of users per cell, $K$, and
the number of antennas per base station, $N$, tend to infinity, with
fixed ratio $\beta = K/N$. These theorems also provide expressions
for the effective bandwidths occupied by users, and the effective interference caused in the adjacent cell, which allow direct
comparisons between the considered schemes.
\end{abstract}


\section{Introduction}
\label{sec:intro}

\subsection{Problem scope}

Consider the downlink (DL) of a cellular system in which a base station (BS)
services many mobiles within the cell. If the BS is
equipped with multiple antennas we have the classic MIMO broadcast
channel (BC), which has been the focus of much attention in the past few
years, including the recent, celebrated characterization of the
capacity region using dirty paper coding \cite{weingarten_it06}. There has also been
substantial interest in devising suboptimal, but practical
approaches based on linear precoding techniques (i.e.
beamforming).

The MIMO BC  is the appropriate model for a single
isolated cell. What happens when we bring several cells together, so
that each is affected by the interference from the others? This
paper analyzes the performance of linear precoding in a multicell setting.

Consideration of interference leads us to examine the system level architectural issue of cooperation between nodes in the network. The
traditional approach to interference is to partition the cells in  time or bandwidth to avoid a strong interference coupling between adjacent cells. However, with multiple antennas at each base
station, this may be suboptimal.  When multiple antennas are incorporated at each BS, we can tradeoff the maximization of the rates of the in-cell users
(ignoring interference), with the minimization of the interference spilled over into the other cells. If enough cooperation is enabled, these two objectives can go hand in hand.

A large body of research has recently dealt with cooperation and coordination in multicell systems. Many papers are concerned with
developing new algorithms, to meet various proposed performance objectives (e.g. transmit power minimization under given SINR
constraints for coordinated beamforming (CBf) \cite{dahrouj_iss08,dahrouj_tw10}, or minimum SINR maximization for network MIMO \cite{karakayali_icc06}). Others provide a performance analysis of a given scheme under a particular channel model.

One can distinguish between scenarios where BSs each serve a
different group of users, and cases
where all the transmitters jointly transmit to all users in the
system, the so-called multicell processing (MCP), macrodiversity or
network MIMO. For MCP, a classical model for performance analysis is
Wyner's model \cite{jing_eurasip08, somekh_it09}. This model was
first used on the DL in \cite{shamai_vtc2001}, where a linear
pre-processing dirty paper coding approach is proposed. In
\cite{somekh_it09}, the sum rate is characterized for the case where
single-antenna base stations pool their antennas together to perform
zero-forcing (ZF) to the users in the system. More precisely, they
consider a circular variant of the infinite linear Wyner model for
both non-fading and fading scenarios, with scheduling based on local
channel statistics. Results are derived for the regime in which the
number of cooperating BSs tends to infinity, and scaling results are
also obtained by letting the number of users per cell do the same.

In this work, we focus on optimizing linear precoding under
different states of CSI and data sharing. We assume that each BS can
simultaneously serve more than one user, and we formulate the
problem of maximizing the minimum network-wide achievable rate, i.e.
rate balancing, under the following three different architectures:
\begin{enumerate}
\item Single cell processing (SCP), in which each BS has perfect CSI about mobiles in its cell, but no knowledge
about the channels to mobiles in other cells. Here, the BS can control interference between the mobiles in its
own cell, but not the interference spill-over to other cells.
\item CBf, in which the beamforming decisions at each BS take into account the impact of interference on
the other cells. In this approach, each BS has CSI about the channels to its own users, but it also
knows the channels to the mobiles in the other cells. This allows the
BS to control the interference it causes to the mobiles in other cells.
\item MCP, in which the BSs cooperate to jointly precode signals to all the mobiles in all
cells. Through cooperation, the BSs know the channels from all BSs to all the mobiles in the whole network,
and hence they can jointly beamform, using their combined resources.
\end{enumerate}
In the first two approaches, user data is routed to a single BS
only, whereas in the MCP case, it is routed to all cooperating BSs.
We specialize our derivations to only two
cells. Nevertheless, such an approach can be extended to larger
networks, if one wishes to perform system-wide optimization of a
large cellular network. This is the object of ongoing research \cite{ZH_isit11}.

Even the two cell model gets very complicated when we have multiple
antennas at the BSs and many mobiles in each cell. With
independent flat fading between each transmitter-receiver pair,
many parameters need to be specified. For this
reason, we assume that they are selected randomly, and we take a
large system approach in which the number of antennas at the base
station, $N$, and the number of mobiles in each cell, $K$, both grow
large together, while the ratio $\frac{K}{N}$, which we refer to as
cell loading, is held at a constant, denoted $\beta$.

\subsection{Related work}\label{intro:related}

Random matrix theory has received a lot of attention in the
communications literature recently \cite{tulino04}, particularly
large system results where some of the system parameters, such as
number of users, the length of the signature sequence (in CDMA
scenarios), or the number of transmit and receive antennas (in MIMO
settings), are allowed to grow large at the same rate. Such
asymptotics often produce a compact characterization of performance
in the large system limit, amenable to system optimization. An
attractive feature of these results is that the asymptotic
expressions turn out to be good estimates for even relatively low
values of the scaling parameters \cite{tse_it1999, dai_sp03,
tulino_it05, aktas_it06, dumont_it2010}.

Most applications of large system analysis have been for the
uplink (UL). Surprisingly, there has been 
 limited work on the
DL until quite recently, despite a well established UL-DL duality theory.
Recently, however,
 a large systems analysis of regularized ZF (RZF)
beamforming was carried out to characterize its limiting performance in a
single cell context, allowing the optimization of the regularization
parameter \cite{nguyen_globecom08}, and \cite{couillet_WDS08} considers ZF and RZF for correlated channel models.
The present paper generalizes \cite{nguyen_globecom08} to the multicell context, and to a
wider class of beamformers, exploiting an UL-DL duality theory.

In the past year or so, there has been further work that
explores the interplay between UL-DL duality, Lagrangian optimization and large systems analysis. In \cite{Couillet_2010}, duality between MAC and BC is used to characterize and optimize asymptotic ergodic capacity for correlated channels. A similar approach is taken in \cite{caire_icc10,caire_ciss10,caire_isit10} to treat a large systems analysis of ergodic, weighted sum-rate maximization. These papers consider maximizing network utility to obtain fairness, under user scheduling, and they consider various forms of base station cooperation (clustering). The aim is to use large system analysis to obtain numerical methods that are much more efficient than Monte-Carlo simulations; in \cite{caire_isit10} they obtain an ``almost closed-form'' numerical analysis tool. Linear ZF beamforming and non-ideal CSIT are considered in \cite{caire_ciss10}. In \cite{lakshminaryana_pimrc10} random matrix theory is applied to a different CBf setup than the one considered here: more particularly, they consider the problem of weighted sum of the transmit powers minimization CBf problem initially formulated in \cite{dahrouj_iss08}, and propose a strategy which also requires instantaneous local CSI and sharing channel statistics.  Monte Carlo simulations are resorted to in order to claim asymptotical optimality of the results; these are however derived for a more general channel model than we use in the present paper. In \cite{zakhour_allerton10}, we provided preliminary results that we now present in more detail.

There has also been some very recent interest in using large systems analysis to succinctly answer questions concerning channel uncertainty, optimal amount of training, and required rates of feedback of channel state information. These questions are addressed for the downlink of a single cell in \cite{wagner_arxiv10}, where deterministic equivalents of SINRs for ZF and RZF are derived, under channel uncertainty, and the resulting expressions are then optimized with respect to the number of users and the number of symbols devoted to training. In \cite{hoydis_arxiv10}, large systems analysis is used to optimize the number of BSs that should be cooperating on the UL, taking into account unreliable backhaul links, and the channel estimation required to measure the channels. Both papers consider much more general channel models than we do in this work.

Many duality results have been established in the context of MIMO
communications. The first UL-DL duality result was obtained for the
point to point MIMO channel in \cite{telatar_99}. Another early
work, \cite{farrokhi_jsac98}, considered joint optimization of power
and beamforming vectors for the DL of a multiple antenna cellular
system employing a simple linear transmission strategy followed by
single user receivers, such that the SINR at each mobile is above a
target value; the proposed algorithm achieves a feasible solution
for the DL if there is one and minimizes the total transmit power
in the network. In \cite{farrokhi_jsac98}, connections were made
between this problem, and the UL power control framework of
\cite{yates95}, and we exploit these connections in the present
paper (as have many other authors). In \cite{visotsky_vtc99}, the
problem of DL power minimization subject to target SINR constraints
is addressed using a Lagrangian duality framework, and the transmit
powers of the dual UL are found to correspond to the Lagrange
coefficients associated with SINR constraints.

In the context of the capacity region of the Gaussian BC, duality
results are provided in
\cite{viswanath_it03,vishwanath_it03,yu_it04,yu_it06}. When linear
beamforming is considered, as in the present paper, several 
duality results have been obtained, 
and applied to designing iterative algorithms for DL beamforming. It was
shown in \cite{tse_isit02} that, under a sum power constraint,  UL and DL achievable rate regions are the same. 
 Effective bandwidth results derived for the UL
were also proven to hold in the DL. 
 A similar approach to 
 duality for the BC is taken in \cite{boche_vtc02}.

Over the past decade, a large body of research has appeared in which
beamformers for the BC are derived via Lagrangian
optimization techniques. Optimization formulations include the
minimization of the sum of mean-squared errors
\cite{vojcic_com98,baretto_icc01,choi_veh04,joham_sp05}, the
minimization of power subject to SINR constraints
\cite{wiesel_sp06}, and the maximization of SINR's subject to power
constraints \cite{schubert_vt04, wiesel_sp06}. In
\cite{boche_icc03}, the problems of maximizing the sum of effective
bandwidths in both the UL and DL (via duality) are considered. For
the two-user case, a closed form solution may be obtained, whereas
the more general problem can be formulated as a convex optimization
problem and solved via an interior point method. Related problems
such as SINR balancing are also considered. In \cite{yu_sp07}, the
Lagrangian duality approach is extended to include per-antenna power
constraints. It is shown that the dual problem corresponds to an UL
with uncertain noise at the receiver, and the focus is the
derivation of various algorithms for solving the power minimization
problem.

Some of the above-mentioned formulations are not directly convex problems;
for example the DL beamforming problem to minimize power subject
to SINR constraints is not a convex problem. However, it can be
transformed into a convex problem, as shown in \cite{wiesel_sp06}, which
shows that strong duality holds in the original non-convex problem \cite{yu_sp07}.
Thus Lagrangian methods can be applied to these non-convex problems, and
this fact is exploited in the present work.

The optimization approach to beamforming provides much insight into
questions of system capacity, but when it comes to practical
beamforming design other considerations also come into play. For
example, using an iterative algorithm to find an optimal beamformer
may not be viable if channels change rapidly over time.\footnote{The classical
Foschini-Miljanic\cite{foschini_miljanic93} power control algorithm  attempts to find the
optimal power levels in a dynamic environment, but with beamforming there
are many more parameters to be determined.}
For this
reason, many authors have focused on beamformers that can
be found without iterative methods, such as the classical ZF
beamformer.\footnote{By ``iterative'' we mean schemes that require
iterative updates of physical parameters, such as transmit powers
and beamforming vectors, not simply numerical methods that can be
implemented within a single network node, {\it e.g.} in the
computation of a matrix inverse.} Of course, rather than adapt to
the instantaneous channel realizations, an alternative approach is
to adapt to changing channel {\it statistics}. Stochastic approximation
methods have a long history, and provide 
 iterative algorithms
for these scenarios. It is of interest to extend the results of the present
paper to smaller (non-asymptotic) systems and such methods may prove useful.

ZF beamforming 
 steers nulls at the other users so that
they each get an interference-free version of their desired signal. However, the signal to noise penalty can be very high if the
matrix to invert is ill-conditioned. RZF beamforming
adds a regularization term to the ZF beamformer to provide
numerical stability and better performance \cite{peel_com05}. As noted earlier, \cite{nguyen_globecom08} undertakes a large system analysis of RZF, exploiting how its
beamformer resembles the LMMSE receiver on the UL. 

A number of works have recognized that the optimal
beamformer can have a RZF structure in some
special cases
\cite{vojcic_com98,baretto_icc01,choi_veh04,schubert_vt04,joham_sp05,wiesel_sp06}.
Note that for the power minimization subject
to SINR constraints problem \cite{wiesel_sp06}, the simple form of the
regularized beamformer only arises when there is a great deal of
symmetry in the channel model. The examples given in Section VI of
\cite{wiesel_sp06}, where such a form emerges, consist of
the diagonal case (i.e. no interference between channels)
and the symmetric case, when the channel matrix has equal diagonal
elements, and equal off-diagonal elements. In the present paper,
channel matrices are randomly selected, so this particular symmetry
condition is not satisfied. However, our model is symmetric in the
average statistics of the channel matrices,
which leads to the RZF beamformer resulting from
the large systems limit. Nevertheless, this form of symmetry  is more general than the very special one considered in \cite{wiesel_sp06}.

In general, \cite{wiesel_sp06} uses iterative techniques to find
numerical solutions for the optimal beamformer using conic
optimization techniques. The latter are exploited in
\cite{yu_sp07} to handle per-antenna power constraints, and extended
further in \cite{dahrouj_iss08,dahrouj_tw10} to provide a duality
theory for multicell systems in which there are per-BS
power constraints. In \cite{dahrouj_iss08,dahrouj_tw10}, CBf, a novel beamforming strategy 
 in which the BS takes into account the interference it creates
in adjacent cells,  is proposed. \cite{dahrouj_tw10} focuses on
developing fast algorithms to find optimal beamformers, and
numerical evidence is provided to show the improved
performance due to coordination, relative to traditional
SCP. 

The RZF structure emerging from our symmetric two cell model
is reminiscent of beamformers 
 designed to optimize
other criteria, such as minimum variance distortionless
response (MVDR) beamformers. In particular, there is a similarity
between the RZF structure and the so-called diagonal loading
structure that has been used to deal with 
 uncertainty in the
noise correlation matrix \cite{A81}. In \cite{ML06}, a large systems
analysis of MVDR beamformers with diagonal loading is undertaken
where the number of antennas, and the number of observations, grow
large together. Deterministic equivalents are obtained, allowing the
diagonal loading factor to be optimized.

In Section~\ref{c5:sec:opt} we review the optimization problems
associated with the classical (non-coordinated) SCP, and the CBf strategy of
\cite{dahrouj_tw10}. We also formulate the optimization problem for
MCP in which the BS antennas jointly
coordinate their transmissions, as in network MIMO
\cite{karakayali_wc06}.

The main contribution of the paper is a large systems analysis of
these three schemes, and optimizations based on the limiting
asymptotics. In particular, we provide a clean characterization of
each of the three schemes in the large systems limit in terms of
effective bandwidths. In the case of SCP, we show that the
RZF beamformer is asymptotically optimal in the large
systems limit.\footnote{the structure that emerges is more general than RZF, since it also applies in regimes in which ZF is not defined.} In the case of CBf, we show that the asymptotic form of the optimal beamformer provides an enhanced
version of regularized ZF, one that we call generalized regularized
zero-forcing (GRZF).\footnote{again, it also applies when ZF is not defined.} This beamformer is a novel contribution of the
present paper.
\textbf{Notation: } Lowercase letters,  boldface lowercase letters and boldface uppercase letters are used to represent scalars, vectors and matrices, respectively. $\mathbf{x}^T$
 and $\mathbf{x}^H$ are the transpose 
 and  conjugate transpose of vector $\mathbf{x}$, respectively. $\mathcal{CN}(\boldsymbol{\mu}, \mathbf{C})$ denotes a circularly symmetric complex Gaussian random vector of mean $\boldsymbol{\mu}$ and covariance matrix $\mathbf{C}$.
Finally $\mathbf{S} \succeq 0$ means $\mathbf{S}$ is a positive semidefinite (PSD) matrix.

\section{System Model}
\begin{figure}[htp]
\centering
\includegraphics[width=2.8in,height=1.8in,viewport=50 375 500 600,clip]{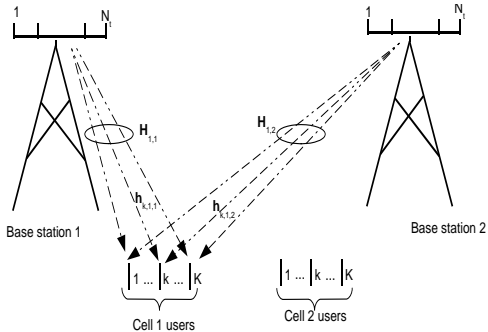}
\vspace{-.2cm}
\caption{System model} \label{fig:Scenario}
\end{figure}

Our model,  illustrated in Figure \ref{fig:Scenario}, has two cells, and each BS is equipped with an
array of $N$ antennas. There are $K$ single-antenna mobiles in each cell. We assume flat fading and
adopt the following notation regarding channel coefficients:
\begin{enumerate}
\item the channel vector from BS $j$ to user $k$ in cell $j'$ is denoted $\mathbf{h}_{k,j', j}$, where $\mathbf{h}_{k, j', j} \in \mathbb{C}^{1\times N}$;
\item the channel vector from all BSs to user $k$ in cell $j'$ is denoted $\mathbf{\tilde{h}}_{k,j'}$; in other words,
$\mathbf{\tilde{h}}_{k, j'} = [\mathbf{h}_{k, j', 1} \quad \mathbf{h}_{k, j', 2}]$.
\end{enumerate}

The data symbols to be transmitted to each user are assumed to
be independent identically distributed (i.i.d.)
$\mathcal{CN}(0,1)$ random variables (r.v.), and the data symbol for user $k$ in cell $j$ is
denoted $s_{kj}$. Let $\mathbf{s}_j = [s_{1j}~ \ldots ~
s_{Kj}]^T$ and $\mathbf{s} = [\mathbf{s}_1^T ~\mathbf{s}_2^T]^T$.
The received signal at user $k$ in cell $j$ is given by
\begin{align}
y_{k,j} =
\sum_{j'=1}^2 \mathbf{h}_{k,j, j'} \mathbf{x}_{j'} +
n_{k,j},
\end{align}
where $\mathbf{x}_{j'} \in \mathbb{C}^{N}$ denotes BS
${j}'$'s transmit signal, which consists of the linearly precoded
symbols of the users it is serving, and which is subject to the
average power constraint $\mathbb{E}
\left[\mathbf{x}_{j'}^H\mathbf{x}_{j'}\right] \le P$ and
$n_{k,j} \sim \mathcal{CN}(0, \sigma^2)$ is the receiver noise. Receivers perform single-user detection, i.e.
treat interfering signals as noise. The way the precoding vector
$\mathbf{x}_{j'}$ is generated depends on the considered
scheme; See Section~\ref{c5:sec:opt}.

We assume the channels between each BS and user are independent.
Moreover, channels between a user and his serving BS are
i.i.d. $\mathcal{CN}(\mathbf{0}, \mathbf{I})$ whereas channels
between a user and an interfering BS are  i.i.d. $\mathcal{CN}(\mathbf{0},  \epsilon\mathbf{I})$. Thus, $\epsilon$
controls the interference level between neighbor cells. This
is a simplified model of a cellular network, yet provides useful insights
\cite{jsac_tute10}. It is the two cell DL version of Wyner's
model \cite{wyner_it94}.

\section{Cooperation and Linear Beamforming Schemes: Primal problems}\label{c5:sec:opt}
We consider the problem of maximizing the network-wide minimum
achievable rate for three different degrees of cooperation and
coordination between the cells. The optimization to be
carried out in each case is presented.
The first problem is the now classic one treated in
\cite{farrokhi_jsac98}, the second is the cooperative scheme
proposed in \cite{dahrouj_tw10}, and the third is new, although
similar approaches have been proposed in several papers (e.g.
\cite{karakayali_wc06}). In the equations below, when we consider a
particular BS $j$, index $\bar{j} = \mod(j, 2)+1$ corresponds to the {\it other} BS.

\subsection{Single cell processing}
This is the conventional case where each BS serves its own users
without worrying about the other cell. Here, we assume full re-use of time and spectrum across cells.
$\mathbf{x}_j$ is of the form:
\begin{align}
\mathbf{x}_j = \sum_{k=1}^K \mathbf{w}_{kj} s_{kj} = \mathbf{W}_j \mathbf{s}_j, \label{eq:SC_bf_tx_signal}
\end{align}
where the symbol for user $k$ in cell $j$, $s_{kj}$, is linearly precoded by
beamforming vector $\mathbf{w}_{kj}$. 
$\mathbf{W}_j$ is the concatenation of the $K$ precoding vectors in cell $j$ into
a $N \times K$ matrix, the $k$th column being $\mathbf{w}_{kj}$.

In cell $j$ the problem to be solved is the following:
\begin{align}
&\textrm{max.}_{\gamma, \mathbf{w}_{kj}, k=1, \ldots, K} \quad \gamma \nonumber \\
& \textrm{s.t. } \quad \textrm{SINR}_{k,j} \ge \gamma, \quad  k = 1, \ldots, K \nonumber \\
& \quad \quad \quad \sum_{k=1}^K \|\mathbf{w}_{kj}\|^2 \le P. \label{eq:MaxMinSINR_SCP}
\end{align}

The SINR at user $k$ in cell $j$ is given by
\begin{align}
\textrm{SINR}_{k,j} = \frac{|\mathbf{h}_{k,j,j}
\mathbf{w}_{kj}|^2}{\sigma_{k,j}^2 + \sum_{k'=1,k'\neq
k}^K |\mathbf{h}_{k,j,j} \mathbf{w}_{k'j}|^2}, \label{SCP:SINR_eq}
\end{align}
$\sigma_{k,j}^2$ is the noise plus other-cell interference power
at that user:
\begin{align}
\sigma_{k,j}^2 = \sigma^2 + \sum_{k'=1}^K |\mathbf{h}_{k,j,\bar{j}} \mathbf{w}_{k'\bar{j}}|^2, \label{eq:c5:sigma_kj}
\end{align}
and needs to be fed back to BS $j$.

Solving \eqref{eq:MaxMinSINR_SCP} may require an iterative procedure, since beamforming at each BS influences the interference, and therefore the transmission design, at the other. We focus on finding the maximum SINR that can be met in \emph{both} cells.
This can be obtained using a bisection method. Thus, for
fixed $\gamma$ we obtain the beamforming vectors by minimizing total
transmit power subject to SINR constraints on the cell's users.
If the optimum is $\le P$ for both BSs, $\gamma$ is
attainable.
We thus focus on solving:
\begin{align}
& \textrm{min.}_{\mathbf{w}_{jk}, j=1, \ldots, K} \quad  \sum_{k=1}^K \|\mathbf{w}_{kj}\|^2 \nonumber \\
& \textrm{s.t.} \quad \quad \textrm{SINR}_{k,j} \ge \gamma, \quad  k = 1, \ldots, K.\label{eq:MaxMinSINR_SCP_feasibility}
\end{align}

 For $\gamma$ achievable with unlimited transmit power,
 its solution will be a set of beamforming vectors that minimizes the total cell transmit power, and achieves SINR $\gamma$. Further, in Theorem~\ref{THEO_BF_SCP}, we will provide
a condition that is both necessary and sufficient for the target SINR to be achievable, given unlimited power.
The maximum $\gamma$ is the target SINR for which the transmit power constraint is met with equality.

\subsection{Coordinated Beamforming}
Here, each BS sends data to its own users
only, as in SCP, but CSI is shared between the BSs so that the interference generated in other cells is
taken into consideration. $\mathbf{x}_j$
will be similar to that in \eqref{eq:SC_bf_tx_signal},
although the precoding  design differs.

The problem formulation in \eqref{eq:MaxMinSINR_SCP} becomes
\begin{align}
& \textrm{max.}_{\gamma, \mathbf{w}_{kj}, k=1, \ldots, K, j = 1, 2} \quad \gamma \nonumber \\
& \textrm{s.t.} \quad \textrm{SINR}_{k,j} \ge \gamma, \quad  k = 1, \ldots, K, j = 1, 2 \nonumber \\
& \quad \quad \sum_{k=1}^K \|\mathbf{w}_{kj}\|^2 \le P, \quad j = 1, 2,
\label{eq:MaxMinSINR_Coord}
\end{align}
which is a joint, two-cell optimization problem, requiring a
coordinated solution. The SINR at user $k$ in cell $j$ is similar to \eqref{SCP:SINR_eq} but can be expanded into
\begin{align}
\textrm{SINR}_{k,j} = \frac{|\mathbf{h}_{k,j,j} \mathbf{w}_{kj}|^2}{\sigma^2 + \sum_{j'=1}^2\sum_{k'=1, (k', j') \neq (k, j)}^K |\mathbf{h}_{k,j,j'} \mathbf{w}_{k'j'}|^2},
\end{align}
since all the channels are centrally known.

Here too the problem may be solved by a bisection method. To
determine feasibility of a given $\gamma$, we solve (as in
\cite{dahrouj_tw10})\footnote{Note that the constant $2 P$ factor in the objective function is included as it leads to $P$ being eliminated from the dual problem formulation.}:
\begin{align}
& \text{min.}_{\phi > 0, \mathbf{w}_{kj}, k=1, \ldots, K, j = 1, 2} \quad  2 P \phi  \nonumber \\
& \text{s.t.}\quad  \textrm{SINR}_{k,j} \ge \gamma,  \quad k = 1, \ldots, K, j = 1, 2\nonumber \\
& \quad \quad  \sum_{k = 1}^K \mathbf{w}_{kj}^H\mathbf{w}_{kj} \le \phi P, \quad \forall j = 1, 2, \label{eq:MaxMinSINR_Coord_feasibility}
\end{align}
where $\phi P$ upper bounds the power expended at each BS, and the aim is to use the minimum power level. Clearly, the maximum network-wide achievable SINR is the $\gamma$ for which the optimal $\phi=1$.

\subsection{Multicell processing}
This is the case where the BSs cooperate fully: both CSI and data is available at all transmitters, who pool their antennas together to serve the users jointly.
The transmitted signal $\mathbf{x} = [\mathbf{x}_1; \mathbf{x}_2]$ will be of the form
\begin{align}
\mathbf{x} = \sum_{j=1}^2 \sum_{k=1}^K \mathbf{w}_{kj} s_{kj} = \mathbf{W} \mathbf{s},
\end{align}
where 
$\mathbf{W} \in \mathbb{C}^{ 2N
\times 2K}$ is the overall precoding matrix. The MCP optimization problem is
\begin{align}
&\textrm{max.}_{\gamma, \mathbf{w}_{kj}, k=1, \ldots, K, j = 1, 2} \quad \gamma \nonumber \\
& \textrm{s.t.} \quad \textrm{SINR}_{k,j} \ge \gamma, \quad k = 1, \ldots, K, j = 1, 2 \nonumber\\
& \quad \quad \sum_{j'=1}^2 \sum_{k'=1}^K
\|\mathbf{E}_{j}\mathbf{w}_{k',j'}\|^2 \le P, ~~j= 1, 2,
\label{eq:MaxMinSINR_JTx}
\end{align}
where matrix $\mathbf{E}_{j}, j = 1, 2$ is diagonal and used to select
the elements of each beamforming vector corresponding to BS $j$ (i.e. its non-zero diagonal elements occupy locations
$(j-1)N+1$ to $jN$). The MCP SINR at user $k$ in cell $j$ is 
\begin{align}
\textrm{SINR}_{k,j} = \frac{|\mathbf{\tilde{h}}_{k,j} \mathbf{w}_{kj}|^2}{\sigma^2 + \sum_{j'=1,k'=1, (k', j') \neq (k, j)}^{2, K} |\mathbf{\tilde{h}}_{k,j} \mathbf{w}_{k'j'}|^2}.
\end{align}

Once again, the above problem may be solved by the bisection method.
To determine feasibility of a given $\gamma$, we solve, as in the CBf case, the following
optimization problem:
\begin{align}
\textrm{min.}\quad & 2 P\phi  \nonumber \\
\textrm{s.t.} \quad & \mathrm{SINR}_{k,j} \ge \gamma, \quad k = 1, \ldots, K, j = 1, 2 \nonumber  \\
& \sum_{j'=1}^2 \sum_{k' = 1}^K \|\mathbf{E}_j
\mathbf{w}_{k'j'}\|^2 \le \phi
P,~~j=1,2.\label{eq:MaxMinSINR_JTx_feasibility}
\end{align}

\section{Solution via Duality theory}
The 
 above optimization problems are non-convex, and apparently nontrivial to solve. Similar problems of power
control and receiver optimization on the UL are easier to solve, since each UL receive vector can be individually optimized. As discussed in Section \ref{intro:related}, motivated by this observation, a number of duality results have
been established connecting DL optimization problems, to corresponding dual UL ones. Most relevant to us is the elegant duality theory developed by Yu and
Lan~\cite{yu_sp07} to handle per-antenna power constraints, exploiting earlier work on conic optimization~\cite{wiesel_sp06}.

Algorithms analogous to those proposed in \cite{yu_sp07} can be applied to solve the above problems, but even after these iterative algorithms have converged to the optimal solutions there is a great deal of complexity in the form of the optimal beamformers. Our goal is to use the duality theory to derive suboptimal (but, in the context of our simplified
model, asymptotically optimal) beamformers that have relatively simple structures. In fact, generalized regularized beamformers will
emerge from our analysis. Our approach will be to apply a large systems analysis to the dual, virtual UL optimization problems.
Following \cite{yu_sp07}, we begin by writing down the dual
problems to \eqref{eq:MaxMinSINR_SCP_feasibility},
\eqref{eq:MaxMinSINR_Coord_feasibility}, and \eqref{eq:MaxMinSINR_JTx_feasibility}, respectively.

\subsection{Dual UL SCP problem}
Let $\lambda_{kj}/N \ge 0$ denote the Lagrange multipliers corresponding to the SINR constraints in
\eqref{eq:MaxMinSINR_SCP_feasibility}. The dual problem is
\begin{align}
& \text{max.}_{\lambda_{kj} \ge 0, k = 1, \ldots, K} \quad
\frac{1}{N} \sum_{k=1}^K \lambda_{kj} \sigma^2_{k,j}  \nonumber \\
& \text{s.t. } \quad
\mathbf{I} - \frac{\lambda_{kj}}{N \gamma}\mathbf{h}_{k,j,j}^H\mathbf{h}_{k,j,j} + \frac{1}{N} \sum_{k' \neq k} \lambda_{k'j}
\mathbf{h}_{k', j,j}^H\mathbf{h}_{k', j,j} \succeq 0, \nonumber \\
&\quad \quad \quad \quad \quad \quad \quad \quad \quad \quad \quad \quad \quad \quad \quad k = 1, \ldots, K. \label{eq:SCP_dual}
\end{align}
As explained in \cite{yu_sp07}, strong duality holds despite non-convexity\footnote{because the non-convex problem can be transformed into a convex problem using the techniques in \cite{wiesel_sp06}}, so the optimal value in the dual problem is equal to that of
the primal DL beamforming problem.

The dual variables $\lambda_{kj}/N$ may be thought of as dual UL transmit powers, in the following dual
UL power control and beamforming problem \cite{yu_sp07}:
\begin{align}
& \text{min.}_{\mathbf{\hat{w}}_{kj}, \lambda_{kj}, k = 1, \ldots, K} \quad
\frac{1}{N} \sum_k \lambda_{kj} \sigma^2_{k,j}  \nonumber \\
& \text{s.t.} \quad  \max_{\mathbf{\hat{w}}_{kj}} \frac{\lambda_{kj}\left|\mathbf{h}_{k,j,j}\mathbf{\hat{w}}_{kj}\right|^2}{
N \mathbf{\hat{w}}_{kj}^H\mathbf{\hat{w}}_{kj} + \sum_{k' \neq k} \lambda_{k'j}
\left|\mathbf{h}_{{k}', j,j}\mathbf{\hat{w}}_{kj}\right|^2} \ge \gamma,  \nonumber \\
& \quad \quad \quad \quad \quad \quad \quad \quad \quad \quad \quad \quad \quad \quad \quad  k = 1, \ldots, K.  \label{eq:SCP_dual_uplink}
\end{align}
Indeed,
\eqref{eq:SCP_dual} can be shown to have the same optimal values as \eqref{eq:SCP_dual_uplink}.
The optimal $\mathbf{\hat{w}}_{kj}$, up to a scalar, are given by
\begin{align}
\left( \mathbf{I} +
\sum_{k' \neq k} \frac{\lambda_{k'j}}{N}
\mathbf{h}_{k',j,j}^H\mathbf{h}_{k', j,j} \right)^{-1}
\mathbf{h}_{k,j,j}^H. \label{eq:SCP_opt_bf}
\end{align}
which we recognize to be MMSE filters for the UL problem.

To see why these problems are equivalent, note that the SINR constraints in the UL problem \eqref{eq:SCP_dual_uplink} become, after substituting for the MMSE filters:
\begin{align}
\frac{\lambda_{kj}}{N} \mathbf{h}_{k,j,j}
\left( \mathbf{I} + \sum_{{k}' \neq k}
\frac{\lambda_{{k}'j}}{N} \mathbf{h}_{k', j,j}^H\mathbf{h}_{k',
j,j} \right)^{-1} \mathbf{h}_{k,j,j}^H \geq \gamma. \label{eq:SCP_uplink_SINR}
\end{align}
The optimal $\lambda_{kj}$s are the unique solution to the fixed
point equation:
\begin{align}
\lambda_{kj} = \frac{\gamma N} {\mathbf{h}_{k,j,j} \left( \mathbf{I} +
 \sum_{k' \neq k} \frac{\lambda_{k'j}}{N}
\mathbf{h}_{k', j,j}^H\mathbf{h}_{k', j,j} \right)^{-1}
\mathbf{h}_{k,j,j}^H}. \label{eq:SCP_int_fn}
\end{align}
Since $\gamma$ is achieved exactly, the SINR constraints in \eqref{eq:SCP_dual_uplink} are achieved with equality. It follows that if the minimization in \eqref{eq:SCP_dual_uplink} is changed to a maximization, and the SINR inequalities are reversed, then the same solution will be found. It is shown in \cite{yu_sp07} that in so doing, one obtains 
 \eqref{eq:SCP_dual}.

The right hand side of \eqref{eq:SCP_int_fn} is an {\it interference
function} of the dual UL powers, in the sense of Yates'
framework on UL power control \cite{yates95}. Thus, iterative
approaches to power control converge to the optimal dual
variables. Indeed, this was the original approach to the multicell
beamforming problem taken in \cite{farrokhi_jsac98}, one of the
first papers to exploit an UL-DL duality. More recently,
more efficient approaches to solving the dual problem have been
proposed \cite{yu_sp07}, but these still require several iterations before the algorithm converges, which could be problematic. The large systems analysis in Section~\ref{c5:sec:LAS} allows us to derive simple yet asymptotically
optimal beamformers.

From the Karush-Kuhn-Tucker (KKT) conditions, one can show that the optimal $\mathbf{w}_{kj}$ for the primal are the same, up to a scaling factor, as $\mathbf{\hat{w}}_{kj}$. Thus, 
they can be written as
\begin{align}
\mathbf{w}_{kj}  
= \sqrt{\frac{p_{kj}}{N}} \frac{\mathbf{\hat{w}}_{kj}}{\|\mathbf{\hat{w}}_{kj}\|}, \label{SCP:bf_vec_form}
\end{align}
where 
$\frac{p_{kj}}{N}$ is the transmit power allocated to beamforming vector
$\mathbf{w}_{kj}$ on the DL. From the DL SINR constraints, for $k=1, \ldots, K$,
\begin{align}
&\frac{p_{kj}}{N \gamma} \frac{\left|\mathbf{h}_{k,
j,j}\mathbf{\hat{w}}_{kj} \right|^2}{\|\mathbf{\hat{w}}_{kj}\|^2} -
\sum_{k' \neq k} \frac{p_{k'j}}{N}
\frac{\left|\mathbf{h}_{k, j,j}
\mathbf{\hat{w}}_{k'j}\right|^2}{\|\mathbf{\hat{w}}_{k'j}\|^2}
= \sigma^2_{k,j},  \label{eq:sinr_scp}
\end{align}
and the $\left(p_{kj}\right)_{k=1}^K$ can be determined from this
set of equations. We undertake a large system analysis of this scheme in Section~\ref{c5:sec:LAS}, and present the results in Theorem~\ref{THEO_BF_SCP}.

As a final remark, we note that the UL-DL duality described in this section is not the same as the celebrated UL-DL duality used to characterize the capacity region of the MIMO BC. In the present section (and indeed throughout this paper), 
 we do not allow so-called dirty paper coding (DPC), which is the coding technique that achieves the capacity of the MIMO BC. The original notion of UL-DL duality was a correspondence between points in the capacity region of the MIMO BC with points in a dual UL multiple access channel with suitably chosen noise covariances. However, it has been shown that the optimization approach developed in \cite{yu_sp07} to obtain the optimal beamforming vectors for the DL power minimization problem (as reviewed in the present section) can be extended to allow DPC, and hence to obtain the MIMO BC capacity results, including the setup with per antenna power constraints (see \cite{yu_sp07}).

\subsection{Dual UL CBf problem}\label{sec:dual:CBf}
Letting $\lambda_{kj}/N \ge 0$ denote the Lagrange multipliers corresponding to the SINR constraints and $\mu_j$ those corresponding to the power constraints in
\eqref{eq:MaxMinSINR_Coord_feasibility}, its Lagrangian dual problem is
\begin{align}
&\text{max.}_{\lambda_{kj}\ge 0, \mu_j \ge 0} \quad
\sum_{j,k} \frac{\lambda_{kj}}{N} \sigma^2 \nonumber \\
& \text{s.t. } \mu_j \mathbf{I} - \frac{\lambda_{kj}}{\gamma
N}\mathbf{h}_{k,j,j}^H\mathbf{h}_{k,j,j}
 \nonumber \\
& \quad
+ \sum_{(k',j') \neq (k,j)} \frac{\lambda_{k'j'} }{N} \mathbf{h}_{k', j', j}^H \mathbf{h}_{k', {j}', j} \succeq 0, \nonumber \\
& \hspace{1in}  \quad k = 1, \ldots, K, j = 1, 2 \nonumber \\
& \quad ~~ \sum_{j=1}^2 (1-\mu_j) P = 0.   \label{eq:coord_dual}
\end{align}

As for the SCP case, this problem can be shown to be equivalent to a dual UL problem with uncertain noise (cf. \cite{yu_sp07}), or equivalently
\begin{align}
&\text{max.}_{\mu_j\ge 0} \text{min.}_{\lambda_{kj}\ge 0, \mathbf{\hat{w}}_{kj}} \quad   \sum_{k,j} \frac{\lambda_{kj}}{N} \sigma^2 \nonumber \\
& \text{s.t.} \quad
\frac{\lambda_{kj}|\mathbf{\hat{w}}_{kj}\mathbf{h}_{k,j,j}|^2}{\mathbf{\hat{w}}_{kj}^H\left[N
\mu_j
\mathbf{I}+\sum_{({k}',{j}') \neq (k,j)} \lambda_{{k}'j'} \mathbf{h}_{k', j', j}^H \mathbf{h}_{k', j', j}\right]
\mathbf{\hat{w}}_{kj}} \geq \gamma, \nonumber \\
& \hspace{2.1in}  \quad k = 1, \ldots, K, j = 1, 2 \nonumber \\
& \quad \quad \sum_{j=1}^2 (1-\mu_j)  = 0.  \label{eq:coord_dual_uplink} 
\end{align}
The Lagrange multipliers $\mu_1, \mu_2$ can be interpreted as noise
levels at the two BSs, respectively, on the virtual UL
\cite{yu_sp07}.

For any choice of dual variables, $(\mu_1, \mu_2)$,
$\left(\frac{1}{N} \lambda_{kj}\right)_{k=1,j=1}^{K,~~2}$, the SINR
achieved on the virtual UL can be shown to be:
\begin{align}
& \frac{\lambda_{kj}}{N} \mathbf{h}_{k,j, j}
\left[\mu_j \mathbf{I} +\sum_{(k',j') \neq
(k,j)} \frac{\lambda_{k'j'} }{N} \mathbf{h}_{k', j', j}^H
\mathbf{h}_{k', j', j}\right]^{-1} \mathbf{h}_{k,j, j}^H.
\label{eq:coord_uplink_SINR}
\end{align}
Using the optimal dual variables, all \eqref{eq:coord_uplink_SINR} equal $\gamma$, which
allows us to write down the optimal
$\left(\lambda_{kj}\right)_{k=1,j=1}^{K,~~2}$ as the unique
solution to a fixed point equation:
\begin{align}
&\lambda_{kj} = \frac{\gamma N}{\mathbf{h}_{k,j, j} \left[\mu_j
\mathbf{I} +\sum_{(k',j') \neq (j,k)}
\frac{\lambda_{k'j'} }{N} \mathbf{h}_{k', j', j}^H
\mathbf{h}_{k', j', j}\right]^{-1} \mathbf{h}_{k,j,
j}^H}. \label{eq:coord_int_fn}
\end{align}
Once $(\mu_1, \mu_2)$ are determined, this equation
 provides an implicit solution to the dual
problem, in that standard iterative methods can be applied to obtain
the dual UL powers. It remains to find $\mu_1$ and $\mu_2$. This is addressed in Section~\ref{c5:sec:LAS}.

The optimal beamforming vectors on the dual UL, $\mathbf{\hat{w}}_{kj}$, assuming
feasibility, are (up to a scaling factor)
\begin{align}
\left[\mu_j \mathbf{I}
+\sum_{(k',j') \neq (k,j)}
\frac{\lambda_{k'j'}}{N} \mathbf{h}_{{k}', {j}', j}^H
\mathbf{h}_{{k}', j', j}\right]^{-1} \mathbf{h}_{k,j, j}^H;
\end{align}
the optimal beamforming vectors on the DL are of the form
$\mathbf{w}_{kj} = 
\sqrt{\frac{p_{kj}}{N}}
\frac{\mathbf{\hat{w}}_{kj}}{\|\mathbf{\hat{w}}_{kj}\|}$,
where $\displaystyle \frac{p_{kj}}{N}$ is the power allocated to the
beamforming vector $\mathbf{w}_{kj}$. From DL SINR constraints, for $j = 1,2, k = 1, \ldots, K$,
\begin{align}
&\frac{p_{kj}}{N \gamma} \frac{\left|\mathbf{h}_{k,
j,j}\mathbf{\hat{w}}_{kj} \right|^2}{\|\mathbf{\hat{w}}_{kj}\|^2} -
\sum_{{j}', {k}', ({k}', {j}') \neq (k,j)}
\frac{p_{{k}'{j}'}}{N} \frac{\left|\mathbf{h}_{k, j,{j}'}
\mathbf{\hat{w}}_{{k}'{j}'}\right|^2}{\|\mathbf{\hat{w}}_{{k}'{j}'}\|^2}
= \sigma^2.  \label{eq:sinr_cbf}
\end{align}

\subsection{Dual UL MCP problem}\label{sec:dual_MCP}
With $\frac{\lambda_{kj}}{N}$'s and $\mu_j$'s defined as in \eqref{sec:dual:CBf}, the Lagrangian dual to problem \eqref{eq:MaxMinSINR_JTx_feasibility} is equivalent to
\begin{align}
& \textrm{max.}_{\mu_j \ge 0} \textrm{min.}_{\lambda_{kj} \ge 0, \mathbf{\hat{w}}_{kj}} \quad \sum_{j=1}^2 \sum_{k=1}^K \frac{\lambda_{kj}}{N} \sigma^2 \label{eq:MCP_DUL_prob} \\
& \textrm{s.t.} ~
\frac{\frac{\lambda_{kj}}{N}|\mathbf{\tilde{h}}_{k,j}\mathbf{\hat{w}}_{kj}|^2}{\mathbf{\hat{w}}_{kj}^H\left[
\sum_{\left(k', j'\right) \neq \left(k, j\right)}
\frac{\lambda_{k'j'}}{N}
\mathbf{\tilde{h}}_{k',j'}^H\mathbf{\tilde{h}}_{k',j'}+ \mathbf{M}
 \right]  \mathbf{\hat{w}}_{kj}} \geq \gamma \label{eq:MCP_DUL_SINR} \\
 & \quad \sum_{j=1}^2 \left(1-\mu_j\right) = 0.  \label{eq:MCP_dual_uplink}
\end{align}
where, to simplify the equations, $\mathbf{M} = \sum_{j'=1}^2 \mu_{j'}
\mathbf{E}_{j'}$ is used.

The optimal $\mathbf{\hat{w}}_{kj}$ is
\begin{align}
\left[ \frac{1}{N} \sum_{\left(k',
{j}'\right) \neq \left(k, j\right)} \lambda_{{k}'j'}
\mathbf{\tilde{h}}_{k',j'}^H\mathbf{\tilde{h}}_{k',j'}
+ \mathbf{M}
 \right]^{-1} \mathbf{\tilde{h}}_{k,j}^H.
\end{align}
Plugging this into the inequality in \eqref{eq:MCP_DUL_SINR}, we
obtain
\begin{align}
\frac{\lambda_{kj}}{N}\mathbf{\tilde{h}}_{k,j}\left[ \sum_{\left(k',
j'\right) \neq \left(k, j\right)} \frac{\lambda_{k'j'}}{N}
\mathbf{\tilde{h}}_{k',j'}^H\mathbf{\tilde{h}}_{k',j'} +
\mathbf{M}\right]^{-1}  \mathbf{\tilde{h}}_{k,j}^H \geq \gamma. \label{eq:SINR_jTx_ul}
\end{align}
%

Define $\mathbf{\breve{H}}_{k,j}$ as 
\vspace{-7mm}
\begin{align}
&\mathbf{\breve{H}}_{k,j} = \left[\begin{array}{c} \mathbf{\breve{h}}_{1,j,j} \\
 \vdots \\ \mathbf{\breve{h}}_{k-1,j,j} \\
  \mathbf{\breve{h}}_{k+1,j,j} \\
\vdots \\
  \mathbf{\breve{h}}_{K,j,j} \\
  \mathbf{\breve{h}}_{1,\bar{j},j} \\
 \vdots \\
 \mathbf{\breve{h}}_{K,\bar{j},j}  \end{array} \right],
\end{align}
where $\mathbf{\breve{h}}_{k,j} = \mathbf{\tilde{h}}_{k,j}\mathbf{M}^{-1/2}$, and let $\mathbf{L}_{k,j}$ is the diagonal matrix whose diagonal entries vector is given by
\begin{align}
\left[\lambda_{1j}; \ldots; \lambda_{(k-1)j}; \lambda_{(k+1)j}; \ldots; \lambda_{Kj}; \lambda_{1\bar{j}}; \ldots; \lambda_{K\bar{j}}\right].
\end{align}
Using this notation, for any choice of dual variables, $(\mu_1, \mu_2)$,
$\left(\frac{\lambda_{kj}}{N} \right)_{k=1,j=1}^{K,~~2}$, the SINR achieved on the virtual UL (i.e. the left-hand side of Eq. \eqref{eq:SINR_jTx_ul}) can be rewritten as
\begin{align}
\frac{\lambda_{kj}}{N} \mathbf{\breve{h}}_{k,j}\left[\frac{1}{N}\mathbf{\breve{H}}_{k,j}^H \mathbf{L}_{k,j} \mathbf{\breve{H}}_{k,j} + \mathbf{I}\right]^{-1} \mathbf{\breve{h}}_{k,j}^H. \label{eq:MCP_uplink_SINR}
\end{align}

With the {\it optimal} dual variables, all \eqref{eq:MCP_uplink_SINR} equal $\gamma$,
which allows us to write down the optimal
$\left(\lambda_{kj}\right)_{k=1,j=1}^{K~~2}$ as the solution of a
fixed point equation:
\begin{align}
\lambda_{kj} &= \gamma N
\left(\mathbf{\breve{h}}_{k,j}\left[\frac{1}{N}\mathbf{\breve{H}}_{k,j}^H
\mathbf{L}_{k,j} \mathbf{\breve{H}}_{k,j} +
\mathbf{I}\right]^{-1} \mathbf{\breve{h}}_{k,j}^H\right)^{-1}.
\label{eq:MCP_int_fn}
\end{align}
As for CBf, once $\mu_1, \mu_2$ are obtained, the
optimal $\left(\lambda_{kj}\right)_{k=1,j=1}^{K~~2}$ are implicitly
given as the unique solution of \eqref{eq:MCP_int_fn}. 
 As the duality gap is zero for a feasible primal, at the optimum,
\begin{align}
2 P \phi = \frac{1}{N} \sum_{j=1}^2 \sum_{k=1}^K \lambda_{kj} \sigma^2.
\end{align}
Moreover, the beamforming vectors in the original problem and in the
dual problem are related as follows:
\begin{align}
\mathbf{w}_{kj} = 
\sqrt{\frac{p_{kj}}{N}}
\frac{\mathbf{\hat{w}}_{kj}}{\|\mathbf{\hat{w}}_{kj}\|}.
\end{align}
Plugging these into the DL SINR constraints provides the solution for the $p_{kj}$:
\begin{align}
\frac{p_{kj}}{N \gamma} \frac{|\mathbf{\tilde{h}}_{k,j}
\mathbf{\hat{w}}_{kj}|^2}{\|\mathbf{\hat{w}}_{kj}\|^2} -\sum_{\left(k', j'\right) \neq
\left(k,j\right)} \frac{p_{k'j'}}{N}
\frac{|\mathbf{\tilde{h}}_{k,j}
\mathbf{\hat{w}}_{k'j'}|^2}{\|\mathbf{\hat{w}}_{k'j'}\|^2}
= \sigma^2. \label{eq:sinr_mcp}
\end{align}

\section{Large system results}\label{c5:sec:LAS}

We proceed to a large system
analysis of the solutions to the above optimization problems, in the
limit as $N,K$ grow large, keeping the cell loading ratio $\beta = \frac{K}{N}$
fixed. The basic idea is to apply large systems analysis techniques
to the dual UL problems, guided by the analysis of similar
UL problems in the literature \cite{tse_it1999,hanly_it2001}.
For example, the SINR expression in \eqref{eq:SCP_uplink_SINR} is
typically analyzed by considering the limit of the empirical
distribution of the eigenvalues of the matrix $\left(\mathbf{I} + \frac{1}{N}
\sum_{k' \neq k} \lambda_{k'j} \mathbf{h}_{k',
j,j}^H\mathbf{h}_{k', j,j} \right)^{-1}$ as $N, K \rightarrow \infty$, with $K/N$ held at $\beta$. Once this is
characterized, a law of large numbers for the trace of this matrix
can be obtained, and from that, the limit of the SINR expression in
\eqref{eq:SCP_uplink_SINR}.

In the present paper, extra technical difficulties arise. To directly apply the known
theorems about large random matrices, we need to assume that the
dual UL powers $\lambda_{kj}$ are {\it independent} of the channel
matrix parameters, and are chosen {\it independently} of each other
from some fixed distribution. However, being solutions of the UL
dual problem, they form a collection of {\it dependent} random
variables which, moreover, depend on the channel matrix parameters.
Our approach to this problem is to provide lower and upper bounds to
the SINR's using {\it deterministic} dual UL powers. We obtain these
bounds by exploiting the underlying monotonicity structure of the UL
power control problem \cite{hanly95,yates95,farrokhi_jsac98}. With
deterministically chosen UL powers, we can perform a large system
analysis, and then provide a sandwich type argument to show that the
optimal dual UL powers must converge to deterministic values. Since
this part of the paper is somewhat technical, we relegate it to the
appendices, and focus
 on the results of the analysis.

\subsection{Asymptotically Optimal Beamformers}

\begin{theorem}[Asymptotically optimal beamforming for SCP]
\label{THEO_BF_SCP} Assume $\beta \left(\frac{\gamma}{1 + \gamma} +
\epsilon \gamma\right) < 1$. Then, asymptotically, SINR $\gamma$ is
achievable at each mobile terminal in the limit as $N \rightarrow
\infty$ with $\frac{K}{N}\rightarrow \beta > 0$. In this case, the
empirical distribution of the (normalized) UL dual power levels
({\it i.e.} the $\lambda_{kj}$s) converges weakly to the constant
$\bal$ given by
\begin{align}
\bal = \frac{\gamma}{1 - \beta \frac{\gamma}{1 + \gamma}},
\label{eq:SCP_bal}
\end{align}
the empirical distribution of the (normalized) DL power per users
({\it i.e.} the $p_{kj}$s) converges weakly to the constant
\begin{align}
\bap = \frac{\sigma^2 \gamma}{1 - \beta \left(\frac{\gamma}{1 +
\gamma} + \epsilon \gamma \right)}. \label{eq:SCP_bap}
\end{align}
The per BS power converges to $\bar{P} = \beta \bap$.
The asymptotically optimal form of the DL beamformer for user $k$ in cell $j$ is
\begin{align}
\mathbf{w}_{kj}^{SCP} &= \sqrt{\frac{\bar{p}}{N}} \frac{\mathbf{\hat{w}}_{kj}^{SCP}}{\|\mathbf{\hat{w}}_{kj}^{SCP}\|},\label{eq:SCP_reg_bf}
\end{align}
\vspace{-.4cm}
\begin{align}
\textrm{where } \mathbf{\hat{w}}_{kj}^{SCP} = \left(
\mathbf{I} + \frac{\bar{\lambda}}{N} \sum_{k' \neq k}
\mathbf{h}_{k', j,j}^H\mathbf{h}_{k', j,j} \right)^{-1} \mathbf{h}_{k,j,j}^H.  \nonumber 
\end{align}
Finally, the asymptotic SINR, $\gamma$, is related to the other
variables via the fixed point equations
\begin{align}
\gamma = \frac{1}{\frac{1}{\bar{\lambda}}+\frac{\beta}{1+\gamma}} =
\frac{1}{\frac{\sigma^2}{\bar{p}} + \epsilon \beta +
\frac{\beta}{1+\gamma} }.
\label{eq:SCP_SINR_FPE}
\end{align}
Conversely, if $\beta \left(\frac{\gamma}{1 + \gamma} + \epsilon
\gamma \right) > 1$, then, asymptotically, the SINR target $\gamma$
is not achievable under the SCP strategy.
\end{theorem}
\begin{IEEEproof} See Appendix \ref{appendix:SCP_dual_LAS}.
\end{IEEEproof}

\begin{corollary}[SCP]
Subject to the per BS power constraint $P$, the maximum
asymptotic network-wide achievable SINR for a given cell loading
factor $\beta$ is the unique positive solution to the following
fixed point equation:
\begin{align}
\gamma^*_{SCP} =  \frac{1}{\beta}~~\frac{1}{\frac{\sigma^2}{P}+\epsilon +\frac{1}{1+\gamma^*_{SCP}}} \label{eq:GammaOpt_SCP},
\end{align}
which has the explicit solution $\gamma^*_{SCP}$ equal to
\begin{align}
 \frac{-\left(\frac{\sigma^2}{P}+\epsilon-\frac{1}{\beta}+1\right)+\sqrt{\left(\frac{\sigma^2}{P}+\epsilon-\frac{1}{\beta}+1\right)^2+4\frac{\frac{\sigma^2}{P}+\epsilon}{\beta}}}{2 \left(\frac{\sigma^2}{P}+\epsilon\right)}. \nonumber 
\end{align}
\end{corollary}

Theorem~\ref{THEO_BF_SCP} is interesting in that it relates the
solution to the optimization problem \eqref{eq:MaxMinSINR_SCP_feasibility} to a notion of {\it
regularized zero-forcing} proposed in \cite{peel_com05} as a {\it practical} approach that is as simple to
implement as ZF, yet with better performance.  It was studied in the asymptotic regime in \cite{nguyen_globecom08}.
The beamformers defined in \eqref{eq:SCP_reg_bf} asymptotically lead to a precoding matrix
\begin{align}
\mathbf{W}_j^{SCP} = c_j \left[\mathbf{I} +\frac{\bar{\lambda} }{N} \mathbf{H}_{j,j}^H\mathbf{H}_{j,j}\right]^{-1}\mathbf{H}_{j,j}^H,
\end{align}
where $c_j$ ensures the power constraint at $BS_j$ is met with equality, and $\mathbf{H}_{j,j}$ is the concatenation of
channels between cell $j$ users and their serving BS.
We should add that the optimal beamformer always exists, even when the 
ZF beamformer does not, so technically we should only refer to it as
RZF 
 in those scenarios where the ZF beamformer exists.

Another interesting observation is that Theorem~\ref{THEO_BF_SCP}
provides a condition that is both necessary and sufficient for the
target SINR, $\gamma$, to be achieveable. We can interpret
$\displaystyle \frac{\gamma}{1 + \gamma}$ as the {\it effective
bandwidth} of a user in cell $j$, and $\epsilon \gamma$ as the
effective bandwidth of an interferer in cell $\baj$. Effective
bandwidths provide a simple metric by which different beamforming
schemes can be compared, as shown in the next two theorems.

\begin{theorem}[Asymptotically optimal beamforming for CBf]
\label{THEO_BF_COORD} Assume $\beta \left(\frac{\gamma}{1 + \gamma}
+  \frac{\epsilon \gamma}{1 + \epsilon \gamma}\right)< 1$. Then,
asymptotically, SINR $\gamma$ is achievable at each mobile terminal,
in the limit as $N \rightarrow \infty$ with $\frac{K}{N}\rightarrow \beta
> 0$. In this case, the empirical distribution of the (normalized)
UL dual power levels ({\it i.e.} the $\lambda_{kj}$s) converges
weakly to the constant $\bal$, given by
\begin{align}
\bal = \frac{\gamma}{1 - \beta \left(\frac{\gamma}{1 + \gamma} +
\frac{\epsilon \gamma}{1 + \epsilon \gamma}\right)},
\label{eq:coord_bal}
\end{align}
the empirical distribution of the (normalized) DL power per users
({\it i.e.} the $p_{kj}$s) converges weakly to the constant
\begin{align}
\bap = \bal \sigma^2 = \frac{\sigma^2 \gamma}{1 - \beta \left(\frac{\gamma}{1 + \gamma} + \frac{\epsilon \gamma}{1 + \epsilon \gamma}\right)}. \label{eq:coord_bap}
\end{align}
The per BS power converges to $\bar{P} = \beta \bap$.
The asymptotically optimal form of the DL beamformer for user $k$ in cell $j$ is
\begin{align}
\mathbf{w}_{kj}^{Coord} &= \sqrt{\frac{\bar{p}}{N}} \frac{\mathbf{\hat{w}}_{kj}^{Coord}}{\|\mathbf{\hat{w}}_{kj}^{Coord}\|}, \label{eq:coord_bf}
\end{align}
\vspace{-.4cm}
\begin{align}
\textrm{where }\mathbf{\hat{w}}_{kj}^{Coord} = \left( \mathbf{I} +
\frac{\bar{\lambda}}{N} \sum_{\left(k', \bar{j}\right) \neq
\left(k,j\right)} \mathbf{h}_{k', j,j}^H\mathbf{h}_{k',
j,j} \right)^{-1} \mathbf{h}_{k,j,j}^H.\nonumber
\end{align}
Finally, the asymptotic SINR, $\gamma$, is related to the other
variables via
\begin{align}
\gamma = \frac{1}{\frac{1}{\bar{\lambda}} + \frac{\epsilon
\beta}{1+\epsilon \gamma} + \frac{\beta}{1+\gamma}} =
\frac{1}{\frac{\sigma^2}{\bar{p}} + \frac{\epsilon \beta}{1+\epsilon
\gamma} + \frac{\beta}{1+\gamma}}.
\label{eq:coord_SINR_FPE}
\end{align}
Conversely, if $\beta \left(\frac{\gamma}{1 + \gamma} +
\frac{\epsilon \gamma}{1 + \epsilon \gamma}\right) > 1$, then
asymptotically the SINR target $\gamma$ is not achievable under the
coordinated beamforming strategy.
\end{theorem}
\begin{IEEEproof} See Appendix \ref{appendix:coord_dual_LAS}.
\end{IEEEproof}

\begin{corollary}[CBf]
Subject to per BS power constraint $P$, the maximum asymptotic network-wide achievable SINR for a given cell loading factor $\beta$ is the unique positive solution to the following fixed point equation:
\begin{align}
\gamma^*_{CBf} =  \frac{1}{\beta}~~\frac{1}{\frac{\sigma^2}{P} +
\frac{\epsilon}{1+ \epsilon \gamma^*_{CBf}} +
\frac{1}{1+\gamma^*_{CBf}}}\label{eq:GammaOpt_Coord}.
\end{align}
In other words, $\gamma^*_{CBf}$ is the root of a cubic equation.
\end{corollary}

Theorem~\ref{THEO_BF_COORD} is interesting in that it provides a {\it novel} form of 
RZF beamformer.\footnote{in the scenarios where the ZF beamformer exists.} This beamformer is not a direct
regularization of the standard ZF 
 beamformer in a single
cell. Rather, it is a regularization of a beamformer that zero
forces the interference it creates at users in the other cell as
well as its own; in other words, it transmits to a {\it subset} of
the users that it is zero forcing. We call this a generalized
RZF beamformer, and it is a novel contribution
of the present paper. It can be used in a finite system, where it is
suboptimal, but relatively straightfoward to implement.
Theorem~\ref{THEO_BF_COORD} shows that it is asymptotically optimal
in the class of coordinated beamforming strategies. Note also the clean characterization of effective bandwidth for this beamformer, and that it provides a significant reduction in the effective bandwidth of the other-cell users, compared to SCP, when $\epsilon$ is non-negligible.

\begin{theorem}[Asymptotically optimal beamforming for MCP]
\label{THEO_BF_MCP} Assume $\beta \frac{\gamma}{1 + \gamma} < 1$.
Then, asymptotically, SINR $\gamma$ is achievable at each mobile
terminal, in the limit as $N \rightarrow \infty$ with
$\frac{K}{N}\rightarrow \beta > 0$. In this case, the empirical
distribution of the (normalized) UL dual power levels ({\it i.e.}
the $\lambda_{kj}$s) converges weakly to the constant $\bal$ given
by
\begin{align}
\bal = \frac{1}{1+\epsilon} ~\frac{\gamma}{(1 - \beta
\frac{\gamma}{1 + \gamma})}, \label{eq:MCP_bal}
\end{align}
the empirical distribution of the (normalized) DL power per user
({\it i.e.} the $p_{kj}$s) converges weakly to the constant
\begin{align}
\bap = \bal \sigma^2 = \frac{1}{1+ \epsilon} ~~\frac{\sigma^2
\gamma}{(1 - \beta \frac{\gamma}{1 + \gamma})}. \label{eq:MCP_bap}
\end{align}
The per BS power converges to $P = \beta \bap$.
The asymptotically optimal form of the DL beamformer for user $k$ in cell $j$ is
\begin{align}
\mathbf{w}_{kj}^{MCP} &= \sqrt{\frac{\bar{p}}{N}} \frac{\mathbf{\hat{w}}_{kj}^{MCP}}{\|\mathbf{\hat{w}}_{kj}^{MCP}\|},
\end{align}
\vspace{-.4cm}
\begin{align}
\textrm{where }  \mathbf{\hat{w}}_{kj}^{MCP} = \left( \mathbf{I} +
\frac{\bar{\lambda}}{N} \sum_{\left(k', j'\right) \neq
\left(k,j\right)} \mathbf{\tilde{h}}_{k',
j}^H\mathbf{\tilde{h}}_{k', j} \right)^{-1}
\mathbf{\tilde{h}}_{k,j}^H. \label{eq:MCP_bf}
\end{align}
Finally, the asymptotic SINR, $\gamma$, is related to the other
variables via
\begin{align}
\gamma = \frac{1}{\frac{1}{(1+\epsilon) \bar{\lambda}} + \frac{\beta}{1+\gamma}} = \frac{1}{\frac{\sigma^2}{\bar{p}(1+\epsilon)} + \frac{\beta}{1+\gamma}}.
\label{eq:MCP_SINR_FPE}
\end{align}
Conversely, if $\beta \frac{\gamma}{1 + \gamma} > 1$, then asymptotically the SINR target $\gamma$ is not achievable under the MCP beamforming strategy.
\end{theorem}
\begin{IEEEproof} See Appendix \ref{appendix:MCP_dual_LAS}.
\end{IEEEproof}

\begin{corollary}[MCP]
Subject to per BS power constraint $P$, the maximum asymptotic network-wide achievable SINR for a given cell loading factor $\beta$ is the unique positive solution to the following fixed point equation:
\begin{align}
\gamma^*_{MCP} =  \frac{1}{\beta}~~\frac{1}{\frac{\sigma^2}{(1+\epsilon)P} + \frac{1}{1+\gamma^*_{MCP}}}\label{eq:GammaOpt_MCP}.
\end{align}
In other words, $\gamma^*_{MCP}$ is equal to
\begin{align}
& \frac{-\left(\frac{\sigma^2}{(1+\epsilon)P}-\frac{1}{\beta}+1\right)+\sqrt{\left(\frac{\sigma^2}{(1+\epsilon)P}-\frac{1}{\beta}+1\right)^2+4\frac{\frac{\sigma^2}{(1+\epsilon)P}}{\beta}}}{2 \left(\frac{\sigma^2}{(1+\epsilon)P}\right)}.\nonumber
\end{align}
\end{corollary}

Although MCP is a complex strategy, in that the cooperation between
BSs is much greater, the reward is better performance than that
attainable in a single isolated cell with no intercell interference.
Note that the power levels in \eqref{eq:MCP_bap} are {\it less} than what they would be in a single
isolated cell. The effective bandwidth of each user is the same as
for a single isolated cell, under SCP, but the power
consumption is reduced by the factor $(1+\epsilon)$, which
corresponds to a power gain from having both BSs involved in the
beamforming, instead of just one.

\subsection{Effective interference}\label{sec:EffI}
The optimal SINR expressions in \eqref{eq:SCP_SINR_FPE}, \eqref{eq:coord_SINR_FPE}, \eqref{eq:MCP_SINR_FPE} are striking in how they capture
the effect of interference for the three different beamformers. Indeed, they supply a simple ``effective interference'' characterization, which can be used to directly check if a particular target SINR can be achieved.

It is natural to try and compare the schemes directly using the
limiting SINR expressions. This is accomplished in the following
theorem, where $SNR$ denotes $\frac{P}{\sigma^2}$.

\begin{theorem}
\label{THEO_EFF_INT} Let $\gamma^*_{SCP}, \gamma^*_{CBf},
\gamma^*_{MCP}$ denote the SINRs under SCP,
CBf, and MCP, respectively.
Then
\begin{align}
\gamma^*_{SCP} < \gamma^*_{CBf} < \gamma^*_{MCP}.
\label{eq:SINR_ordering}
\end{align}
At signal to noise ratio $SNR$ and interference level $\epsilon$,
denote the {\it effective interference} at {\it target SINR}
$\gamma$ by
\begin{align}
I_{\textrm{eff~}}(SNR,\epsilon,\gamma) = \left\{ \begin{array}{ll}
\beta \left(1 +
\frac{SNR}{1+\gamma}+ \epsilon SNR\right) & \textrm{~~~~~SCP} \\
\beta \left(1 + \frac{SNR}{1+\gamma} + \frac{\epsilon SNR}{1+
\epsilon \gamma} \right) & \textrm{~~~~~CBf} \\
\beta \left(1 + \frac{SNR}{1+\gamma} + \frac{\epsilon SNR}{1+
\gamma} \right) & \textrm{~~~~~MCP} \end{array} \right.
\nonumber 
\end{align}
Then the feasibility of $\gamma$ in the case of SCP, or CBf, is
equivalent to satisfaction of the inequality $\displaystyle
\frac{SNR}{I_{\textrm{eff~}}(SNR, \epsilon, \gamma)} > \gamma$, and
in the MCP case, it is equivalent to $\displaystyle
\frac{(1+\epsilon) SNR}{I_{\textrm{eff~}}(SNR, \epsilon, \gamma)} >
\gamma.$
\end{theorem}
\begin{IEEEproof}
Follows closely that of Proposition 3.2 in \cite{tse_it1999}.
\end{IEEEproof}

We note here the close parallel with the effective
interference arising in the large system analysis of linear
UL multiuser receivers \cite{tse_it1999}. This is due to
 the underlying UL-DL duality.

\subsection{Asymptotically optimal cell loading}
The above theorems characterize the optimal SINR for fixed cell
loading $\beta$ under SCP, CBf, and MCP, respectively.
Our next step is to characterize this optimum loading: this
determines the optimal number of users to serve given a number of
antennas at the BS. This corresponds to finding the $\beta$ that
maximizes the normalized (by the number of antennas) rate per cell
$r$, i.e. the optimizing $\beta$ that solves the following problem:
\begin{align}
\textrm{maximize}_{\beta~} r = \beta \log(1+\gamma^*)
\label{eq:OptRate}
\end{align}
with $\gamma^*$ characterized by the appropriate fixed point equation (cf. Eqs \eqref{eq:GammaOpt_SCP}, \eqref{eq:GammaOpt_Coord} and \eqref{eq:GammaOpt_MCP}).

\begin{proposition}[Characterization of the optimum $\beta$ for SCP]\label{THEO_BETA_SCP}
If
\begin{align}
\epsilon + \frac{\sigma^2}{P} \ge 1 \label{eq:InfBeta_SCP}
\end{align}
then $r(\beta)$ is an increasing function. Otherwise, the optimum
occurs at a finite $\beta^*$ which may be found by a line search.
\end{proposition}
\begin{IEEEproof} Refer to Appendix \ref{appendix:proofTheoBETA_SCP}.
\end{IEEEproof}

\begin{proposition}[Optimal cell loading for CBf]\label{THEO_BETA_COORD} If $\frac{\sigma^2}{P} + \epsilon - 2\epsilon^2-1 \ge 0$
then $r(\beta)$ is an increasing function. Otherwise, there is a
finite value of $\beta$ at which $r$ is maximized.
\end{proposition}
\begin{IEEEproof}
Refer to Appendix \ref{appendix:proofTheo_BETA_COORD}.
\end{IEEEproof}

\begin{proposition}[Characterization of the optimum $\beta$ for MCP]\label{THEO_BETA_MACRO}
If
\begin{align}
\frac{\sigma^2}{P} \ge (1 + \epsilon) \label{eq:beta_JTx}
\end{align}
then $r(\beta)$ is an increasing function. Otherwise, the optimum
occurs at a finite $\beta^*$ which may be found by a line search.
\end{proposition}
\begin{IEEEproof} Comparing \eqref{eq:GammaOpt_MCP} and \eqref{eq:GammaOpt_SCP}, we see that the former is the same as the latter with $\epsilon + \frac{\sigma^2}{P} \leftarrow \frac{\sigma^2}{P (1+\epsilon)}$. Performing this substitution in \eqref{eq:InfBeta_SCP} yields the result.
\end{IEEEproof}

The above results define for each scheme a noise-limited region, in
which cell loading can be increased indefinitely; however, this
leads to ever decreasing rates per user, not to mention
that more user channels would have to be learned.

\section{Performance Results}
\label{sec-performance}
How do these schemes compare with each other and with other
approaches from the literature? Clearly, CBf requires more CSI than SCP, and MCP involves much more BS cooperation,
so it is not surprising that the SINRs are ordered as in
\eqref{eq:SINR_ordering}.
In this section, we obtain numerical results to provide a quantitative comparison between these schemes
in different scenarios. Throughout this paper, we have assumed full re-use of time and spectrum across cells; however, intercell interference can be avoided
altogether by applying the classic principle of re-use partitioning: we thus also consider in our simulations the time division (TD) scheme in which
each BS is given a separate time-slot, which we shall also call
``$1/2$-reuse''.
We also consider two forms of pure ZF in the context of single cell
processing. SCP-ZF zero forces the same-cell
interference, with the BS oblivious to other-cell interference.
Generalized zero-forcing (GZF) is when the BSs independently zero
force the interference in the two-cell system. Finally, in the MCP setting, the two BSs can jointly zero-force all the interference in the two-cell
system, and we denote this case by ``MCP-ZF''.

\subsection{When is half re-use SCP better than CBf?}
Let $\beta_{TD}$ be the cell loading in the half reuse scheme in
which each BS transmits half the time. To compare with coordinated
beamforming (a full re-use scheme), let $\beta:= \beta_{CBf} :=
\beta_{TD}/2.$ Then the rate for the TD scheme is
\begin{align}
r_{TD}(\beta) = \beta \log\left(1+\gamma^*_{TD}(\beta)\right),
\end{align}
\begin{align}
\text{with} \quad \gamma^*_{TD}(\beta) &= \frac{1}{2\beta}~~
\frac{1}{\frac{\sigma^2}{2P} + \frac{1}{1+\gamma^*_{TD}(\beta)}} =
\frac{1}{\beta} ~~\frac{1}{\frac{\sigma^2}{P} +
\frac{2}{1+\gamma^*_{TD}(\beta)}}.
\end{align}
The rate using coordinated beamforming is given by
\begin{align}
r_{CBf}(\beta) = \beta \log\left(1+\gamma^*_{CBf}(\beta)\right)
\end{align}
\begin{align}
\text{with} \quad \gamma^*_{CBf} = \frac{1}{\beta}~~
\frac{1}{\frac{\sigma^2}{P}+\frac{1}{1+\gamma^*_{CBf}(\beta)}+\frac{\epsilon}{1+\epsilon
\gamma^*_{CBf}(\beta)}}.
\end{align}
Thus, $\displaystyle r_{CBf}(\beta) > r_{TD}(\beta)$ if $\epsilon <
1$, and $\displaystyle r_{CBf}(\beta) < r_{TD}(\beta)$ if $\epsilon
> 1.$ It follows that coordinated beamforming is only useful when $\epsilon < 1$;
otherwise, it is better to partition the cells with a reuse factor
of $1/2$. Of course, if BS association is performed properly, $\epsilon$ should be less than 1.

\subsection{Numerical results}

\begin{figure}[t]
\centering
\includegraphics[width=3in,height=2.3in,viewport=40 160 550 590,clip]{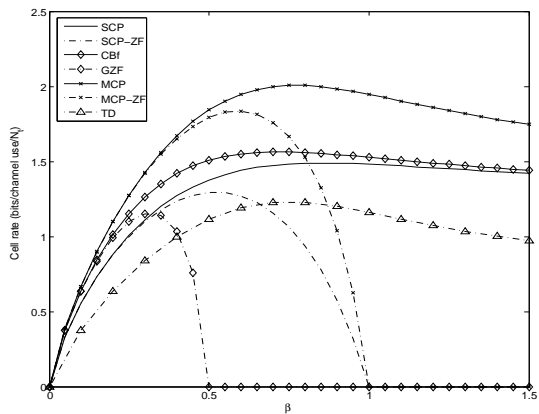}
\vspace{-.7cm}
\caption{Effect of cell loading $\beta$ on rate achieved for SNR = 10dB, $\epsilon = .1$} \label{fig:Rates10dBe01}
\end{figure}

\begin{figure}[htp]
\centering
\includegraphics[width=3in,height=2.5in,viewport=80 160 550 680,clip]{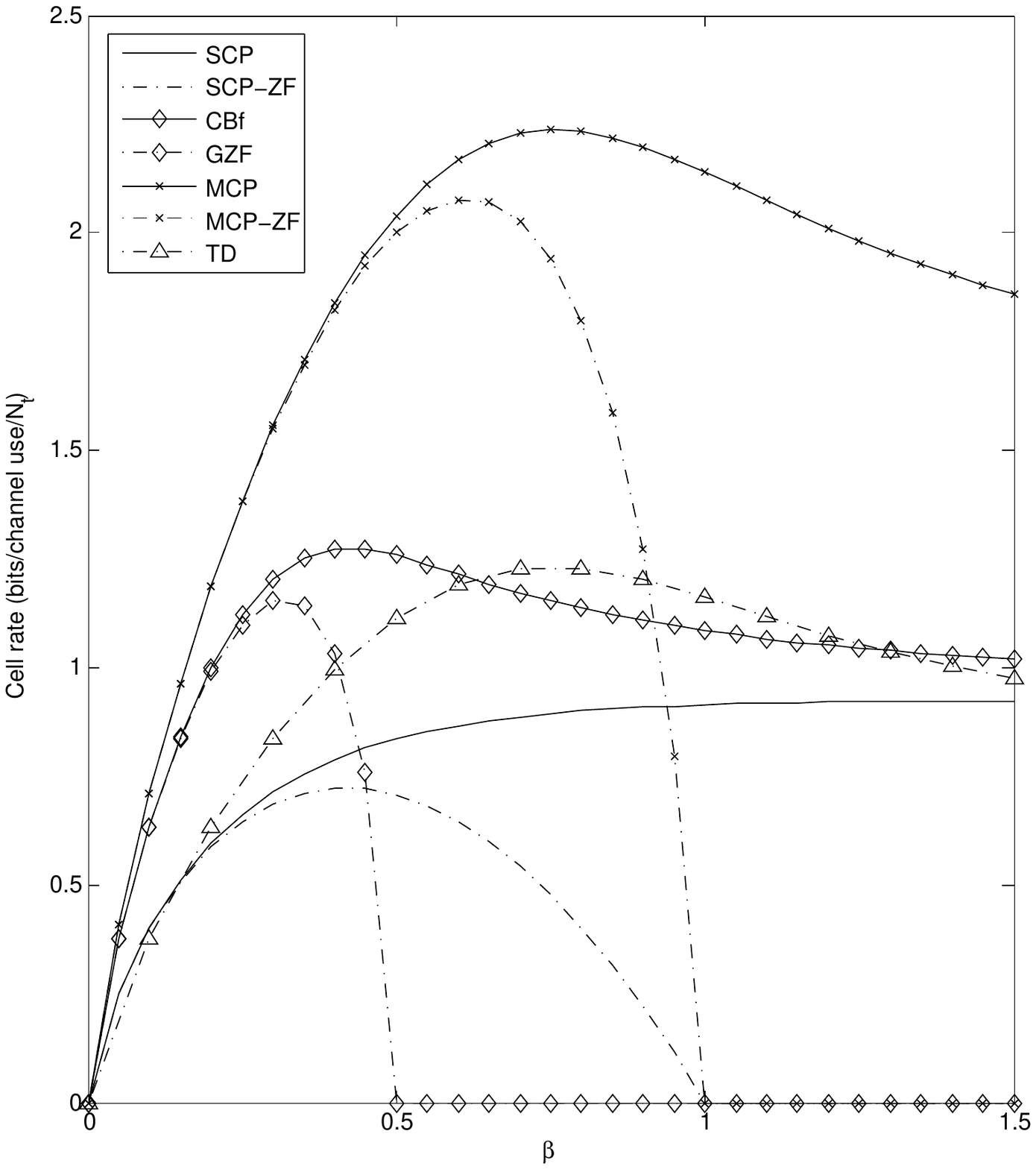}
\caption{Effect of cell loading on rate achieved for SNR = 10dB, $\epsilon = .5$} \label{fig:Rates10dBe05ZF}
\end{figure}

\begin{figure}[htp]
\centering
\includegraphics[width=3in,height=2.1in,viewport=40 110 600 645,clip]{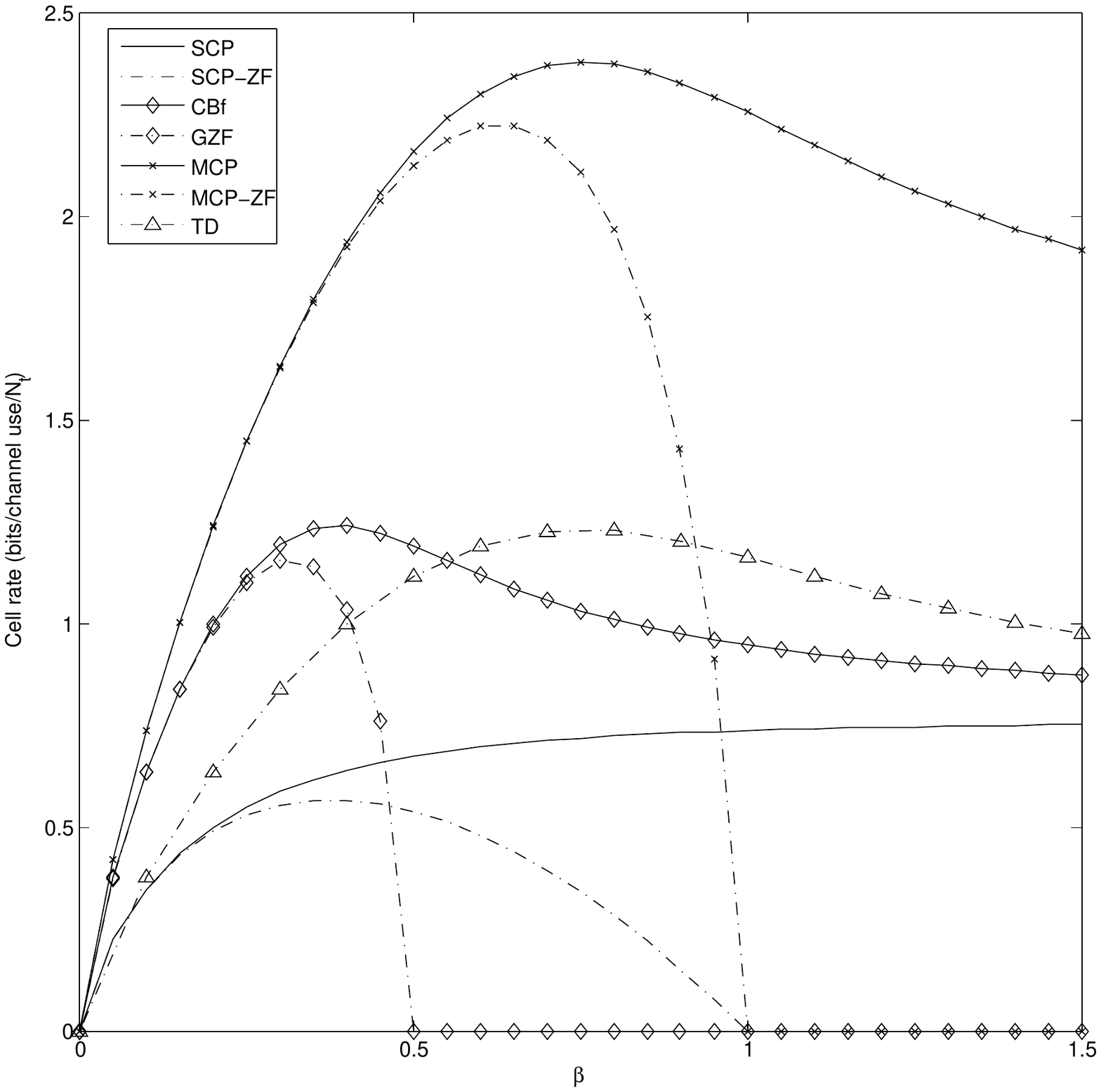}
\vspace{-.5cm}
\caption{Effect of cell loading on rate achieved for SNR = 10dB, $\epsilon = .8$} \label{fig:Rates10dBe08ZF}
\end{figure}

Figures~\ref{fig:Rates10dBe01}-\ref{fig:Rates10dBe08ZF} compare the
different schemes by varying the cell loading $\beta$. We notice that when
$\epsilon$ is small, CBf gains little over
SCP, but offers significant gains compared to pure ZF or
$1/2$-reuse. When $\epsilon$ is small, SCP-ZF is superior to GZF, as
expected. When $\epsilon$ is large, but $< 1$ (e.g.
$\epsilon = 0.8$), CBf gains significantly over
SCP, but does not gain much over $1/2$-reuse, or to GZF (when
the loading is perfectly optimized). Note that in this case, the
relevant comparison is with GZF.

When $\epsilon < 1$, CBf is always better, for appropriately selected $\beta$, than SCP, SCP-ZF, and
GZF. If one were to insist on using pure ZF, and could choose the appropriate ZF scheme, and the exact optimal loading for that scheme, then it can get reasonably close to the performance of the optimized CBf. Similarly, if one could select between SCP or $1/2$-reuse, the performance can be quite close to that of CBf. The advantage of the latter is that it is universally good, if no joint transmission is allowed, across all levels of inter-cell interference. Compared to
the ZF schemes, it performs better across a wider range of
cell loadings. In large networks, it avoids the intractable frequency planning problem associated with fractional re-use schemes. When $\epsilon > 1$,
MCP offers very significant gains over the $1/2$-reuse
scheme (not depicted).

In the two-cell model, MCP offers the most gain when $\epsilon$ is
large. Even when $\epsilon$ is small, as in
Figure~\ref{fig:Rates10dBe01}, the gains over CBf, $1/2$-reuse, and
single cell ZF schemes, respectively, are significant, and in
Figure~\ref{fig:Rates10dBe08ZF} they are higher still, because
$\epsilon$ is larger in that case. Unlike the other schemes, MCP
improves with increasing $\epsilon$, but it requires significant cooperation
between BSs, including full data sharing, whose cost in terms of backhaul capacity is not
accounted for here.

\begin{figure}[htp]
\centering
\includegraphics[width=3in,height=2.1in,viewport=70 200 530 580,clip]{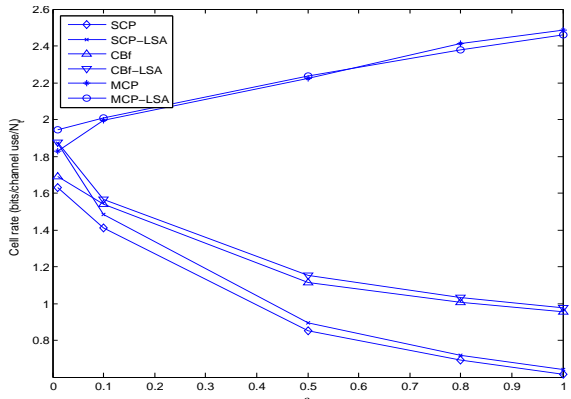}
\vspace{-.5cm}
\caption{Average normalized cell rates for the optimization problems for $K = 3, N = 4$ and LSA rates for $\beta = .75$, both at SNR = 10 dB.} \label{fig:LAS_Optimized}
\end{figure}

Finally, we investigate the applicability of the asymptotic results
to a finite system. In a first step, for $K = 3, N = 4$,
$\frac{P}{\sigma^2} = 10$ and $\epsilon$ taking values in $[.01; .1;
.5; .8; 1]$, we solve the optimization problems described in Section
\ref{c5:sec:opt} for different independent samples of the channel
and obtain the corresponding average rates. Even for such a small
number of antennas, the large system analysis (LSA) results provide
quite a good approximation. The results are shown in Figure
\ref{fig:LAS_Optimized}.

\begin{figure}[htp]
\centering
\includegraphics[width=3in,height=2.1in,viewport=80 200 520 580,clip]{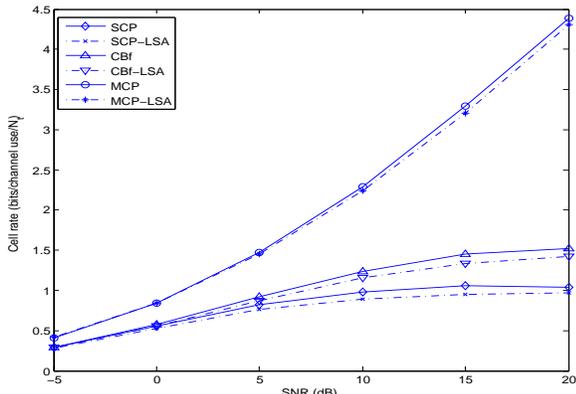}
\vspace{-.5cm}
\caption{Average normalized cell rates from using the asymptotically optimal beamformers to the finite system case for $K = 3, N = 4$ and LSA results for $\beta = .75$, both with $\epsilon = 0.5$.} \label{fig:LAS_UsingAsymptoticBF_05}
\end{figure}

The optimizations in Section~\ref{c5:sec:opt} are
time-consuming, particularly for the SCP case, which requires
iterations between the optimization at the two transmitters until
convergence. Thus, we consider applying the asymptotically optimal
beamforming vectors from Section \ref{c5:sec:LAS} (slightly modified
so as not to break the per transmitter power constraint for the MCP
scheme) in the finite system case. The results are shown in Figure
\ref{fig:LAS_UsingAsymptoticBF_05}. 

\section{Conclusions}

This paper has provided an asymptotic analysis of a two cell
interfering network in which the number of antennas at the BSs, and that of users per cell, both grow large
together. Schemes that balance rates across users in the
system were compared for three levels of cooperation, namely single cell
processing, coordinated beamforming and multicell
processing. MCP offers significant rate gains, if
we can accommodate the additional coordination and
communication between BSs, not accounted for here.
We characterized and compared the limiting SINRs of
the three schemes, in particular, using the notion of {\it
effective interference}, which can also be used to establish
 if a given SINR target is feasible. 
The validity of the obtained results was verified via
Monte Carlo simulations.

Note that we have  assumed the users' channels
are selected randomly. It is important to emphasize that our conclusions
do {\it not} hold if users have been selected based on their
channels, as the conditional distributions change drastically.
Indeed, it is well known that for a large enough pool of users, with careful scheduling, the performance of ZF can be almost optimal. 

\appendices

\section{Useful Theorems}\label{app:theorems}

We start by reproducing a few lemmas, which play an important role in our derivations.


\begin{lemma}[Lemma 6.3.3 in \cite{hachem_08}]\label{lemm:633}
Let $\rho > 0$, $\mathbf{A}$ and $\mathbf{B}$ $N \times N$ matrices with $\mathbf{B}$ Hermitian, $\tau \in \mathbb{R}$, and $\mathbf{q} \in \mathbb{C}^N$. Then
\begin{align}
\left|\textrm{tr} \left(\left(\left(\mathbf{B} + \rho \mathbf{I}\right)^{-1} - \left(\mathbf{B} + \mathbf{q}\mathbf{q}^H + \rho \mathbf{I}\right)^{-1}
\right)
 \mathbf{A}\right)\right| \le \frac{\|\mathbf{A}\|}{\rho}
\end{align}
\end{lemma}

We will also be making use of results from \cite{kammoun_it09} (some themselves reproduced from \cite{hachem_07}). \cite{kammoun_it09} derives a central limit theorem for the SINR at the receiver in a multiple-access MIMO system, where the dimensions of the system (number of users and the size of the received random vector) grows large: the considered asymptotic regime satisfies
\begin{align}
\tilde{K} \rightarrow \infty, \lim \inf \frac{\tilde{K}}{\tilde{N}} > 0, \quad \lim \sup \frac{\tilde{K}}{\tilde{N}} < \infty,
\end{align}
and the quantity studied, i.e. the UL SINR, is equal to
\begin{align}
\Gamma_{\tilde{K}} = \mathbf{y}^H \left(\mathbf{Y} \mathbf{Y}^H + \rho \mathbf{I}_{\tilde{N}} \right)^{-1} \mathbf{y} \label{eq:Q1},
\end{align}
where the sequence of matrices $\boldsymbol{\Sigma}(\tilde{K}) = \left[\mathbf{y}(\tilde{K}) \mathbf{Y}(\tilde{K})\right]$ is given by
\begin{align}
\boldsymbol{\Sigma}(\tilde{K})  =
\left(\Sigma_{nk}(\tilde{K})\right)_{n=1, k = 0}^{\tilde{N},\tilde{K}} = \left(\frac{\tilde{\sigma}_{nk}(\tilde{K})}{\sqrt{\tilde{K}}}w_{nk}\right)_{n=1, k = 0}^{\tilde{N},\tilde{K}}, \label{eq:varProf}
\end{align}
and is such that the following assumptions hold: \\
\textbf{A1}: the complex r.v.'s $(w_{nk}: n \ge 1, k \ge 0)$ are i.i.d., $\mathbb{E} w_{10} = 0$, $\mathbb{E} w_{10}^2 = 0$, $\mathbb{E} |w_{10}|^2 = 1$ and $\mathbb{E} |w_{10}|^8 < \infty$\footnote{For the channel model considered here, the entries are i.i.d. $\mathcal{CN}(0,1)$, so they clearly satisfy this condition.} \\
\textbf{A2}: the variance profile is such that there exists a real number $\tilde{\sigma}_{\max} < \infty$ such that
\begin{align}
\sup_{\tilde{K} \ge 1} \max_{1 \le n \le \tilde{N}, 0 \le k \le \tilde{K}} |\tilde{\sigma}_{nk}(\tilde{K})| \le \tilde{\sigma}_{\max},
\end{align}
The dual UL SINR expressions in the present paper for $\lambda_{kj}$'s in each cell held at some constant value, will be of the same form as \eqref{eq:Q1}, and the matrices considered will satisfy assumptions \textbf{A1} and \textbf{A2}. In fact, they
constitute a special case of the above model, since we consider the asymptotic regime where $\frac{K}{N} = c$ as $K, N \rightarrow \infty$, and the variance profiles of the random matrices considered (see \eqref{eq:varProf}) are either scaled all ones matrices, or obtained by the regular sampling of a piece-wise continuous function.  Note that the expressions for asymptotic SINR on the dual UL's may equivalently be obtained from earlier results in the literature, \cite{dai_sp03} for example.

We thus reproduce below the most relevant results from \cite{kammoun_it09}.

%
\begin{theorem}[Parts 1 and 3 of Theorem 1 in \cite{kammoun_it09}] \label{T_tildeT}
The following statements hold true.
\begin{itemize}
\item Let $\left(\tilde{\sigma}^2_{nk}(\tilde{K}): 1 \le n \le \tilde{N}, 1 \le k \le \tilde{K}\right)$ be a sequence of arrays of real numbers and consider matrices $\mathbf{D}_k(\tilde{K})$ 
\begin{align}
\mathbf{D}_k(\tilde{K}) = \textrm{diag}\left(\tilde{\sigma}_{1k}^2(\tilde{K}), \ldots, \tilde{\sigma}_{\tilde{N}k}^2(\tilde{K})\right), 0 \le k \le \tilde{K}. 
\end{align}
 The system of $\tilde{N}$ functional equations
\begin{align}
t_{n, \tilde{K}}(z) = \frac{1}{-z+\frac{1}{\tilde{K}} \sum_{k=1}^{\tilde{K}} \frac{\tilde{\sigma}_{nk}^2(\tilde{K})}{1+\frac{1}{\tilde{K}} \textrm{tr} {\mathbf{D}}_k(\tilde{K}) \mathbf{T}_{\tilde{K}}(z)},
}
\end{align}
for $1 \le n \le \tilde{N}$ and where
\begin{align}
  \mathbf{{T}}_{\tilde{K}}(z) = \textrm{diag}\left(t_{1,\tilde{K}}(z), \ldots, t_{\tilde{N}, \tilde{K}}(z)\right)
\end{align}
admits a unique solution $\mathbf{T}$ 
 among the diagonal matrices for which the $t_{n,\tilde{K}}$ 
 belong to class $\mathcal{S}$\footnote{A complex function $t(z)$ belongs to class $\mathcal{S}$ if $t(z)$ is analytical in the upper half plane $\mathbb{C}_{+} = \{\textrm{im}(z) > 0\}$, if $t(z) \in \mathbb{C}_{+}$ for all $z\in \mathbb{C}_{+}$, and if $\textrm{im} z \left|t(z)\right|$ is bounded over the upper half plane $\mathbb{C}_{+}$.}. Moreover, the functions admit analytical continuations over $\mathbb{C}-[0, \infty)$ which are real and positive for $z \in \left(-\infty, 0\right)$.
\item  Assume \textbf{A1} and \textbf{A2} hold true. Consider the sequence of random matrices $\mathbf{Y}(\tilde{K})\mathbf{Y}(\tilde{K})^H$, where $\mathbf{Y}_{nk} = \frac{\tilde{\sigma}_{nk}}{\sqrt{\tilde{K}}} w_{nk}$. For every sequence $\mathbf{S}_K$ of $\tilde{N} \times \tilde{N}$ diagonal matrices  with
\begin{align}
\sup_{\tilde{K}} \|\mathbf{S}_{\tilde{K}}\| < \infty
\end{align}
the following limits hold true almost surely (a.s.):
\begin{align}
\lim_{\tilde{K} \rightarrow \infty} \frac{1}{\tilde{K}} \textrm{tr} \mathbf{S}_{\tilde{K}}\left(\mathbf{Q}_{\tilde{K}}(z) - \mathbf{T}_{\tilde{K}}(z)\right) = 0, \forall z \in \mathbb{C}-\mathbb{R}_{+}, \label{eq:conv_tr}
\end{align}
where $\mathbf{Q}_{\tilde{K}}(z)$ denotes the resolvent of $\mathbf{Y}(\tilde{K})\mathbf{Y}(\tilde{K})^H$, i.e. the $\tilde{N} \times \tilde{N}$ matrix defined by
\begin{align}
\mathbf{Q}_{\tilde{K}}(z) = \left(\mathbf{Y}(\tilde{K})\mathbf{Y}(\tilde{K})^H - z\mathbf{I}_{\tilde{N}}\right)^{-1}. \label{eq:resolvent}
\end{align}
\end{itemize}
\end{theorem}
\begin{corollary} \label{theo:Xi}
Assume \textbf{A1} and \textbf{A2} hold true.
Let $\Xi_{\tilde{K}} = \frac{1}{\tilde{K}} \textrm{tr} \left(\mathbf{S}_{\tilde{K}} \mathbf{Q}^2_{\tilde{K}}(-\rho)\right)$ where $\rho \in \mathbb{R}_{+}$,  $\mathbf{S}_{\tilde{K}}$,
 $\bar{\Xi}_{\tilde{K}} = -\frac{1}{\tilde{K}} \textrm{tr} \mathbf{S}_{\tilde{K}} \frac{d}{d\rho}\left(\mathbf{T}_{\tilde{K}}(-\rho)\right)$,
  and $\mathbf{T}_{\tilde{K}}$ be as given by Theorem \ref{T_tildeT}.
Then
\begin{align}
\Xi_{\tilde{K}} - \bar{\Xi}_{\tilde{K}} \stackrel{\tilde{K} \rightarrow \infty}{\longrightarrow}  0 \textrm{ a.s. }
\end{align}
\end{corollary}
\begin{IEEEproof}
Differentiating \eqref{eq:conv_tr} with respect to $z$, we get
\begin{align}
\lim_{\tilde{K} \rightarrow \infty} \frac{1}{\tilde{K}} \textrm{tr} \left( \mathbf{S}_{\tilde{K}} \mathbf{Q}^2_{\tilde{K}}(z)\right) - \frac{1}{\tilde{K}} \textrm{tr}  \mathbf{S}_{\tilde{K}} \frac{d}{dz}\left(\mathbf{T}_{\tilde{K}}(z)\right) = 0, \nonumber \\
\forall z \in \mathbb{C}-\mathbb{R}_{+} \label{eq:DiffQ}
\end{align}
$\frac{d}{dz}\left(\mathbf{T}_{\tilde{K}}(z)\right)$ exists since $t_{n,\tilde{K}}(z)$ admit analytical continuations over the range considered.
This yields the result.
\end{IEEEproof}

\begin{theorem}[Theorem 2 in \cite{kammoun_it09}]\label{theo:UL_SIR_conv}
Let $\bar{\Gamma}_{\tilde{K}} = \frac{1}{\tilde{K}} \textrm{tr} \left(\mathbf{D}_0(\tilde{K}) \mathbf{T}_{\tilde{K}}(-\rho)\right)$ where $\rho \in \mathbb{R}_{+}$, and $\mathbf{T}_{\tilde{K}}$ is given by Theorem \ref{T_tildeT}. Assume \textbf{A1} and \textbf{A2}
then
\begin{align}
\Gamma_{\tilde{K}} - \bar{\Gamma}_{\tilde{K}} \stackrel{\tilde{K} \rightarrow \infty}{\longrightarrow} 0 \textrm{ a.s. }
\end{align}
\end{theorem}

\begin{proposition}[\cite{kammoun_it09}]\label{prop:Ul}
Introduce the r.v.'s $U_l = \frac{1}{\tilde{K}} \textrm{tr} \mathbf{D}_0 \mathbf{Q} \mathbf{D}_l \mathbf{Q}$, for $0 \le l \le \tilde{K}$. The $U_l$ satisfy the following system of equations:
\begin{align}
U_l = \sum_{k=1}^K c_{lk} U_k + \frac{1}{\tilde{K}} \textrm{tr} \mathbf{D_0} \mathbf{D}_l \mathbf{T}^2 + \epsilon_l, 0 \le l \le \tilde{K},
\end{align}
where
\begin{align}
c_{lk} = \frac{1}{\tilde{K}} \frac{\frac{1}{\tilde{K}}\textrm{tr} \mathbf{D}_l \mathbf{D}_k \mathbf{T} \left(-\rho\right)^2}{\left(1+\frac{1}{\tilde{K}}\textrm{tr} \mathbf{D}_k \mathbf{T}\left(-\rho\right)\right)^2},
\end{align}
and the perturbations $\epsilon_l$ satisfy $\mathbb{E} \left|\epsilon_l \right| \le C \tilde{K}^{-1/2}$, where $C$ is independent of $l$.
\end{proposition}

\section{Useful Asymptotic expressions}\label{app:bounds}
In this appendix, we derive asymptotic expressions for quantities of interest in the large asymptotic analysis of SCP, CBf and MCP for the considered two cell symmetric channel model. This relies on the application of the theorems in Appendix \ref{app:theorems}, noting that for our model the $\mathbf{T}$ diagonal matrix in Theorem \ref{T_tildeT} collapses to one or two variables, as many of its entries will be equal.
\subsection{SCP}\label{app:SCP_UL}
Clearly, our Rayleigh channel model satisfies the conditions in Theorem \ref{theo:UL_SIR_conv}, so that for $\lambda_{k,j}$'s fixed for all users in cell $j$ at bounded\footnote{This is required for the results in Appendix \ref{app:theorems} to be applicable. On the other hand, if this was not the case, in the studied optimization problem, the dual objective would be unbounded and the primal unfeasible.} $\lambda_j$, the quantity of interest for any user is
\begin{align}
\frac{\lambda_j}{N} \mathbf{h}_{k,j,j} \left(\frac{\lambda_j}{N}\sum_{k' \neq k}\mathbf{h}_{k',j,j}^H\mathbf{h}_{k',j,j} + \mathbf{I} \right)^{-1} \mathbf{h}_{k,j,j}^H.
\end{align}
From Theorem \ref{theo:UL_SIR_conv}, this quantity will converge a.s. as $N, K \rightarrow \infty$ with $\frac{K}{N} \rightarrow \beta$, to\footnote{The $\mathbf{T}_{\tilde{K}}$ matrix in the theorem is simply a scaled identity with scaling factor converging to $t_{SCP}(-1, \lambda_j)$, as given below.}
\begin{align}
\gamma^{SCP, UL}\left(\lambda_j\right) =  \lambda_j t_{SCP}(-1, \lambda_j),
\end{align}
where
\begin{align}
t_{SCP}(z, \lambda_j) = \frac{1}{-z + \frac{\beta \lambda_j}{1+\lambda_j t_{SCP}(z, \lambda_j)}}.\label{t_SCP}
\end{align}
Thus, $\gamma^{SCP, UL}\left(\lambda_j\right)$ satisfies
\begin{align}
\gamma^{SCP, UL}\left(\lambda_j\right) =  \frac{\lambda_j}{1 + \frac{\beta \lambda_j}{1+\gamma^{SCP, UL}\left(\lambda_j\right)}}. \label{eq:GammaSCP_UL}
\end{align}

Applying \eqref{eq:conv_tr}, we can show that
\begin{align}
\frac{1}{N}\textrm{tr} \left[\left(\frac{\lambda_j}{N}\sum_{k'}\mathbf{h}_{k',j,j}^H\mathbf{h}_{k',j,j} + \mathbf{I} \right)^{-1}\right] \stackrel{a.s.}{\longrightarrow} t_{SCP}(-\rho, \lambda_j). \label{eq:SCP_tr_limit}
\end{align}

Furthermore, applying Corollary \ref{theo:Xi}, we get that
\begin{align}
\frac{1}{N} \textrm{tr} \left[\left(\frac{\lambda_j}{N}\sum_{k'}\mathbf{h}_{k',j,j}^H\mathbf{h}_{k',j,j} + \mathbf{I} \right)^{-2}\right]
\end{align}
converges a.s. as $N, K \rightarrow \infty, \frac{K}{N} \rightarrow \beta$, to
\begin{align}
-\frac{d}{d\rho}t_{SCP}(-\rho, \lambda_j)\Big|_{\rho = 1}.
\end{align}
From \eqref{t_SCP}, $t_{SCP}(-\rho, \lambda_j) = \frac{1}{\rho + \frac{\beta \lambda_j}{1+\lambda_j t_{SCP}(-\rho, \lambda_j)}},$
so that
\begin{align}
\frac{d}{d\rho}t_{SCP}(-\rho, \lambda_j) = -\frac{t_{SCP}(-\rho, \lambda_j)}{\left[\rho + \frac{\beta \lambda_j}{(1+\lambda_j t_{SCP}(-\rho, \lambda_j))^2}
\right]}.  \label{eq:SCP_tr2_limit}
\end{align}

\subsection{CBf}\label{app:CBf_UL}
Here too, we may apply Theorem  \ref{theo:UL_SIR_conv} when $\lambda_{k,j}$'s for all users in cell $j$ fixed at $\lambda_j$, and the values of $\mu_j$ also held constant in both cells.
For any user in cell $j$, we will need to characterize
\begin{align}
\frac{\lambda_j}{N} \mathbf{h}_{k,j,j} \left(\mathbf{Y}_{k,j}\mathbf{Y}_{u,j}^H+ \mu_j \mathbf{I} \right)^{-1} \mathbf{h}_{k,j,j}^H,
\end{align}
where
\begin{align}
&\mathbf{Y}_{k,j}\mathbf{Y}_{k,j}^H\nonumber \\&= \frac{\lambda_j}{N}\sum_{k' \neq k}\mathbf{h}_{k',j,j}^H\mathbf{h}_{k',j,j} + \frac{\lambda_{\bar{j}}}{N}\sum_{k' }\mathbf{h}_{k',\bar{j},j}^H\mathbf{h}_{k',\bar{j},j}.
\end{align}
From Theorem  \ref{theo:UL_SIR_conv}, this quantity will converge a.s. as $N, K \rightarrow \infty$ with $\frac{K}{N} \rightarrow \beta$, to\footnote{This uses the fact that $\frac{K-1}{K} \rightarrow 1$ and $\frac{K}{N} \rightarrow \beta$ as $K \rightarrow \infty$.}
\begin{align}
\gamma^{CBf, UL}\left(\mu_j, \lambda_j, \lambda_{\bar{j}}\right) =  \lambda_j t_{CBf}(-\mu_j, \lambda_j, \lambda_{\bar{j}}),
\end{align}
where
\begin{align}
&t_{CBf}(z, \lambda_j, \lambda_{\bar{j}}) \nonumber \\
&= \frac{1}{-z+  \frac{\beta \lambda_j}{\left(1+ \lambda_j t_{CBf}(z, \lambda_j, \lambda_{\bar{j}})\right)}+
 \frac{\beta \epsilon \lambda_{\bar{j}}}{\left(1+ \epsilon \lambda_{\bar{j}} t_{CBf}(z, \lambda_j, \lambda_{\bar{j}})\right)}}.\label{eq:t_CBf}
\end{align}
Thus, $\gamma^{CBf, UL}\left(\mu_j, \lambda_j, \lambda_{\bar{j}}\right)$ satisfies
\begin{align}
&\gamma^{CBf, UL}\left(\mu_j, \lambda_j, \lambda_{\bar{j}}\right) \nonumber \\
&=  \frac{\lambda_j }{\mu_j +  \frac{\beta \lambda_j}{1+ \gamma^{CBf, UL}\left(\mu_j, \lambda_j, \lambda_{\bar{j}}\right)}+
 \frac{\beta \epsilon \lambda_{\bar{j}}}{1+ \epsilon \frac{\lambda_{\bar{j}}}{\lambda_j} \gamma^{CBf, UL}\left(\mu_j, \lambda_j, \lambda_{\bar{j}}\right)}}.
\end{align}

Applying \eqref{eq:conv_tr}, we can show that
\begin{align}
&\frac{1}{N} \textrm{tr} \left(\sum_{j'=1}^2 \frac{\lambda_{j'}}{N}\sum_{k'}\mathbf{h}_{k',j',j}^H\mathbf{h}_{k',j',j} + \mu_j \mathbf{I} \right)^{-1} \nonumber \\
&~~~~~~~ \stackrel{a.s.}{\longrightarrow} t_{CBf}(-\rho, \lambda_j, \lambda_{\bar{j}}). \label{eq:CBf_tr_limit}
\end{align}

Moreover, by Corollary \ref{theo:Xi}, we can also show that
\begin{align}
\frac{1}{N} \textrm{tr} \left(\sum_{j'=1}^2 \frac{\lambda_{j'}}{N}\sum_{k'}\mathbf{h}_{k',j',j}^H\mathbf{h}_{k',j',j}+ \mu_j \mathbf{I} \right)^{-2}
\end{align}
converges a.s. as $N, K \rightarrow \infty, \frac{K}{N} \rightarrow \beta$, to
\begin{align}
-\frac{d}{d\rho}t_{CBf}(-\rho, \lambda_j, \lambda_{\bar{j}})\Big|_{\rho = \mu_j}. \label{eq:Xi_CBF}
\end{align}
From \eqref{eq:t_CBf},
\begin{align}
&\frac{d}{d\rho}t_{CBf}(-\rho, \lambda_j, \lambda_{\bar{j}})\nonumber \\
&= \frac{- t_{CBf}(-\rho, \lambda_j, \lambda_{\bar{j}})}{\rho+  \frac{\beta \lambda_j}{\left(1+ \lambda_j t_{CBf}(-\rho, \lambda_j, \lambda_{\bar{j}})\right)^2}+
 \frac{\beta \epsilon \lambda_{\bar{j}}}{\left(1+ \epsilon \lambda_{\bar{j}} t_{CBf}(-\rho, \lambda_j, \lambda_{\bar{j}})\right)^2}}. 
\end{align}

\subsection{MCP}\label{app:MCP_UL}
For $\lambda_{k,j}$'s in cell $j$ equal to a constant $\lambda_j$ and the $\mu_j$'s also fixed held fixed, for any user in cell $j$, we will need an asymptotic expression for (cf. Eq. \eqref{eq:MCP_uplink_SINR})
\begin{align}
\frac{\lambda_{j}}{N} \mathbf{\breve{h}}_{k,j}\left[
\frac{\lambda_j}{N} \sum_{k' \neq k}  \mathbf{\breve{h}}_{k',j}^H  \mathbf{\breve{h}}_{k',j}
+ \frac{\lambda_{\bar{j}}}{N} \sum_{k'}  \mathbf{\breve{h}}_{k',{\bar{j}}}^H  \mathbf{\breve{h}}_{k',{\bar{j}}}
+ \mathbf{I}\right]^{-1} \mathbf{\breve{h}}_{k,j}^H.\label{eq:MCP_1}
\end{align}

Introducing the short-hand $\boldsymbol{\eta} = (\mu_j, \mu_{\bar{j}}, \lambda_j, \lambda_{\bar{j}})$ , applying Theorem  \ref{theo:UL_SIR_conv}, we can show that \eqref{eq:MCP_1} converges a.s. as $N, K \rightarrow \infty$ with $\frac{K}{N} \rightarrow \beta$, to
$\gamma_1^{MCP, UL}\left(\mu_j, \mu_{\bar{j}}, \lambda_j\right)$ equal to
\begin{align}
&\gamma_1^{MCP, UL}\left(\boldsymbol{\eta}\right) =  \lambda_j \left(\frac{t_{1,MCP}(-1, \boldsymbol{\eta})}{\mu_j} + \frac{\epsilon  t_{2, MCP}(-1, \boldsymbol{\eta})}{\mu_{\bar{j}}}\right), \label{eq:G1_MCP}
\end{align}
where $t_{1,MCP}(z, \boldsymbol{\eta})$ and $t_{2,MCP}(z, \boldsymbol{\eta})$ are given by \eqref{eq:t_MCP}.
\begin{figure*}
\begin{align}
&t_{1, MCP}(z, \boldsymbol{\eta}) = \frac{1}{-z+\frac{\beta \frac{\lambda_j}{\mu_j}}{1+ \frac{\lambda_j}{\mu_j} t_{1, MCP}(z, \boldsymbol{\eta})+ \frac{\epsilon}{\mu_{\bar{j}}} \lambda_j t_{2, MCP}(z, \boldsymbol{\eta})}+\frac{\epsilon \beta \frac{\lambda_{\bar{j}}}{\mu_j}}{1+\epsilon \frac{\lambda_{\bar{j}}}{\mu_j} t_{1, MCP}(z, \boldsymbol{\eta})+ \frac{1}{\mu_{\bar{j}}} \lambda_{\bar{j}} t_{2, MCP}(z, \boldsymbol{\eta})}}, \nonumber \\
&t_{2, MCP}(z, \boldsymbol{\eta}) = \frac{1}{-z+\frac{\beta \epsilon \frac{\lambda_j}{\mu_{\bar{j}}}}{1+ \frac{1}{\mu_j} \lambda_j t_{1, MCP}(z, \boldsymbol{\eta})+\epsilon \frac{\lambda_j}{\mu_{\bar{j}}} t_{2, MCP}(z, \boldsymbol{\eta})}+\frac{\beta \frac{\lambda_{\bar{j}}}{\mu_{\bar{j}}}}{1+ \frac{\epsilon}{\mu_j} \lambda_{\bar{j}} t_{1, MCP}(z, \boldsymbol{\eta})+ \frac{\lambda_{\bar{j}}}{\mu_{\bar{j}}} t_{2, MCP}(z, \boldsymbol{\eta})}}. \label{eq:t_MCP}
\end{align}
\end{figure*}

For a user in cell $\bar{j}$,  one can verify that \eqref{eq:MCP_1} (replace $j$ by $\bar{j}$ and vice versa)  converges to
\begin{align}
\gamma_2^{MCP, UL}\left(\boldsymbol{\eta}\right) =  \lambda_{\bar{j}} \left(\epsilon \frac{t_{1,MCP}(-1, \boldsymbol{\eta})}{\mu_{j}} + \frac{t_{2, MCP}(-1, \boldsymbol{\eta})}{\mu_{\bar{j}}}\right). \label{eq:G2_MCP}
\end{align}

Plugging in \eqref{eq:G1_MCP} and \eqref{eq:G2_MCP} in the expressions for $t_{1, MCP}(-1, \boldsymbol{\eta})$ and $t_{2, MCP}(-1, \boldsymbol{\eta})$, (cf. \eqref{eq:t_MCP}), we get
\begin{align}
&t_{1, MCP}(-1, \boldsymbol{\eta}) = \frac{1}{1+\frac{\beta \frac{\lambda_j}{\mu_j}}{1+ \gamma_1^{MCP, UL}\left(\boldsymbol{\eta}\right)}+\frac{\epsilon \beta \frac{\lambda_{\bar{j}}}{\mu_j}}{1+\gamma_2^{MCP, UL}\left(\boldsymbol{\eta}\right)}}, \nonumber \\
&t_{2, MCP}(-1, \boldsymbol{\eta}) = \frac{1}{1+\frac{\beta \epsilon \frac{\lambda_j}{\mu_{\bar{j}}}}{1+ \gamma_1^{MCP, UL}\left(\boldsymbol{\eta}\right)}+\frac{\beta \frac{\lambda_{\bar{j}}}{\mu_{\bar{j}}}}{1+ \gamma_2^{MCP, UL}\left(\boldsymbol{\eta}\right)}}. \label{eq:t_MCP_1}
\end{align}

Now using \eqref{eq:t_MCP_1} in \eqref{eq:G1_MCP} and \eqref{eq:G2_MCP}, $\gamma_1^{MCP, UL}\left(\boldsymbol{\eta}\right)$ and $\gamma_2^{MCP, UL}\left(\boldsymbol{\eta}\right)$ will satisfy
\begin{align}
&\gamma_1^{MCP, UL}\left(\boldsymbol{\eta}\right) \nonumber \\
&= \lambda_j \left[\frac{1}{\mu_j+\frac{\beta \lambda_j}{1+ \gamma_1^{MCP, UL}\left(\boldsymbol{\eta}\right)}+\frac{\epsilon \beta \lambda_{\bar{j}}}{1+\gamma_2^{MCP, UL}\left(\boldsymbol{\eta}\right)}} \right. \nonumber \\
& \quad\quad\quad \left.
+ \frac{\epsilon}{
\mu_{\bar{j}}+\frac{\beta \epsilon \lambda_j}{1+ \gamma_1^{MCP, UL}\left(\boldsymbol{\eta}\right)}+\frac{\beta \lambda_{\bar{j}}}{1+ \gamma_2^{MCP, UL}\left(\boldsymbol{\eta}\right)}
} \right], \nonumber \\
&\gamma_2^{MCP, UL}\left(\boldsymbol{\eta}\right) \nonumber \\
&= \lambda_{\bar{j}} \left[\frac{\epsilon}{\mu_j+\frac{\beta \lambda_j}{1+ \gamma_1^{MCP, UL}\left(\boldsymbol{\eta}\right)}+\frac{\epsilon \beta \lambda_{\bar{j}}}{1+\gamma_2^{MCP, UL}\left(\boldsymbol{\eta}\right)}}
\right. \nonumber \\
& \quad\quad\quad \left. +
\frac{1}{
\mu_{\bar{j}}+\frac{\beta \epsilon \lambda_j}{1+ \gamma_1^{MCP, UL}\left(\boldsymbol{\eta}\right)}+\frac{\beta \lambda_{\bar{j}}}{1+ \gamma_2^{MCP, UL}\left(\boldsymbol{\eta}\right)}
} \right].\label{eq:G1G2}
\end{align}

Now define
\begin{align}
\mathbf{D}_{k,1} = \left[\begin{array}{cc} \frac{\lambda_1}{\mu_1} \mathbf{I}_N & \mathbf{0}_{N \times N} \\ \mathbf{0}_{N \times N} & \epsilon \frac{\lambda_{1}}{\mu_2} \end{array} \right] \label{eq:MCP_D1} \\
\mathbf{D}_{k,2} = \left[\begin{array}{cc} \epsilon \frac{\lambda_2}{\mu_1} \mathbf{I}_N & \mathbf{0}_{N \times N} \\ \mathbf{0}_{N \times N} & \frac{\lambda_{2}}{\mu_2} \end{array} \right].  \label{eq:MCP_D2}
\end{align}
Applying \eqref{eq:conv_tr}, we can show that
\begin{align}
&\frac{1}{N} \textrm{tr} \mathbf{D}_{k,j} \left(\sum_{j'=1}^2 \frac{\lambda_{j'}}{N} \sum_{k'}  \mathbf{\breve{h}}_{k',j'}^H  \mathbf{\breve{h}}_{k',j'}
+ \mathbf{I} \right)^{-1} \nonumber \\
&~~~~~\stackrel{a.s.}{\longrightarrow} \gamma_j^{MCP, UL}\left(\boldsymbol{\eta}\right). \label{eq:MCP_tr_limit}
\end{align}

We also need to characterize
\begin{align}
\frac{1}{N} \textrm{tr} \mathbf{D}_{k,j} \left[
\sum_{j'=1}^2 \frac{\lambda_{j'}}{N} \sum_{k'}  \mathbf{\breve{h}}_{k',j'}^H  \mathbf{\breve{h}}_{k',j'}
+ \mathbf{I}\right]^{-2}, \label{eq:MCP_2}
\end{align}
 for the special case where $\mu_j = \mu_{\bar{j}}= \mu$ and $\lambda_j = \lambda_{\bar{j}} = \lambda$, i.e. $\boldsymbol{\eta} = \boldsymbol{\eta}_{sym} = \left[\mu, \mu, \lambda, \lambda\right]$.
In this case,
\begin{align}
&t_{1, MCP}(-\rho, \boldsymbol{\eta}_{sym}) = t_{2, MCP}(-\rho, \boldsymbol{\eta}_{sym}) = t_{MCP}(-\rho, \boldsymbol{\eta}_{sym})\nonumber \\
&= \frac{1}{\rho+\frac{\left(1+\epsilon\right) \beta \lambda}{\mu+ \left(1+\epsilon\right) \lambda t_{MCP}(-\rho, \boldsymbol{\eta}_{sym})}} .
 \label{eq:t_MCP_eq}
\end{align}
Thus,
\begin{align}
&\frac{d}{d\rho}t_{MCP}(-\rho, \boldsymbol{\eta}_{sym}) = -\frac{t_{MCP}(-\rho, \boldsymbol{\eta}_{sym})}{\rho+\frac{\left(1+\epsilon\right) \beta \lambda \mu}{\left(\mu+ \left(1+\epsilon\right) \lambda t_{MCP}(-\rho, \boldsymbol{\eta}_{sym})\right)^2}},
\end{align}
and, by Corollary \ref{theo:Xi}, \eqref{eq:MCP_2} converges a.s. as $N, K \rightarrow \infty$ with $\frac{K}{N} \rightarrow \beta$ to
\begin{align}
- \frac{(1+\epsilon) \lambda}{\mu} \frac{d}{d\rho}t_{MCP}(-\rho, \boldsymbol{\eta}_{sym})\Big|_{\rho = 1}.
\end{align}

Finally, we will need to characterize
\begin{align}
\frac{1}{N} \textrm{tr} \mathbf{D}_{k,j} \mathbf{A} \mathbf{D}_{k',j'} \mathbf{A} \label{eq:MCP_trD1D2}
\end{align}
with $\mathbf{A}$ as defined in \eqref{eq:A_MCP}, where in $\lambda_1$ and $\lambda_2$ in $\mathbf{D}_{k,1}$, $\mathbf{D}_{k',2}$ are equal to $\bal$ for all $k \le K$. 
For any sequence of diagonal matrices $\mathbf{S}$ with bounded diagonal entries, and taking into account the fact that the $\mathbf{D}_{k,j}$'s are equal for all users in the same cell, define $V_{j}\left(\mathbf{S}\right) = \frac{1}{N} \textrm{tr} \mathbf{S} \mathbf{A} \mathbf{D}_{1,j} \mathbf{A}$, for $j = 1, 2$.
Applying Proposition \ref{prop:Ul}, we get that
\begin{align}
V_1\left(\mathbf{S}\right)  &= \beta
\frac{t_{MCP}^2 \bal^2 \left(1+\epsilon^2\right)}{\left(1+(1+\epsilon) \bal t_{MCP} \right)^2} V_{1}\left(\mathbf{S}\right) \nonumber \\
&+
\beta \frac{2 \epsilon t_{MCP}^2 \bal^2}{\left(1+(1+\epsilon) \bal t_{MCP} \right)^2} V_2\left(\mathbf{S}\right) \nonumber \\
&+ \frac{t_{MCP}^2}{N} \textrm{tr} \mathbf{S} \mathbf{D}_{1, 1}
+ \eta_1, \\
V_2\left(\mathbf{S}\right) &=
\beta \frac{2 \epsilon t_{MCP}^2 \bal^2}{\left(1+(1+\epsilon) \bal  t_{MCP} \right)^2}
 V_1\left(\mathbf{S}\right) \nonumber \\
&+ \beta \frac{t_{MCP}^2 \bal^2 \left(1+\epsilon^2\right)}{\left(1+(1+\epsilon) \bal t_{MCP} \right)^2} V_2\left(\mathbf{S}\right) \nonumber \\
&+  \frac{t_{MCP}^2}{N} \textrm{tr} \mathbf{S} \mathbf{D}_{1, 2}  + \eta_2,
\end{align}
where $\eta_1$ and $\eta_2$ satisfy $\mathbb{E} |\eta_j| \le C N^{-1/2}$, $j = 1, 2$, and $t_{MCP} = t_{MCP}(1, 1, \bal, \bal)$, for some constant $C$.

For $\mathbf{S} = \mathbf{D}_{1,1}$, we obtain \eqref{V1D1} and \eqref{V2D1}, where $\phi_1$ and $\phi_2$ denote deviation terms such that
$\mathbb{E} |\phi_j| \le C N^{-1/2}$.
%

\begin{figure*}
\begin{align}
V_1\left(\mathbf{D}_{1,1}\right) &=  t_{MCP}^2 \bal^2 \left(1+\epsilon\right)^2\frac{
\frac{1+\epsilon^2}{(1+\epsilon)^2}
- \beta \frac{(1-\epsilon)^2 \bal^2 t_{MCP}^2}{\left(1+(1+\epsilon) \bal t_{MCP} \right)^2}
}{\left[1-\beta
 \frac{(1+\epsilon)^2 \bal^2 t_{MCP}^2}{\left(1+(1+\epsilon) \bal t_{MCP} \right)^2}\right]
\left[1- \beta \frac{(1-\epsilon)^2 \bal^2 t_{MCP}^2}{\left(1+(1+\epsilon) \bal t_{MCP} \right)^2}\right] }
+ \phi_1
 \label{V1D1}\\
V_2\left(\mathbf{D}_{1,1}\right) &= t_{MCP}^2 \bal^2 \left(1+\epsilon\right)^2 \frac{
\frac{2\epsilon}{(1+\epsilon)^2}}{\left[1- \beta
 \frac{(1+\epsilon)^2 \bal^2 t_{MCP}^2}{\left(1+(1+\epsilon) \bal t_{MCP} \right)^2}\right]
\left[1-\beta \frac{(1-\epsilon)^2 \bal^2 t_{MCP}^2}{\left(1+(1+\epsilon) \bal t_{MCP} \right)^2}\right]
}
+ \phi_2.\label{V2D1}
\end{align}
\end{figure*}
Note that by definition, $V_1\left(\mathbf{D}_{1,2}\right) = V_2\left(\mathbf{D}_{1,1}\right)$, and $V_2\left(\mathbf{D}_{1,2}\right) = V_1\left(\mathbf{D}_{1,1}\right)$.

\section{A simple monotonicity result} \label{appendix:monotonicity}

Let $I(\bol)$ be a standard interference function for the UL, in the sense of Yates~\cite{yates95}, where we denote the transmit power vector by $\bol$. Suppose it is of the form
\begin{align}
I_k(\bol) &= \gamma_k F_k(\bol)~~~k=1, 2, \ldots, K
\end{align}
where $\gamma_k$ is the SINR target for user $k$, and $F(\bol)$ is a standard interference function. The vector $\bol$ is called {\it feasible} if
\begin{align}
\lambda_k &\geq \gamma_k F_k(\bol)~~~k=1, 2, \ldots, K.
\end{align}
It is shown in \cite{yates95}, Theorem 1, that if a feasible solution exists, then function $I$ has a unique, positive, fixed point.
\cite{yates95} also shows that, starting at any power vector $\bol$, the iterative power control $I^n(\bol)$, $n=1,2, \ldots$ converges to it.
Using the fixed-point powers, all users achieve exactly their SINR target.

Now consider two different vectors of SINR targets $\bog^{(1)}$ and $\bog^{(2)}$, and let $\bol^{(1)}$ and $\bol^{(2)}$ denote the corresponding fixed points. Lemma \ref{lem:monotonicity} is a simple corollary of Lemma~1 in \cite{yates95}.

\begin{lemma}
If $\bog^{(1)} \leq \bog^{(2)}$ then $\bol^{(1)} \leq \bol^{(2)}$. \label{lem:monotonicity}
\end{lemma}
\begin{IEEEproof}
$\bol^{(2)}$ is feasible for the power control problem with SINR targets given by $\bog^{(1)}$. By \cite{yates95}, Lemma 1, $I^n\left(\bol^{(2)}\right)$, $n=1, 2, \ldots$ is a monotone decreasing sequence of feasible power vectors that converges to $\bol^{(1)}$.
\end{IEEEproof}

\section{Proof of Theorem~\ref{THEO_BF_SCP}} \label{appendix:SCP_dual_LAS}
Throughout this section, let
\begin{align}
\mathbf{A}_{j} = \left(\mathbf{I} + \frac{\bal}{N} \sum_{l = 1}^K
\mathbf{h}_{l, j,j}^H\mathbf{h}_{l, j, j}\right)^{-1} \\
\mathbf{A}_{k,j} = \left(\mathbf{I} + \frac{\bal}{N} \sum_{l \neq k}
\mathbf{h}_{l, j,j}^H\mathbf{h}_{l, j, j}\right)^{-1} \\
\mathbf{A}_{k, k', j} = \left(\mathbf{I} + \frac{\bal}{N} \sum_{l \neq (k, k')}
\mathbf{h}_{l, j,j}^H\mathbf{h}_{l, j, j}\right)^{-1},
\end{align}
where $\bal > 0$ will be defined later. \\

{\it Asymptotic analysis of the dual problem}

We start by considering the dual problem at each of the BSs: this will yield the asymptotically optimal dual variables and beamforming directions in both dual and primal problems. Note that even though the value of the dual objective function in one cell depends on the primal beamforming decisions through the $\sigma_{k,j}$s in \eqref{eq:SCP_dual}, the optimal dual variables themselves are fully determined by the constraints and will be the unique strictly positive solutions to \eqref{eq:SCP_int_fn}. Since the analysis is identical in both cells, without loss of generality, assume the cell index $j$ is $j = 1$. 
 Assume also that $\displaystyle \beta
\left(\frac{\gamma}{1+\gamma} + \epsilon \gamma\right) < 1$. As
noted in Section~\ref{c5:sec:LAS}, we cannot immediately apply
standard large system analysis to \eqref{eq:SCP_uplink_SINR} as
optimal dual variables $\lambda_{k,1}$'s are not 
 independent of the channel vectors $\left(\mathbf{h}_{k,1,1}\right)_{k=1}^K$.

Rather than directly analyze the asymptotic performance of the SCP
system using {\it optimal} $\lambda_{k,1}$'s, consider any constant $\bal > 0$ (later, we will assign it a particular value
, but for now it is arbitrary) and consider the large system
regime in which all users have the same transmit UL power of
$\bal/N$, as $N,K\rightarrow \infty$, with fixed ratio $ K/N = \beta$.
This is a virtual UL with noise of unit power, so
the SINR for user $k$ 
 is given by
\begin{align}
\frac{\bal}{N} \mathbf{h}_{k,1,1} \left(
\mathbf{I} + \frac{\bal}{N} \sum_{k' \neq k}
\mathbf{h}_{k', 1,1}^H\mathbf{h}_{k', 1,1} \right)^{-1}
\mathbf{h}_{k,1,1}^H.  \label{eq:SCP_uplink_SINR_app}
\end{align}
Appendix \ref{app:SCP_UL} shows how this converges a.s. to a constant in the considered asymptotic regime.

 In particular,
if $\delta$ is small enough so that $\displaystyle \beta
\frac{\gamma+\delta}{1+ \gamma + \delta} < 1$, and we assign the
following particular value to $\bal$:
\begin{align}
\bald := \frac{\gamma+\delta}{1 - \beta \frac{\gamma+\delta}{1+
\gamma + \delta}},
\end{align}
which we denote by
$\bald$ to make explicit its dependence on the parameter $\delta$,
then \eqref{eq:SCP_uplink_SINR_app} will converge a.s. to $\gamma + \delta$.
Theorem 3 in \cite{kammoun_it09} shows that for this (suboptimal) system, the value of \eqref{eq:SCP_uplink_SINR_app} is
\begin{align}
\gamma_k(\delta) = \gamma + \delta + \mathcal{O}\left(\sqrt{\frac{1}{N}}\right), ~~~\forall k = 1, 2, \ldots, K, \label{eq:suboptimal_SINR_limit}
\end{align}
where $\gamma_k(\delta)$ denotes the dual UL SINR of user $k$ under this suboptimal power allocation, and the last term on the right hand side is $1/\sqrt{N}$ times a r.v. that converges weakly to a zero-mean Gaussian distributed r.v. This holds since $\frac{K}{N} \rightarrow \beta < \infty$, for the considered channel model and in the notation of the theorem, $\Gamma_{\tilde{K}}$ and $\Theta_{\tilde{K}}$ will converge a.s. to bounded limits.

Denote by $\bobld$ the vector of UL powers in this suboptimal system, where all entries have the same value $\bld$. For $\delta > 0$, it follows from
\eqref{eq:suboptimal_SINR_limit} that for any particular $k$,
$\displaystyle {\mathbb P}(\gamma_k(-\delta) > \gamma \mbox{~or~}
\gamma_k(\delta) < \gamma)$ decays to 0 exponentially in
$N$ (it is a large deviation event). Applying the union bound,
we obtain that
\begin{equation}
\gamma_k(-\delta) \leq \gamma \leq \gamma_k(\delta) ~~\forall k=1,
2, \ldots K \label{eq:gamma_bounds}
\end{equation}
will hold with probability tending to $1$ as $N \uparrow \infty$. Let $\boldsymbol{\gamma}(-\delta)$ and $\boldsymbol{\gamma}(\delta)$ denote the vectors grouping the left and right-hand sides of \eqref{eq:gamma_bounds}, respectively.

Denote by $\bol$ the vector of {\it optimal} UL powers for the dual problem, which achieves SINR of $\gamma$ for each user.
On the other hand, $\boldsymbol{\bar{\lambda}}(-\delta)$ and $\boldsymbol{\bar{\lambda}}(\delta)$ (defined above) are the vectors of \emph{optimal} UL powers for the virtual UL problem with target SINR vectors $\boldsymbol{\gamma}(-\delta)$ and $\boldsymbol{\gamma}(\delta)$, respectively (instead of the all-$\gamma$ vector in the original dual problem).
 Thus, by
Lemma~\ref{lem:monotonicity} in Appendix~\ref{appendix:monotonicity}, 
\begin{align}
\boblmd \leq \bol \leq \bobld
\end{align}
will hold with probability tending to $1$ as $N \uparrow \infty$.
Since this holds for any sufficiently small $\delta >0$, the empirical distribution of the components of the
optimal $\bol$ will converge weakly to the constant $\bal$,
given in \eqref{eq:SCP_bal}. Let \eqref{eq:SCP_bal} provide the
particular value 
of $\bal$ in the
rest of this section.
This establishes the asymptotic optimality of having the $\lambda_{k,j}$'s in both cells all equal to $\bar{\lambda}$, and consequently 
that of the beamforming vectors
\begin{align}
\mathbf{\hat{w}}_{k,j} = \mathbf{A}_{k,j} \mathbf{h}_{k,j,j}^H. \label{app:scp_bf_vec}
\end{align}
By UL-DL duality and from the KKT conditions, these are, up to a scale factor, also the optimal DL beamforming vectors  \cite{viswanath_it03}. 
Thus, this analysis shows that, asymptotically, the DL beamforming {\it directions} in one cell do not depend on the beamforming {\it directions} used in the other cell, although it is clear that the optimal DL power levels (which modulate the beamforming directions) {\it will} depend on the power levels used in the other cell, even in the limit. Finally, also note that the dual objective function value $\frac{\bal}{N} \sum_{k=1}^K \sigma^2_{k,j}$ is an upper bound on the primal objective function, i.e. on the total transmit power in the cell; denote the latter by $\bar{P}_j$ for cell $j$.\\

{\it Asymptotic analysis of the primal problems}

We now turn to the DL primal problems, and fix the beamforming directions in both cells to be those given by \eqref{app:scp_bf_vec}.
Thus, only the DL power levels $p_{kj}$ in \eqref{SCP:bf_vec_form} need to be determined.
These must satisfy, for all users in both cells,
\begin{align}
&p_{kj} = \frac{\sigma^2_{k,j} + \sum_{k' \neq k} \frac{p_{k'j}}{N} \frac{\left|\mathbf{h}_{k, j,j} \mathbf{\hat{w}}_{k'j}\right|^2}{\|\mathbf{\hat{w}}_{k'j}\|^2}}{\frac{1}{N \gamma} \frac{\left|\mathbf{h}_{k,
j,j}\mathbf{\hat{w}}_{kj} \right|^2}{\|\mathbf{\hat{w}}_{kj}\|^2} },  \label{app:scp_pkj} \\
&\sigma_{k,j}^2 = \sigma^2 + \sum_{k'=1}^K \frac{p_{k'\bar{j}}}{N} \frac{|\mathbf{h}_{k,j,\bar{j}} \mathbf{\hat{w}}_{k'\bar{j}}|^2}{\|\mathbf{\hat{w}}_{k'\bar{j}}\|^2}. \label{app:scp_sigmakj}
\end{align}
Assuming feasibility, one can verify using similar standard interference function arguments as for the dual problem, that the set of equations in both cells will have a unique power minimizing solution \cite{farrokhi_jsac98}.

As was the case with asymptotic analysis of the dual problem, it is easier to first fix the DL powers to constants and study the resulting limiting regime.
Thus, assume $p_{kj}$'s in cell $j$ are held fixed at a constant value $\bar{p}_j~~j=1,2$, and note that this implies that $\bar{P}_j = \beta \bar{p}_j$. After analyzing this limiting regime, we optimize the choice of the constants, $\bar{p}_1$ and $\bar{p}_2$ and finally show that the optimal constants are asymptotically optimal with respect to the primal optimization problem \eqref{eq:MaxMinSINR_SCP_feasibility}.

Under the regime in which $p_{kj} = \bar{p}_j$ for all $k$ in cell $j$, the following lemmas hold.

\begin{lemma}\label{asymptotics_SCP}
With $\mathbf{\hat{w}}_{kj}$  $= \mathbf{A}_{kj} \mathbf{h}_{k,j, j}^H$, such that $\bal = \frac{\gamma}{1-\beta \frac{\gamma}{1+\gamma}}$, the following holds, as $K, N \rightarrow \infty$, $\frac{K}{N} = \beta$:
\begin{align}
\max_{j = 1, 2, k \le K} \left|
\frac{
 \left|\mathbf{h}_{k,j,j}\mathbf{\hat{w}}_{kj}\right|^2
}{N \|\mathbf{\hat{w}}_{kj}\|^2}- \left[1 - \frac{\beta \gamma^2}{\left(1+\gamma\right)^2}\right]
\right|
\stackrel{a.s.}{\longrightarrow} 0 \\
\max_{j = 1, 2, k, l \le K} \left|\frac{\left|\mathbf{h}_{k,j,\bar{j}} \mathbf{\hat{w}}_{l\bar{j}}\right|^2}{\|\mathbf{\hat{w}}_{l\bar{j}}\|^2}-
\epsilon
\right| \stackrel{a.s.}{\longrightarrow} 0 \label{eq:SCP_OtherInter} \\
\max_{j = 1, 2, k, l \le K, l \neq k} \left|\frac{\left|\mathbf{h}_{k,j,j} \mathbf{\hat{w}}_{lj}\right|^2}{\|\mathbf{\hat{w}}_{lj}\|^2}-
\frac{1}{\left(1+\gamma\right)^2}
\right| \stackrel{a.s.}{\longrightarrow} 0.  \label{eq:SCP_OwnInter}
\end{align}
\end{lemma}
\begin{IEEEproof}
Applying Lemma 5.1 in \cite{liang_it07},
\begin{align}
\max_{j = 1, 2, k \le K} \left|\frac{\bal}{N} \|\mathbf{\hat{w}}_{kj}\|^2 - \frac{1}{N} \textrm{tr} \mathbf{D}_{k,j,j} \mathbf{A}_{k,j}^2\right| \stackrel{a.s.}{\longrightarrow} 0 \label{eq:Norm_SCP} \\
\max_{j = 1, 2, k \le K} \left|\frac{\bal}{N} \mathbf{h}_{k,j,j}\mathbf{\hat{w}}_{kj} - \frac{1}{N}  \textrm{tr} \mathbf{D}_{k,j,j} \mathbf{A}_{k,j}\right| \stackrel{a.s.}{\longrightarrow} 0, \label{eq:UsefulSignal_SCP}
\end{align}
where\footnote{We introduce the $\mathbf{D}_{k,j,i}$ matrices to allow for more general formulations.} $\mathbf{D}_{k,j, j} = \bal \mathbf{I}$. Later on in this section, $\mathbf{D}_{k,j, \bar{j}} = \epsilon \bal \mathbf{I}$.
Now applying Lemma \ref{lemm:633}  twice to \eqref{eq:Norm_SCP} and once in \eqref{eq:UsefulSignal_SCP}, we get
\begin{align}
\max_{j=1, 2, k \le K} \left|\frac{\bal}{N} \|\mathbf{\hat{w}}_{kj}\|^2 - \frac{1}{N} \textrm{tr} \mathbf{D}_{k,j, j} \mathbf{A}_{j}^2\right| \stackrel{a.s.}{\longrightarrow} 0 \label{eq:Norm_SCP_b} \\
\max_{j=1, 2, k \le K} \left|\frac{\bal}{N} \mathbf{h}_{k,j,j}\mathbf{\hat{w}}_{kj} - \frac{1}{N} \textrm{tr} \mathbf{D}_{k,j,j} \mathbf{A}_{j}\right| \stackrel{a.s.}{\longrightarrow} 0 \label{eq:UsefulSignal_SCP_b}.
\end{align}

Now consider the interference terms, with $i = 1, 2$, and $(i,l) \neq (k,j)$,
\begin{align}
&\frac{\left|\mathbf{h}_{k,j,i} \mathbf{\hat{w}}_{l i}\right|^2}{\|\mathbf{\hat{w}}_{l i}\|^2} = \frac{1}{\bal} \frac{\frac{\bal^2}{N} \mathbf{h}_{k,j,i} \mathbf{A}_{l, i}\mathbf{h}_{l,i,i}^H
\mathbf{h}_{l,i,i} \mathbf{A}_{l, i} \mathbf{h}_{k,j,i}
^H}{\frac{\bal}{N} \|\mathbf{\hat{w}}_{li}\|^2}. \label{eq:interf_SCP}
\end{align}
Two different cases arise, depending on whether $i = j$ or $i = \bar{j}$. In the latter case, we may apply Lemma 5.1 in \cite{liang_it07} to the numerator of the right-hand side of \eqref{eq:interf_SCP},
\begin{align}
\max_{k,l \le K}&\left|\frac{\bal^2}{N} \mathbf{h}_{k,j,\bar{j}} \mathbf{A}_{l, \bar{j}}\mathbf{h}_{l,\bar{j},\bar{j}}^H
\mathbf{h}_{l,\bar{j},\bar{j}} \mathbf{A}_{l, \bar{j}} \mathbf{h}_{k,j,\bar{j}}
^H \right. \nonumber \\
&\left.~~~-  \frac{\bal}{N} \mathbf{h}_{l,\bar{j},\bar{j}} \mathbf{A}_{l, \bar{j}} \mathbf{D}_{k,j,\bar{j}} \mathbf{A}_{l, \bar{j}}\mathbf{h}_{l,\bar{j},\bar{j}}^H
 \right| \stackrel{a.s.}{\longrightarrow} 0.
\end{align}
Applying Lemma 5.1 in \cite{liang_it07} once again yields,
\begin{align}
\max_{k, l \le K}&\left|\frac{\bal}{N} \mathbf{h}_{l,\bar{j},\bar{j}} \mathbf{A}_{l, \bar{j}} \mathbf{D}_{k,j,\bar{j}} \mathbf{A}_{l, \bar{j}}\mathbf{h}_{l,\bar{j},\bar{j}}^H
 \right. \nonumber \\
&\left.~~~-  \frac{1}{N} \textrm{tr} \mathbf{D}_{l,\bar{j},\bar{j}} \mathbf{A}_{l, \bar{j}} \mathbf{D}_{k,j,\bar{j}} \mathbf{A}_{l, \bar{j}}
 \right| \stackrel{a.s.}{\longrightarrow} 0.
\end{align}
Finally applying Lemma \ref{lemm:633} twice, we get
\begin{align}
&\left|\frac{1}{N} \textrm{tr} \mathbf{D}_{l,\bar{j},\bar{j}} \mathbf{A}_{l, \bar{j}} \mathbf{D}_{k,j,\bar{j}} \mathbf{A}_{l, \bar{j}}
 - \frac{1}{N} \textrm{tr} \mathbf{D}_{l,\bar{j},\bar{j}} \mathbf{A}_{l, \bar{j}} \mathbf{D}_{k,j,\bar{j}} \mathbf{A}_{\bar{j}}
\right| \nonumber \\
&~~\le \frac{1}{N} \left\|\mathbf{D}_{l,\bar{j},\bar{j}} \mathbf{A}_{l, \bar{j}} \mathbf{D}_{k,j,\bar{j}}\right\| \le \frac{
\left\|\mathbf{D}_{l,\bar{j},\bar{j}}\right\| \left\|\mathbf{A}_{l, \bar{j}}\right\| \left\|\mathbf{D}_{k,j,\bar{j}}\right\|
}{N} \nonumber \\
&\le \frac{\epsilon \bal^2}{N}, \\
&\left|\frac{1}{N} \textrm{tr} \mathbf{D}_{l,\bar{j},\bar{j}} \mathbf{A}_{l, \bar{j}} \mathbf{D}_{k,j,\bar{j}} \mathbf{A}_{\bar{j}} -
\frac{1}{N} \textrm{tr} \mathbf{D}_{k,j,\bar{j}} \mathbf{A}_{\bar{j}} \mathbf{D}_{l,\bar{j},\bar{j}} \mathbf{A}_{\bar{j}}\right| \nonumber \\
&\le \frac{1}{N} \left\|\mathbf{D}_{l,\bar{j},\bar{j}} \mathbf{A}_{\bar{j}} \mathbf{D}_{k,j,\bar{j}}\right\|\le \frac{\epsilon \bal^2}{N}.
\end{align}
Thus,
\begin{align}
\max_{k,l \le K}&\left|\frac{\bal^2}{N} \mathbf{h}_{k,j,\bar{j}} \mathbf{A}_{l, \bar{j}}\mathbf{h}_{l,\bar{j},\bar{j}}^H
\mathbf{h}_{l,\bar{j},\bar{j}} \mathbf{A}_{l, \bar{j}} \mathbf{h}_{k,j,\bar{j}}
^H \right. \nonumber \\
&\left.~~~-  \frac{1}{N} \textrm{tr} \mathbf{D}_{k,j,\bar{j}} \mathbf{A}_{\bar{j}} \mathbf{D}_{l,\bar{j},\bar{j}} \mathbf{A}_{\bar{j}}
 \right| \stackrel{a.s.}{\longrightarrow} 0.
\end{align}

When $i$ in \eqref{eq:interf_SCP} is the same as $j$, we cannot apply Lemma 5.1 in \cite{liang_it07} directly, since $\mathbf{A}_{l, j}$ and $\mathbf{h}_{k,j,j}$ are not independent. Thus, we apply the matrix inversion lemma first to get
\begin{align}
&\frac{\bal^2}{N} \mathbf{h}_{k,j,j} \mathbf{A}_{l, {j}}\mathbf{h}_{l,{j},{j}}^H
\mathbf{h}_{l,{j},{j}} \mathbf{A}_{l, {j}} \mathbf{h}_{k,j,{j}}^H \nonumber \\
&= \frac{\frac{\bal^2}{N} \mathbf{h}_{k,j,j} \mathbf{A}_{k, l, {j}}\mathbf{h}_{l,{j},{j}}^H
\mathbf{h}_{l,{j},{j}} \mathbf{A}_{k, l, {j}} \mathbf{h}_{k,j,{j}}
^H}{\left(1 + \frac{\bal}{N} \mathbf{h}_{k,j,j} \mathbf{A}_{k, l, {j}}\mathbf{h}_{k,{j},{j}}^H \right)^2}. \label{eq:interf_SCP_samecell}
\end{align}
We can now consider the numerator in \eqref{eq:interf_SCP_samecell} and show that
\begin{align}
\max_{k,l \le K, (k \neq l)}&\left|\frac{\bal^2}{N} \mathbf{h}_{k,j,{j}} \mathbf{A}_{k, l, {j}}\mathbf{h}_{l,{j},{j}}^H
\mathbf{h}_{l,{j},{j}} \mathbf{A}_{l, {j}} \mathbf{h}_{k,j,{j}}
^H \right. \nonumber \\
&\left.~~~-  \frac{1}{N} \textrm{tr} \mathbf{D}_{k,j,{j}} \mathbf{A}_{{j}} \mathbf{D}_{l,{j},{j}} \mathbf{A}_{{j}}
 \right| \stackrel{a.s.}{\longrightarrow} 0.
\end{align}
We can also show that
\begin{align}
\max_{k,l \le K, (k, \neq l)} \left|\frac{\bal}{N} \mathbf{h}_{k,j,j} \mathbf{A}_{k, l, {j}}\mathbf{h}_{k,{j},{j}}^H-
\frac{1}{N}\textrm{tr} \mathbf{D}_{k,j,j} \mathbf{A}_{{j}} \right| \stackrel{a.s.}{\longrightarrow} 0.
\end{align}

The proof of the lemma is concluded by using the limits of the trace terms as derived in Appendix \ref{app:SCP_UL}, with $\lambda_j = \bal$, noting that for this specific case, all the $\mathbf{D}_{k,j,j}$ and $\mathbf{D}_{k,j,\bar{j}}$ matrices are equal, and are simply scaled identities, so that
\begin{align}
&\frac{1}{N} \textrm{tr} \mathbf{D}_{k,j,{j}} \mathbf{A}_{{j}} = \frac{\bal}{N} \textrm{tr} \mathbf{A}_{{j}} \stackrel{a.s.}{\longrightarrow} \gamma \\
&\frac{1}{N} \textrm{tr} \mathbf{D}_{k,j,{j}} \mathbf{A}_{{j}}^2 = \frac{\bal}{N} \textrm{tr} \mathbf{A}_{{j}}^2
\stackrel{a.s.}{\longrightarrow} \frac{\gamma}{1 + \frac{\beta \bal}{\left(1+\gamma\right)^2}} =  \frac{1}{\bal} \frac{\gamma^2}{1 - \frac{\beta \gamma^2}{\left(1+\gamma\right)^2}}
\\
&\frac{1}{N} \textrm{tr} \mathbf{D}_{k,j,{j}} \mathbf{A}_{{j}} \mathbf{D}_{l,{j},{j}} \mathbf{A}_{{j}} = \frac{\bal^2}{N} \textrm{tr} \mathbf{A}_{{j}}^2
\stackrel{a.s.}{\longrightarrow} \frac{\gamma^2}{1 - \frac{\beta \gamma^2}{\left(1+\gamma\right)^2}}
\\
&\frac{1}{N} \textrm{tr} \mathbf{D}_{k,\bar{j},{j}} \mathbf{A}_{{j}} \mathbf{D}_{l,{j},{j}} \mathbf{A}_{{j}} = \frac{\epsilon \bal^2}{N} \textrm{tr} \mathbf{A}_{{j}}^2
\stackrel{a.s.}{\longrightarrow} \epsilon \frac{\gamma^2}{1 - \frac{\beta \gamma^2}{\left(1+\gamma\right)^2}}.
\end{align}

\end{IEEEproof}

\begin{lemma}\label{lemm:4}
With $\mathbf{\hat{w}}_{kj}$  $= \mathbf{A}_{kj} \mathbf{h}_{k,j, j}^H$, such that $\bal = \frac{\gamma}{1-\beta \frac{\gamma}{1+\gamma}}$, and with $p_{kj} = \bar{p}_j$ for $k = 1, \ldots, K$, $j = 1, 2$, it follows that, with probability 1,
\begin{align}
\sigma^2_{k,j} \stackrel{a.s.}{\longrightarrow} \sigma^2 + \epsilon \bar{P}_{\bar{j}},
\end{align}
$\bar{P}_{\bar{j}}$ denotes the total transmit power of BS $\bar{j}$, i.e. $P_{\bar{j}} = \beta \bar{p}_{\bar{j}}$.
\end{lemma}
\begin{IEEEproof}

Since DL $p_{kj}$s are all fixed to the same value, $\bar{p}_j$, we have
\begin{align}
\sigma_{k,j}^2 &= \sigma^2 +  \frac{\bar{p}_{\bar{j}}}{N} \sum_{\bar{k}=1}^K \frac{|\mathbf{h}_{k,j,\bar{j}} \mathbf{\hat{w}}_{\bar{k}\bar{j}}|^2}{\|\mathbf{\hat{w}}_{\bar{k}\bar{j}}\|^2}.
\end{align}
Applying \eqref{eq:SCP_OtherInter} of Lemma \ref{asymptotics_SCP}, for $k = 1, \ldots, K$, $j = 1, 2$,
\begin{align}
\sigma_{k,j}^2 &= \sigma^2 + \epsilon \bar{P}_{\bar{j}} + o(1).
\end{align}
\end{IEEEproof}

We conclude from Lemma~\ref{lemm:4} that $\sigma_{k,j}^2$ is asymptotically independent of the user index $k$, in the regime considered ($p_{kj} = \bar{p}_j$) and therefore write the limiting value as $\sigma_j^2$. This deals with the asymptotics of the RHS of \eqref{app:scp_sigmakj}, and we now provide two lemmas to deal with those of the RHS of \eqref{app:scp_pkj}.

\begin{lemma}\label{lemma:OwnC_interf_SCP}
With $\mathbf{\hat{w}}_{kj} = \mathbf{A}_{k,j} \mathbf{h}_{k,j, j}^H$, such that $\bal = \frac{\gamma}{1-\beta \frac{\gamma}{1+\gamma}}$, and with $p_{kj} = \bar{p}_j$ for $k = 1, \ldots, K$, $j = 1, 2$. The following holds for any $k'$ in cell $j$:
\begin{align}
& \sum_{k' \neq k} \frac{p_{k'j}}{N} \frac{\left|\mathbf{h}_{k, j,j} \mathbf{\hat{w}}_{k'j}\right|^2}{\left\|\mathbf{\hat{w}}_{k'j}\right\|^2}  \stackrel{a.s.}{\longrightarrow} \bar{p}_j \frac{\beta}{\left(1+\gamma \right)^2} = \frac{\bar{P}_j}{\left(1+\gamma \right)^2},
\end{align}
as $K, N \rightarrow \infty$, $\frac{K}{N} = \beta$; $\bar{P}_j = \beta \bar{p}_j$.
\end{lemma}
\begin{IEEEproof}
\begin{align}
&\sum_{k' \neq k} \frac{p_{k'j}}{N} \frac{\left|\mathbf{h}_{k, j,j} \mathbf{\hat{w}}_{k'j}\right|^2}{\left\|\mathbf{\hat{w}}_{k'j}\right\|^2} = \frac{\bar{p}_{j}}{N} \sum_{k' \neq k} \frac{\left|\mathbf{h}_{k, j,j} \mathbf{\hat{w}}_{k'j}\right|^2}{\left\|\mathbf{\hat{w}}_{k'j}\right\|^2} \nonumber \\
&= \frac{K-1}{N} \frac{\bar{p}_{j}}{(1+\gamma)^2} + o(1),
\end{align}
where we used \eqref{eq:SCP_OwnInter} in Lemma \ref{asymptotics_SCP}. Noting that as $K, N \rightarrow \infty$, $\frac{K-1}{N} \rightarrow \beta$ completes the proof.
\end{IEEEproof}

To find the {\it minimal} pair of constants $(\bar{p}_1, \bar{p}_2)$ for the two cells, we therefore solve the following set of equations,
\begin{align}
\bar{p}_1 = \frac{\gamma}{1-\frac{\beta \gamma^2}{(1+\gamma)^2}} \left[\sigma^2 + \epsilon \beta \bar{p}_{2} + \bar{p}_1 \frac{\beta}{\left(1+\gamma \right)^2}\right], \\
\bar{p}_2 = \frac{\gamma}{1-\frac{\beta \gamma^2}{(1+\gamma)^2}} \left[\sigma^2 + \epsilon \beta \bar{p}_{1} + \bar{p}_2 \frac{\beta}{\left(1+\gamma \right)^2}\right],
\end{align}
to obtain
\begin{align}
\bar{p}_1 = \bar{p}_2 = \bar{p} = \frac{\sigma^2 \gamma}{1-\frac{\beta \gamma}{(1+\gamma)}-\epsilon \beta \gamma}. \label{eq:Pj_converges}
\end{align}
This implies that
\begin{align}
\bar{P}_1 = \bar{P}_2 = \bar{P} := \beta \bar{p}.
\end{align}
Note that such a choice of transmit powers is guaranteed to meet the SINR constraints as $K, N \rightarrow \infty$, $\frac{K}{N} = \beta$.

We confirm the asymptotic optimality of this deterministic power allocation, together with the beamforming directions found from analysis of the dual problem, by verifying that the duality gap is tending to zero in both cells. Indeed, in cell $j$, the primal objective value converges to $\bar{P}$, whereas the dual objective value converges to $\beta \bar{\lambda} \left(\sigma^2 + \epsilon \bar{P}\right)$. Recalling the definition of $\bar{\lambda}$, one can easily verify that these two quantities are the same, thereby completing the proof.

Equation \eqref{eq:Pj_converges} shows that the coupled primal problems have a solution when effective bandwidth condition
$\beta \left(\frac{\gamma}{1 + \gamma} + \epsilon
\gamma \right) < 1$ is satisfied. Conversely, if this condition does not hold, there is asymptotically no feasible solution to the coupled primal
problems, in the limit as $N \uparrow \infty$. The latter observation follows from monotonicity: The optimal powers are increasing functions of $\gamma$, but as the denominator of \eqref{eq:Pj_converges} decreases to zero, the optimal total power from either BS, $\bar{P}$, must diverge to infinity, and for higher values of $\gamma$ there can be no feasible solution.

\section{Proof of Theorem~\ref{THEO_BF_COORD}}  \label{appendix:coord_dual_LAS}
Throughout this section, let
\begin{align}
\mathbf{A}_{j} &= \left(\mathbf{I} + \frac{\bal}{N} \sum_{i=1}^2 \sum_{l = 1}^K
\mathbf{h}_{l, i,j}^H\mathbf{h}_{l, i, j}\right)^{-1} \\
\mathbf{A}_{k, j} &= \left(\mathbf{I} + \frac{\bal}{N} \sum_{i=1}^2 \sum_{(i, l) \neq (j,k)}
\mathbf{h}_{l, i,j}^H\mathbf{h}_{l, i, j}\right)^{-1} \\
\mathbf{A}_{k, j, k', j', j} &= \left(\mathbf{I} + \frac{\bal}{N} \sum_{i=1}^2 \sum_{(i, l) \neq (j,k), (j',k')}
\mathbf{h}_{l, i,j}^H\mathbf{h}_{l, i, j}\right)^{-1},
\end{align}
where $\bal > 0$ will be defined later.

Moreover, the following lemma will be useful in proving the theorem.

\begin{lemma}\label{lem:coord}
Let $(\mu_1, \mu_2)$ satisfy $0 \leq \mu_i \leq 2$, $i=1,2$, and define the function $F(\bal_1, \bal_2)$ by
\begin{align}
F(\bal_1, \bal_2) &= (F_1(\bal_1, \bal_2), F_2(\bal_1, \bal_2)), \nonumber \\
F_j(\bal_1, \bal_2) &= \gamma \left(\mu_j + \frac{\beta}{1 +
\gamma}~ \bal_j + \frac{\beta \epsilon}{1 +
\frac{\bal_{\baj}}{\bal_j} \gamma \epsilon} ~\bal_{\baj}\right)
~~j=1,2. \label{eq:def_F}
\end{align}
Then
\begin{enumerate}
\item F is an interference function \cite{yates95}. 
\item If $\displaystyle \beta \left(\frac{\gamma}{1 + \gamma} + \frac{\epsilon \gamma}{1 + \epsilon \gamma}\right) < 1$ then there exists a unique solution to the fixed point equation
\begin{align}
(\bal_1, \bal_2) = F(\bal_1, \bal_2).   \label{eq:coord_FPE}
\end{align}
\end{enumerate}
\end{lemma}
\begin{IEEEproof}
(i) is easily verified. For (ii), let the value $\tilde{\lambda}$
be
\begin{align}
\tilde{\lambda} = \frac{2 \gamma}{1 - \beta \left(\frac{\gamma}{1 +
\gamma} + \frac{\epsilon \gamma}{1 + \epsilon \gamma}\right)}.\label{eq:lambda_tilde_coord}
\end{align}
Since $\mu_1, \mu_2 \leq 2$, it is easy to verify that
\begin{align}
F(\tilde{\lambda}, \tilde{\lambda}) \leq (\tilde{\lambda}, \tilde{\lambda}) ~~\textrm{componentwise}
\end{align}
and hence the inequality $\displaystyle (\lambda_1, \lambda_2) \geq F(\lambda_1, \lambda_2)$ has a feasible solution. It follows from Theorem 1 in \cite{yates95} that \eqref{eq:def_F} has a unique fixed point $(\bal_1, \bal_2)$. The latter depends on $(\mu_1, \mu_2)$.
\end{IEEEproof}


{\it Asymptotic analysis of the dual problem} \\

Assume that $\displaystyle \beta \left(\frac{\gamma}{1 + \gamma} + \frac{\epsilon \gamma}{1 + \epsilon \gamma}\right) < 1$, let $(\mu_1, \mu_2)$ be feasible for the dual \eqref{eq:coord_dual_uplink} {\it i.e.} $\mu_1, \mu_2 \geq 0$ and $\mu_1 + \mu_2  = 2$. Consider the suboptimal dual power vector that assigns power level $\bal_j/N$ to all the users in cell $j$, $j=1,2$, respectively. The dual UL SINR is equal to
\begin{align}
\frac{\bal_j}{N} \mathbf{h}_{k,j, j} \left[\mu_j \mathbf{I}
+ \sum_{j'=1}^2 \frac{\bal_{j'}}{N} \sum_{k', (j', k')\neq (j,k)} \mathbf{h}_{k', j', j}^H \mathbf{h}_{k', j' , j} 
\right]^{-1} \mathbf{h}_{k,j, j}^H. \label{eq:UL_SINR_CBf}
\end{align}
We show in Appendix \ref{app:CBf_UL} that this quantity converges a.s. to a constant equal to $\bar{\lambda}_j t_{CBf}(-\mu_j, \bar{\lambda}_j, \bar{\lambda}_{\bar{j}})$ (cf. Eq. \eqref{eq:t_CBf}). 

Letting $(\bal_1, \bal_2)$ be the unique solution to \eqref{eq:coord_FPE},
$\bar{\lambda}_j t_{CBf}(-\mu_j, \bar{\lambda}_j, \bar{\lambda}_{\bar{j}})$ will be equal to $\gamma$ for $j = 1, 2$.
Thus, with this set of suboptimal dual UL powers, all SINR's (cf. Eq. \eqref{eq:UL_SINR_CBf}) converge to $\gamma$.
Note that the condition that $0 \leq \mu_i \leq 2,~~i=1,2,$ is weaker than the condition that $(\mu_1, \mu_2)$ satisfy the dual feasibility condition that $\mu_1 + \mu_2 = 2$, but it certainly includes this condition.

We now arrive at a particular choice of $(\mu_1, \mu_2)$ by solving the following optimization problem:
\begin{align}
&\textrm{max.}_{\mu_1, \mu_2 \ge 0} \quad \beta \sigma^2 \left(\bar{\lambda}_1 + \bar{\lambda}_2\right) \label{eq:CBf_opt_lim} \\
&\textrm{s.t.} 
\quad \mu_1 + \mu_2 = 2 \label{eq:coord_summu} \\
& \quad \quad (\bal_1, \bal_2) \textrm{~is the unique fixed point of~}
\eqref{eq:coord_FPE}.
\label{eq:coord_bal_fpe}
\end{align}
Adding up 
the two equations in \eqref{eq:coord_FPE}, and
taking account of \eqref{eq:coord_summu}, we obtain
\begin{align}
&\left(1 - \frac{\beta}{1 + \gamma}\right) (\bal_1 + \bal_2) \nonumber \\
&= 2
\gamma + \beta \epsilon \left(\frac{\bal_1 \bal_2}{\bal_1 + \gamma
\epsilon \bal_2} + \frac{\bal_1 \bal_2}{\bal_2 + \gamma \epsilon
\bal_1}\right). \label{eq:l1l2}
\end{align}
Let $t = \bar{\lambda}_1 + \bar{\lambda}_2$ for $(\bal_1, \bal_2)$ optimal.
For fixed $t$, the $(\bal_1, \bal_2)$ that maximize the RHS of \eqref{eq:l1l2} can be verified to be $(t/2, t/2)$. The optimal $\bal_1$ and $\bal_2$ must thus be equal; we denote the common value by $\bal$. \eqref{eq:coord_FPE} 
 then implies $\mu_1 = \mu_2 = 1$.
As a result, $\bal$ is as given in \eqref{eq:coord_bal}.
We conclude that the optimal choice (with respect to the optimization problem \eqref{eq:CBf_opt_lim}-\eqref{eq:coord_bal_fpe}) is to take $(\mu_1, \mu_2) = (1,1)$.

Note also that if we fix $(\mu_1, \mu_2) = (1,1)$ and use the corresponding deterministic $(\bal_1, \bal_2) = (\bar{\lambda}, \bar{\lambda})$ (with $\bar{\lambda}$ satisfying \eqref{eq:coord_bal}) to generate dual UL powers in each cell, then the dual objective function will converge to the solution of the optimization problem \eqref{eq:CBf_opt_lim}-\eqref{eq:coord_bal_fpe}.


Now consider the {\it optimal} dual variables $(\mu_1, \mu_2)$, and
the vector $\bobl$ of
$\left(\lambda_{kj}\right)_{k=1,j=1}^{K,~~2}$, where optimality here refers to the dual CBf optimization problem \eqref{eq:coord_dual_uplink}. Due to the dual
feasibility constraints, the sequence of $(\mu_1, \mu_2)$
is contained in a compact set and so the probability distribution function of $\mu_1$, $F_1^{(N)}$, forms a tight sequence \cite{billingsley99}. Let $F_1$ denote a limit point, so that $F_1^{(N)} \Rightarrow F_1$ along a convergent subsequence.

For the purpose of obtaining a contradiction, let $\bamu_1$ be such that $F_1(\bamu_1 - \delta, \bamu_1 + \delta) > 0$ for all $\delta > 0$, and let $\bamu_2 = 2 - \bamu_1$. We will assume $0<\bamu_1<2$.\footnote{The cases $\bamu_1 = 0$ or $\bamu_1 = 2$ can be considered separately, in a similar manner, which we mention at the conclusion of the proof.} Roughly speaking, there is non-negligible probability that $\mu_1$ will be close to $\bamu_1$ when $N$ is large, along the subsequence. Indeed, for $\delta > 0$, let us define $B_1(\delta)$ be the event that $\mu_1 \in (\bamu_1-\delta, \bamu_1 + \delta)$, then by the second Borel-Cantelli lemma, event $B_1(\delta)$ will occur infinitely often. Due to the feasibility constraint that $\mu_1 + \mu_2 = 2$, we can equivalently write
\begin{equation}
B_1(\delta) \!=\! \left\{ \!\begin{array}{ll} \bamu_1 \!-\! \delta \!\leq\! \mu_1 \!\leq \!\bamu_1 \!+ \! \delta, \bamu_2 \!- \!\delta \! \leq \! \mu_2 \! \leq \! \bamu_2 \! + \! \delta & \delta \!> \! 0  \\
\bamu_1 \!+\! \delta \!\leq\! \mu_1 \!\leq \!\bamu_1 \!- \! \delta, \bamu_2 \!+ \!\delta \! \leq \! \mu_2 \! \leq \! \bamu_2 \! - \! \delta & \delta \!<\! 0. \end{array} \right.
\label{eq:def_B_1_event}
\end{equation}

To compare the performance of the optimal scheme with a
deterministic power scheme, we modify the SINR target in
\eqref{eq:def_F} from $\gamma$ to $\gamma+\delta$ for some small
constant $\delta$ (positive or negative), and we replace $(\bamu_1, \bamu_2)$ by $(\bamu_1+\delta,
\bamu_2+\delta)$. The latter change violates the constraint that the sum of the $\mu$s should be $2$, but it nonetheless provides a valid pair of noise values for a virtual UL. Provided that $|\delta|$ is sufficiently small, the condition $0 < \bamu_i + \delta < 2~~i=1,2$, will be met.

Denote the corresponding solution to \eqref{eq:coord_FPE} (with $(\mu_1, \mu_2)$ replaced by $(\bamu_1 + \delta, \bamu_2 + \delta)$)
by $(\bloned,\bltwod)$, and let $\bobld$ denote the vector of UL powers, where cell $j$ users use power level $\bljd$. Note that
$(\bloned,\bltwod)$ (and hence $\bobld$) depends on $(\bamu_1,
\bamu_2)$, and $\delta$.
The analysis in Appendix \ref{app:CBf_UL} shows that this power
allocation will asymptotically achieve a SINR of $\gamma+\delta$ for
all users, under external noise levels of $\bamu_1 + \delta$, and $\bamu_2+\delta$, respectively, and for finite $N$, let it achieve SINR $\gamma_{kj}(\delta)$ for user $k$ in cell $j$.

 As in the SCP case, Theorem 3 in \cite{kammoun_it09} can be applied to show that for this (suboptimal) system, the value of \eqref{eq:UL_SINR_CBf} is
\begin{align}
\gamma_{kj}(\delta) = \gamma + \delta + \mathcal{O}\left(\sqrt{\frac{1}{N}}\right), ~~~\forall k = 1, 2, \ldots, K, j = 1, 2 \label{eq:suboptimal_SINR_limit_cbf}
\end{align}
where the last term on the right hand side is $1/\sqrt{N}$ times a r.v. that tends weakly to a zero-mean Gaussian r.v. This holds since $\frac{K}{N} \rightarrow \beta < \infty$, for the considered channel model and in the notation of the theorem, $\Gamma_{\tilde{K}}$ and $\Theta_{\tilde{K}}$ will converge a.s. to bounded limits, though these will be different from the ones in the SCP case.
As in the proof of Theorem~\ref{THEO_BF_SCP} in Appendix \ref{appendix:SCP_dual_LAS}, a union bound may be applied to show that
\begin{align}
\gamma_{kj}(-\delta) \le \gamma \le \gamma_{kj}(\delta)~~~\forall k = 1, 2, \ldots, K, j = 1, 2 \label{eq:gamma_bounds_cbf}
\end{align}
will hold whenever $B_1(\delta)$ occurs, in the limit as $N \uparrow \infty$.

Now compare the dual optimal power levels with this suboptimal, deterministic power allocation, on the event that $B_1(\delta)$ occurs. Consider the case that $\delta > 0$: Noting \eqref{eq:def_B_1_event}, we see that, on this event, the dual UL powers required to achieve $\gamma_{k,j}(\delta)$ will only decrease if we replace the enhanced noise levels, $(\bamu_1+\delta, \bamu_2 + \delta)$,  by $(\mu_1, \mu_2)$. But by
\eqref{eq:suboptimal_SINR_limit_cbf}
and the monotonicity result in Lemma~\ref{lem:monotonicity} in
Appendix~\ref{appendix:monotonicity}, it follows that these decreased UL powers must upper bound $\bol$ since the latter is the optimal vector of powers in each cell to achieve the all-$\gamma$ SINR vector under noise levels $(\mu_1, \mu_2)$. Similarly, consider the case $\delta < 0$: On the event $B_1(\delta)$, the dual UL powers required to achieve $\gamma_{k,j}(\delta)$, will only increase if we replace the lower noise levels, $(\bamu_1+\delta, \bamu_2 + \delta)$,  by $(\mu_1, \mu_2)$, but by
\eqref{eq:suboptimal_SINR_limit_cbf}
and the monotonicity result in Lemma~\ref{lem:monotonicity} in
Appendix~\ref{appendix:monotonicity}, it follows that these increased UL powers lower bound $\bol$. We conclude that for $N$ sufficiently large, and on the event $B_1(\delta)$, we have that
\begin{align}
\boblonemd \leq \bol_{1} \leq \bobloned \label{eq:lambda1_bounds} \\
\bobltwomd \leq \bol_{2} \leq \bobltwod \label{eq:lambda2_bounds}
\end{align}
and hence that the optimal dual objective value lies in the interval $(\beta \sigma^2 (\blone(-\delta) + \bltwo(-\delta)), \beta \sigma^2 (\blone(\delta) + \bltwo(\delta)))$. But $\delta$ can be taken arbitrarily small, so the left and right endpoints of the interval can be made arbitrarily close to $\beta \sigma^2 (\blone + \bltwo)$, where $(\blone, \bltwo)$ are the corresponding deterministic UL powers when we set $\delta = 0$: These are the unique solution to  \eqref{eq:coord_FPE}  
 for the given $(\bamu_1, \bamu_2)$. Thus, when $N$ is sufficiently large, along the chosen subsequence, and on the event $B_1(\delta)$ (which occurs infinitely often) we have the dual objective value getting as close as we like to the value $\beta \sigma^2 (\blone + \bltwo)$, which is a feasible value to the optimization problem \eqref{eq:CBf_opt_lim}-\eqref{eq:coord_bal_fpe}. Thus if $(\bamu_1, \bamu_2) \neq (1,1)$ and $0 < \bamu_1 < 2$, then we get a contradiction of the optimality of the optimal dual power levels, whenever $B_1(\delta)$ occurs, since we can beat the purported optimal value using the deterministic UL power levels obtained for $(\mu_1, \mu_2) = (1,1)$, i.e. \eqref{eq:coord_bal}: this is the case since as discussed earlier the corresponding dual objective converges to the solution of \eqref{eq:CBf_opt_lim}-\eqref{eq:coord_bal_fpe}. A similar argument can be used to show that $(\bamu_1, \bamu_2) = (0,2)$ or $(\bamu_1, \bamu_2) = (2,0)$ lead to similar contradictions. Since these are contradictions, it follows that $(\bamu_1, \bamu_2) = (1,1)$ and hence all the mass of the distribution $F_1$ must be concentrated at $(1,1)$.

We can conclude from the above analysis that for any positive $\delta$, the event $B_1(\delta)$ will occur with probability tending to $1$ as $N$ tends to infinity, and therefore that \eqref{eq:lambda1_bounds}-\eqref{eq:lambda2_bounds} hold with probability also tending to $1$. Taking $\delta$ to zero, we obtain that the empirical distributions of both the $\lambda_{1k}$s and the $\lambda_{2k}$s tend to the same constant, namely $\bar{\lambda}$, as $N$ tends to infinity, where $\bar{\lambda}$ given in \eqref{eq:coord_bal}.

We conclude that the asymptotically optimal dual variables are $\mu_1 = \mu_2 = 1$ and all $\lambda_{kj}$s converge to $\bar{\lambda}$ as given in \eqref{eq:coord_bal}. Thus the optimal downlink beamforming vectors asymptotically point in the directions of the vectors given in \eqref{eq:coord_bf}. It remains to find the optimal power levels, $p_{kj}$, to use for each user. \\

{\it Asymptotic analysis of the primal problem}

As in Appendix \ref{appendix:SCP_dual_LAS}, we start by assuming that the $p_{kj}$'s in each cell are fixed at some common constant value $\bar{p}_j$, so that the total transmit power of cell $j$ is $\bar{P}_j = \beta \bar{p}_j$. Once this regime is analyzed in the large system limit, we optimize the constants $\bar{p}_1$ and $\bar{p}_2$ (to meet the DL SINR constraints) and show that the optimal constants are asymptotically optimal with respect to the primal optimization problem
\eqref{eq:MaxMinSINR_JTx_feasibility}.

Under this regime in which $p_{kj} = \bar{p}_j$ for all $k$ in cell $j$, the following lemmas hold.

\begin{lemma}\label{asymptotics_CBf}
With $\mathbf{\hat{w}}_{kj} = \mathbf{A}_{k,j} \mathbf{h}_{k,j, j}^H$, $\bal$ satisfying \eqref{eq:coord_bal},
\begin{align}
&\max_{j = 1, 2, k \le K} \left|
\frac{
 \left|\mathbf{h}_{k,j,j}\mathbf{\hat{w}}_{kj}\right|^2
}{N \|\mathbf{\hat{w}}_{kj}\|^2}- \left[1 - \beta \left(\frac{\gamma^2}{(1 + \gamma)^2} +
\frac{\epsilon^2 \gamma^2}{(1 + \epsilon \gamma)^2}\right)\right]
\right| \nonumber \\
&~~~~~~~~~~~~~~~~~~~~~~~~~~~~~~~~~~~~~~~~~~~~~~~~~~~\stackrel{a.s.}{\longrightarrow} 0 \\
&\max_{j = 1, 2, k, l \le K} \left|\frac{\left|\mathbf{h}_{k,j,\bar{j}} \mathbf{\hat{w}}_{l\bar{j}}\right|^2}{\|\mathbf{\hat{w}}_{l\bar{j}}\|^2}-
\frac{\epsilon}{\left(1+\epsilon \gamma\right)^2}
\right| \stackrel{a.s.}{\longrightarrow} 0 \label{eq:coord_OtherInter} \\
&\max_{j = 1, 2, k, l \le K, l \neq k} \left|\frac{\left|\mathbf{h}_{k,j,j} \mathbf{\hat{w}}_{lj}\right|^2}{\|\mathbf{\hat{w}}_{lj}\|^2}-
\frac{1}{\left(1+ \gamma\right)^2}
\right| \stackrel{a.s.}{\longrightarrow} 0.  \label{eq:coord_OwnInter}
\end{align}
as $K, N \rightarrow \infty$, $\frac{K}{N} \rightarrow \beta$. 
\end{lemma}
\begin{IEEEproof}
The proof is similar to that of Lemma \ref{asymptotics_SCP}, the main difference lying in the fact that there is only one case to consider when looking at the interference terms (cf. \eqref{eq:interf_SCP}), since under coordinated beamforming, $\mathbf{h}_{k,j,\bar{j}}$ is no longer independent of $\mathbf{\hat{w}}_{k', \bar{j}}$. Thus, defining $\mathbf{D}_{k,j,j}=\bal \mathbf{I}$ and $\mathbf{D}_{k,j,\bar{j}} = \epsilon \bal \mathbf{I}$, exactly as in the proof of Lemma \ref{asymptotics_SCP}, we can show that
\begin{align}
\max_{j =1, 2, k \le K} \left|\frac{\bal}{N} \|\mathbf{\hat{w}}_{kj}\|^2 - \frac{1}{N} \textrm{tr} \mathbf{D}_{k,j,j} \mathbf{A}_{j}^2\right| \stackrel{a.s.}{\longrightarrow} 0 \label{eq:Norm_CBf} \\
\max_{j = 1, 2, k \le K} \left|\frac{\bal}{N} \mathbf{h}_{k,j,j}\mathbf{\hat{w}}_{kj} - \frac{1}{N}  \textrm{tr} \mathbf{D}_{k,j,j} \mathbf{A}_{j}\right| \stackrel{a.s.}{\longrightarrow} 0. \label{eq:UsefulSignal_CBf}
\end{align}
When studying the interference terms, the numerator in the RHS of  \eqref{eq:interf_SCP}, becomes, for the CBf case,
\begin{align}
&\frac{\bal^2}{N} \mathbf{h}_{k,j,i} \mathbf{A}_{l, i} \mathbf{h}_{l,i,i}^H
\mathbf{h}_{l,i,i} \mathbf{A}_{l, i} \mathbf{h}_{k,j,i}^H \nonumber \\
&= \frac{\frac{\bal^2}{N} \mathbf{h}_{k,j,i} \mathbf{A}_{l,i,k, j, i}\mathbf{h}_{l,i,i}^H
\mathbf{h}_{l,i,i} \mathbf{A}_{i, l, k,j, i} \mathbf{h}_{k,j,i}
^H}{\left(1 + \frac{\bal}{N} \mathbf{h}_{k,j,i} \mathbf{A}_{l,i,k, j, i}\mathbf{h}_{k,{j},i}^H \right)^2}. \label{eq:interf_CBf}
\end{align}

Considering the numerator in \eqref{eq:interf_CBf}, we can show that
\begin{align}
\max_{j, i = 1,2, k,l \le K, (k,j) \neq (l,i)}&\left|\frac{\bal^2}{N} \mathbf{h}_{k,j,i} \mathbf{A}_{l,i,k, j, i}\mathbf{h}_{l,i,i}^H
\mathbf{h}_{l,i,i} \mathbf{A}_{l,i,k, j, i} \mathbf{h}_{k,j,i}
^H \right. \nonumber \\
&\left.~~~-  \frac{1}{N} \textrm{tr} \mathbf{D}_{k,j,i} \mathbf{A}_{i} \mathbf{D}_{l,i,i} \mathbf{A}_{i}
 \right| \stackrel{a.s.}{\longrightarrow} 0.
\end{align}
We can also show that
\begin{align}
\max_{j, i = 1,2, k,l \le K, (k,j) \neq (l,i)}& \left|\frac{\bal}{N} \mathbf{h}_{k,j,i} \mathbf{A}_{i,l,k,j,i}\mathbf{h}_{k,{j},i}^H \right. \nonumber \\
&~~~~~~~\left.-
\frac{1}{N}\textrm{tr} \mathbf{D}_{k,j,i} \mathbf{A}_{i} \right| \stackrel{a.s.}{\longrightarrow} 0.
\end{align}

The proof of the lemma is concluded by using the limits of the trace terms as derived in Appendix \ref{app:CBf_UL}, with $\lambda_j = \bal$, noting that for this specific case, all the $\mathbf{D}_{k,j,j}$ and $\mathbf{D}_{k,j,\bar{j}}$ matrices are equal, and are simply scaled identities, so that
\begin{align}
&\frac{1}{N} \textrm{tr} \mathbf{D}_{k,j,{j}} \mathbf{A}_{j} = \frac{\bal}{N} \textrm{tr} \mathbf{A}_{{j}} \stackrel{a.s.}{\longrightarrow} \gamma \\
&\frac{1}{N} \textrm{tr} \mathbf{D}_{k,\bar{j},{j}} \mathbf{A}_{j} = \frac{\epsilon \bal}{N} \textrm{tr} \mathbf{A}_{{j}} \stackrel{a.s.}{\longrightarrow} \epsilon \gamma \\
&\frac{1}{N} \textrm{tr} \mathbf{D}_{k,j,j} \mathbf{A}_{j}^2 = \frac{\bal}{N} \textrm{tr} \mathbf{A}_{j}^2
\nonumber \\
& ~\stackrel{a.s.}{\longrightarrow} \frac{1}{\bal}\frac{\gamma}{\frac{1}{\bal} + \frac{\beta}{\left(1+\gamma\right)^2}+\frac{\beta \epsilon}{\left(1+\epsilon \gamma\right)^2}
} 
=  \frac{1}{\bal}\frac{\gamma^2}{
1 - \beta \left(\frac{\gamma^2}{(1 + \gamma)^2} +
\frac{\epsilon^2 \gamma^2}{(1 + \epsilon \gamma)^2}\right)}
\\
&\frac{1}{N} \textrm{tr} \mathbf{D}_{k,j,{j}} \mathbf{A}_{{j}} \mathbf{D}_{l,{j},{j}} \mathbf{A}_{{j}} = \frac{\bal^2}{N} \textrm{tr} \mathbf{A}_{{j}}^2
\nonumber \\
& ~~~~~~~~~~~~~~~~~~~~\stackrel{a.s.}{\longrightarrow} \frac{\gamma^2}{1 - \beta \left(\frac{\gamma^2}{(1 + \gamma)^2} + \frac{\epsilon^2 \gamma^2}{(1 + \epsilon \gamma)^2}\right)}
\\
&\frac{1}{N} \textrm{tr} \mathbf{D}_{k,\bar{j},{j}} \mathbf{A}_{{j}} \mathbf{D}_{l,{j},{j}} \mathbf{A}_{{j}} = \frac{\epsilon \bal^2}{N} \textrm{tr} \mathbf{A}_{{j}}^2 \nonumber \\
& ~~~~~~~~~~~~~~~~~~~~\stackrel{a.s.}{\longrightarrow}  \frac{\epsilon \gamma^2}{1 - \beta \left(\frac{\gamma^2}{(1 + \gamma)^2} +\frac{\epsilon^2 \gamma^2}{(1 + \epsilon \gamma)^2}\right)}.
\end{align}
\end{IEEEproof}

\begin{lemma}\label{lemm:coord_interf}
With $\mathbf{\hat{w}}_{kj} = \mathbf{A}_{k,j} \mathbf{h}_{k,j, j}^H$, such that $\bal$ satisfies \eqref{eq:coord_bal}, and with $p_{kj} = \bar{p}_j$ for $k = 1, \ldots, K$, $j = 1, 2$, it follows that, with probability 1,
\begin{align}
\sum_{{j}', {k}', ({k}', {j}') \neq (k,j)}
\frac{p_{{k}'{j}'}}{N} \frac{\left|\mathbf{h}_{k, j,{j}'}
\mathbf{\hat{w}}_{k'j'}\right|^2}{\|\mathbf{\hat{w}}_{{k}'{j}'}\|^2} \stackrel{a.s.}{\longrightarrow}
\frac{\bar{P}_{j}}{(1+\gamma)^2} + \frac{\epsilon \bar{P}_{\bar{j}}}{(1+\epsilon \gamma)^2},
\end{align}
$\bar{P}_{j}$ denotes the total transmit power of BS $j$, i.e. $P_{j} = \beta \bar{p}_j$, $j = 1, 2$.
\end{lemma}
\begin{IEEEproof}
Since DL $p_{kj}$s are all fixed to the same value, $\bar{p}_j$, we have
\begin{align}
&\sum_{{j}', {k}', ({k}', {j}') \neq (k,j)}
\frac{p_{{k}'{j}'}}{N} \frac{\left|\mathbf{h}_{k, j,{j}'}
\mathbf{\hat{w}}_{{k}'{j}'}\right|^2}{\|\mathbf{\hat{w}}_{{k}'{j}'}\|^2} \nonumber \\
&= \frac{\bar{p}_j}{N}\sum_{k' \neq k} \frac{\left|\mathbf{h}_{k, j,{j}}
\mathbf{\hat{w}}_{{k}'{j}}\right|^2}{\|\mathbf{\hat{w}}_{{k}'{j}}\|^2} +
\frac{\bar{p}_{\bar{j}}}{N}\sum_{k'} \frac{\left|\mathbf{h}_{k, j,\bar{j}}\mathbf{\hat{w}}_{{k}'\bar{j}}\right|^2}{\|\mathbf{\hat{w}}_{{k}'\bar{j}}\|^2}
 \label{eqCBF_InterfLemmaProof}
\end{align}
Applying \eqref{eq:coord_OwnInter} and \eqref{eq:coord_OtherInter} of Lemma \ref{asymptotics_CBf}, \eqref{eqCBF_InterfLemmaProof} becomes for $k = 1, \ldots, K$, $j = 1, 2$,
\begin{align}
\frac{K-1}{N} \frac{\bar{p}_{j}}{(1+\gamma)^2} + \epsilon \beta \frac{\bar{p}_{\bar{j}}}{(1+\epsilon \gamma)^2} + o(1).
\end{align}
Noting that as $K, N \rightarrow \infty$, $\frac{K-1}{N} \rightarrow \beta$ completes the proof.
\end{IEEEproof}

Referring to \eqref{eq:sinr_cbf}, and using results from Lemmas \ref{asymptotics_CBf} and \ref{lemm:coord_interf}, to find the {\it minimal} pair of constants $(\bar{p}_1, \bar{p}_2)$ for the two cells, we therefore solve the following set of equations,
\begin{align}
&\bar{p}_1 \frac{1-\frac{\beta \gamma^2}{(1+\gamma)^2} + \frac{\beta \epsilon^2 \gamma^2}{(1+\epsilon \gamma)^2}}{\gamma}\nonumber \\
&=  \sigma^2 +  \frac{\epsilon  \beta}{\left(1+\epsilon \gamma \right)^2} \bar{p}_{2} + \bar{p}_1 \frac{\beta}{\left(1+\gamma \right)^2}, \\
&\bar{p}_2 \frac{1-\frac{\beta \gamma^2}{(1+\gamma)^2} + \frac{\beta \epsilon^2 \gamma^2}{(1+\epsilon \gamma)^2}}{\gamma} \nonumber \\
&= \sigma^2 + \frac{\epsilon  \beta}{\left(1+\epsilon \gamma \right)^2} \bar{p}_{1} + \bar{p}_2 \frac{\beta}{\left(1+\gamma \right)^2},
\end{align}
to obtain
\begin{align}
\bar{p}_1 = \bar{p}_2 = \bar{p} = \frac{\sigma^2 \gamma}{1-\frac{\beta \gamma}{(1+\gamma)}-\epsilon \frac{\beta \gamma}{1+\epsilon \gamma}}. \label{eq:Pj_converges_coord}
\end{align}
This implies that
\begin{align}
\bar{P}_1 = \bar{P}_2 = \bar{P} := \beta \bar{p}.
\end{align}
This choice of downlink transmit powers is guaranteed to meet the SINR constraints as $K, N \rightarrow \infty$, $\frac{K}{N} = \beta$.

We confirm the asymptotic optimality of this deterministic power allocation, together with the beamforming directions found from analysis of the dual problem, by verifying that the duality gap is tending to zero. Indeed, the primal objective value converges to $2 \bar{P}$ ($\phi = 1$, since the power consumption in both cells is the same), whereas the dual objective value converges to $2 \beta \bar{\lambda} \sigma^2$. Recalling the definition of $\bar{\lambda}$, one can easily verify that these two quantities are the same, thereby completing the proof.

Equation \eqref{eq:Pj_converges_coord} shows that the primal problem has a solution when effective bandwidth condition
$\displaystyle \beta
\left(\frac{\gamma }{1+\gamma} + \frac{\epsilon \gamma }{1+\epsilon
\gamma} \right) < 1$ is satisfied. Conversely, if this condition does not hold, there is asymptotically no feasible solution to the primal
problem, in the limit as $N \uparrow \infty$. The latter observation follows from monotonicity: The optimal powers are increasing functions of $\gamma$, but as the denominator of \eqref{eq:Pj_converges_coord} decreases to zero, the optimal total power from either BS, $\bar{P}$, must diverge to infinity, and for higher values of $\gamma$ there can be no feasible solution.

\section{Proof of Theorem~\ref{THEO_BF_MCP}} \label{appendix:MCP_dual_LAS}
Throughout this section, let
\begin{align}
\mathbf{A} =  \left[\frac{\bar{\lambda}}{N}\sum_{j'=1}^2 \sum_{k'=1}^K \mathbf{\tilde{h}}_{k',j'}^H\mathbf{\tilde{h}}_{k',j'} + \mathbf{I} \right]^{-1} \label{eq:A_MCP}\\
\mathbf{A}_{k,j} =  \left[\frac{\bar{\lambda}}{N}\sum_{\left(k', j'\right) \neq \left(k, j\right)} \mathbf{\tilde{h}}_{k',j'}^H\mathbf{\tilde{h}}_{k',j'} + \mathbf{I} \right]^{-1} \\
\mathbf{A}_{k,j, l,i} =  \left[\frac{\bar{\lambda}}{N}\sum_{\left(k', j'\right) \neq \left(k, j\right), (l,i)} \mathbf{\tilde{h}}_{k',j'}^H\mathbf{\tilde{h}}_{k',j'} + \mathbf{I} \right]^{-1}
\end{align}
where $\bal$ will be specified later.

The following lemmas will be useful in proving the theorem.

\begin{lemma} \label{lem:MCP_F}
Let $(\mu_1, \mu_2)$ satisfy $0 \leq \mu_i \leq 2$, $~i=1,2$,
and define the function $F(\lambda_1, \lambda_2)$
by
\begin{align}
F(\lambda_1, \lambda_2) &= (F_1(\lambda_1, \lambda_2), F_2(\lambda_1, \lambda_2)) \nonumber \\
F_j(\lambda_1, \lambda_2) &= \gamma \left[\left(\mu_j +
\frac{\beta}{1+\gamma} (\lambda_j + \lambda_{\baj}
\epsilon)\right)^{-1} \right. \nonumber \\
&\left.+ \epsilon \left(\mu_{\baj} +
\frac{\beta}{1+\gamma} (\lambda_{\baj} + \lambda_j
\epsilon)\right)^{-1}\right]^{-1} ~~j=1,2. \label{eq:def_F_MCP}
\end{align}
Then
\begin{enumerate}
\item F is an interference function \cite{yates95}. 
\item If $\displaystyle \beta \frac{\gamma}{1 + \gamma} < 1$ then there exists a unique solution $(\bal_1, \bal_2)$ to the fixed point equation
\begin{align}
(\lambda_1, \lambda_2) = F(\lambda_1, \lambda_2).
\label{eq:MCP_FPE}
\end{align}
\end{enumerate}
\end{lemma}
\begin{IEEEproof}
Similar to the proof of Lemma \ref{lem:coord} with $\tilde{\lambda}$ in \eqref{eq:lambda_tilde_coord} replaced by
\begin{align}
\tilde{\lambda} = \frac{2 \gamma}{1+\epsilon} \left(1 - \beta
\frac{\gamma}{1+\gamma}\right)^{-1}.
\end{align}
\end{IEEEproof}

\begin{lemma} \label{lem:MCP_G}
Given positive $(\mu_1, \mu_2)$, and $(\lambda_1, \lambda_2)$, define function $G(\gamma_1, \gamma_2)$ by
\begin{align}
G(\gamma_1, \gamma_2) &= (G_1(\gamma_1, \gamma_2), G_2(\gamma_1, \gamma_2)) \nonumber \\
G_j(\gamma_1, \gamma_2) &= \lambda_j \left(\left(\mu_j + \frac{\beta \lambda_j}{1+\gamma_j} + \frac{\epsilon \beta \lambda_{\baj}}{1+\gamma_{\baj}}\right)^{-1} \right. \nonumber \\
&+ \left. \epsilon \left(\mu_{\baj} + \frac{\beta
\lambda_{\baj}}{1+\gamma_{\baj}} + \frac{\epsilon \beta \lambda_j}{1
+ \gamma_j}\right)^{-1}\right)~~j=1,2. \label{eq:def_G_MCP}
\end{align}
Then
\begin{enumerate}
\item G is an interference function \cite{yates95}. 
\item If the inequalities
\begin{align}
\gamma_1 \geq G_1(\gamma_1, \gamma_2), \quad
\gamma_2 \geq G_2(\gamma_1, \gamma_2)
\end{align}
have a solution, then $G$ has a unique fixed point.
\end{enumerate}
\end{lemma}
\begin{IEEEproof}
(i) is easily verified, and (ii) follows from \cite{yates95},
Theorem 1.
\end{IEEEproof}

\begin{corollary} \label{cor:MCP_FG}
Let $(\mu_1, \mu_2)$ satisfy $0 \leq \mu_i \leq 2$, $~i=1,2$.
Assume that $\displaystyle \beta
\frac{\gamma}{1 + \gamma} < 1$, and let $(\bal_1, \bal_2)$ be the
unique fixed point in \eqref{eq:def_F_MCP}, as identified in
Lemma~\ref{lem:MCP_F}. Then the function $G$ in
Lemma~\ref{lem:MCP_G} has a unique fixed point, namely $(\gamma,
\gamma)$.
\end{corollary}

{\it Asymptotic analysis of the dual problem} \\

Assume that $\displaystyle \beta \frac{\gamma}{1 + \gamma}  < 1$, let $(\mu_1, \mu_2)$ be
feasible for the dual \eqref{eq:MCP_dual_uplink} {\it i.e.} $\mu_1, \mu_2 \geq 0$ and $\mu_1 + \mu_2  = 2$, and let $(\bal_1, \bal_2)$
be the unique solution to \eqref{eq:MCP_FPE}. Consider the suboptimal dual power vector that assigns power level $\bal_j/N$ to
all the users in cell $j$, $j=1,2$, respectively. 
The dual UL SINR for any user $k$ in cell $j$ is given by Equation \eqref{eq:MCP_uplink_SINR}.
The derivations leading up to \eqref{eq:G1G2} in Appendix \ref{app:MCP_UL} show that with the given suboptimal dual power vector assignment \eqref{eq:MCP_uplink_SINR} will converge a.s. to $(\gamma_1, \gamma_2)$ that is a fixed point of $G$ as defined in Lemma~\ref{lem:MCP_G}.
 However, by Corollary~\ref{cor:MCP_FG}, $G$ has a unique fixed point, namely $(\gamma, \gamma)$. It follows that the SINRs of all users in the system tend to the common value $\gamma$.

We can follow the same reasoning as that in Appendix \ref{appendix:coord_dual_LAS}, replacing the optimization problem  \eqref{eq:CBf_opt_lim} - \eqref{eq:coord_bal_fpe} by:
\begin{align}
& \textrm{max.}_{\mu_1, \mu_2 \ge 0} \quad \beta \sigma^2 \left(\bar{\lambda}_1 + \bar{\lambda}_2\right) \label{eq:MCP_reduced_dual} \\
&\textrm{s.t.}  \quad  \mu_1 + \mu_2  = 2 \\
& \quad \quad (\bal_1, \bal_2) \textrm{~is the unique fixed point of~}\eqref{eq:MCP_FPE}. \label{eq:MCP_bal_fpe}
\end{align}
The solution\footnote{Note that the function $F$ in \eqref{eq:def_F_MCP} actually depends on the choice of $(\mu_1, \mu_2)$.} to \eqref{eq:MCP_reduced_dual}-\eqref{eq:MCP_bal_fpe}
can be shown to be $(\mu_1, \mu_2) = (1,1)$, with corresponding $(\bal_1, \bal_2) = (\bal, \bal)$, where $\bal$ is given in \eqref{eq:MCP_bal}; The details of the proof are skipped due to space constraints.

We thus conclude that the asymptotically optimal dual variables are $\mu_1 = \mu_2 = 1$ and that all components of the optimal $\bol$ corresponding to cell $j$ users must converge to $\bal$, as given in \eqref{eq:MCP_bal}. Thus the optimal downlink beamforming vectors asymptotically point in the directions of the vectors given in \eqref{eq:MCP_bf}.
Fixing the beamforming to lie in these directions, we now find the optimal power levels, $p_{kj}$, to use for each user.

{\it Asymptotic analysis of the primal problem} \\

With the beamforming directions fixed, only the DL power levels $p_{kj}$'s still need to be determined. As in the proofs of Theorems \ref{THEO_BF_SCP} and \ref{THEO_BF_COORD}, we first fix the DL powers to constants, and study the resulting limiting regime. Thus, we assume $p_{kj}$'s to be fixed at a constant $\bar{p}_j$, $j= 1, 2$. Unlike the SCP and CBf cases, this does not imply that the total transmit power of BS $j$ is equal to $\beta \bar{p}_j$, but rather that the total transmit power of BS $j$ is equal to
\begin{align}
\bar{P}_j = \sum_{j'=1}^2 \frac{\bar{p}_{j'}}{N} \sum_{k=1}^K \frac{\|\mathbf{E}_j \mathbf{\hat{w}}_{kj'}\|^2}{\|\mathbf{\hat{w}}_{kj'}\|^2}.  \label{MCP:P_j}
\end{align}

Now turning to the DL power levels, we focus on the transmit strategy that allocates fixed power levels $\bar{p}_j$ to all users in cell $j$.
The following lemmas can be used to show that the left-hand side of \eqref{eq:sinr_mcp} converges \emph{in probability} to $\sigma^2$ with all users being allocated equal power as given by \eqref{eq:MCP_bap}. This power allocation also results in zero duality gap.

\begin{lemma}\label{asymptotics_MCP}
With $\mathbf{\hat{w}}_{kj} = \mathbf{A}_{k,j}  \mathbf{\tilde{h}}_{k,j}^H$, with $\bar{\lambda}$ as given by \eqref{eq:MCP_bal}:
\begin{align}
&\max_{j=1, 2, k \le K} \left|\frac{\left|\mathbf{\tilde{h}}_{k,j}\mathbf{\hat{w}}_{kj} \right|^2}{N\|\mathbf{\hat{w}}_{kj}\|^2} -
(1+\epsilon) \left[1- \frac{ \beta \gamma^2}{\left(1+ \gamma\right)^2}\right]\right|
\rightarrow 0 \textrm{ a.s. } \label{eq:num_MCP}
\end{align}
\begin{align}
&\max_{j=1, 2, k, k' \le K, k' \neq k} \left|
\frac{\left|\mathbf{\tilde{h}}_{k', j} \mathbf{\hat{w}}_{kj}\right|^2}{\|\mathbf{\hat{w}}_{kj}\|^2} \right. \nonumber \\
&\left.-
\frac{1}{1+\gamma^2}\left(
1+\epsilon - \frac{\frac{2\epsilon}{1+\epsilon}}{1-\beta \frac{(1-\epsilon)^2}{(1+\epsilon)^2} \frac{\gamma^2}{(1+\gamma)^2}}
\right)
\right|  \rightarrow 0, \textrm{ i.p.} \label{eq:denom1_MCP}
\end{align}
\begin{align}
&\max_{j=1, 2, k, k' \le K, k' \neq k}\left|\frac{\left|\mathbf{\tilde{h}}_{k', \bar{j}} \mathbf{\hat{w}}_{kj}\right|^2}{\|\mathbf{\hat{w}}_{kj}\|^2} \right. \nonumber \\
&\left. - \frac{1}{1+\gamma^2}\frac{2\epsilon}{1+\epsilon} \frac{1}{1-\beta \frac{(1-\epsilon)^2}{(1+\epsilon)^2} \frac{\gamma^2}{(1+\gamma)^2}}\right| \rightarrow 0,  \textrm{ i.p.}\label{eq:denom2_MCP}
\end{align}
as $K, N \rightarrow \infty$, $\frac{K}{N} \rightarrow \beta$.
\end{lemma}
\begin{IEEEproof}
The proof is similar to that of Lemmas \ref{asymptotics_SCP} and \ref{asymptotics_CBf}. Thus, defining $\mathbf{D}_{k,j}$ and $\mathbf{D}_{k,j}$, as in \eqref{eq:MCP_D1} and \eqref{eq:MCP_D2}, $\mu_1 = \mu_2 = 1$ and $\lambda_1 = \lambda_2 = \bal$, we can show, by applying Lemma 5.1 in \cite{liang_it07} and Lemma \ref{lemm:633} that
\begin{align}
\max_{j =1, 2, k \le K} \left|\frac{\bal}{N} \|\mathbf{\hat{w}}_{kj}\|^2 - \frac{1}{N} \textrm{tr} \mathbf{D}_{k,j} \mathbf{A}^2\right| \stackrel{a.s.}{\longrightarrow} 0 \label{eq:Norm_MCP} \\
\max_{j = 1, 2, k \le K} \left|\frac{\bal}{N} \mathbf{\tilde{h}}_{k,j}\mathbf{\hat{w}}_{kj} - \frac{1}{N}  \textrm{tr} \mathbf{D}_{k,j} \mathbf{A}\right| \stackrel{a.s.}{\longrightarrow} 0. \label{eq:UsefulSignal_MCP}
\end{align}
Now consider the interference terms, $\frac{\left|\mathbf{\tilde{h}}_{k,j}\mathbf{\hat{w}}_{k'j'} \right|^2}{\|\mathbf{\hat{w}}_{k'j'}\|^2}$, with $(k',j') \neq (k,j)$. We have that
\begin{align}
&\frac{\left|\mathbf{\tilde{h}}_{k,j}\mathbf{\hat{w}}_{k'j'} \right|^2}{\|\mathbf{\hat{w}}_{k'j'}\|^2} \nonumber \\
&= \frac{1}{\bal} \frac{\frac{\bal^2}{N}\mathbf{\tilde{h}}_{k,j}\mathbf{A}_{k',j', k, j}\mathbf{\tilde{h}}_{k',j'}^H
\mathbf{\tilde{h}}_{k',j'} \mathbf{A}_{k',j', k, j}\mathbf{\tilde{h}}_{k,j}^H}{\left(1+ \frac{\bal}{N}\mathbf{\tilde{h}}_{k,j}\mathbf{A}_{k',j', k,j} \mathbf{\tilde{h}}_{k,j}^H\right)^2
}\frac{1
}{\frac{\bal}{N} \|\mathbf{\hat{w}}_{k'j'}\|^2}
\end{align}

Similarly to \eqref{eq:Norm_MCP} and \eqref{eq:UsefulSignal_MCP}, we can show that
\begin{align}
\max_{j, i = 1,2, k,l \le K, (k,j) \neq (l,i)}&\left|\frac{\bal^2}{N}\mathbf{\tilde{h}}_{k,j}\mathbf{A}_{k',j', k, j}\mathbf{\tilde{h}}_{k',j'}^H
\mathbf{\tilde{h}}_{k',j'} \mathbf{A}_{k',j', k, j}\mathbf{\tilde{h}}_{k,j}^H \right. \nonumber \\
&\left.~~~-  \frac{1}{N} \textrm{tr} \mathbf{D}_{k,j} \mathbf{A} \mathbf{D}_{l,i} \mathbf{A}
 \right| \stackrel{a.s.}{\longrightarrow} 0.
\end{align}
We can also show that
\begin{align}
\max_{j, i = 1,2, k,l \le K, (k,j) \neq (l,i)}& \left|\frac{\bal}{N}\mathbf{\tilde{h}}_{k,j}\mathbf{A}_{l,i, k,j} \mathbf{\tilde{h}}_{k,j}^H \right. \nonumber \\
&~~~~~~~\left.-
\frac{1}{N}\textrm{tr} \mathbf{D}_{k,j} \mathbf{A} \right| \stackrel{a.s.}{\longrightarrow} 0.
\end{align}

The proof of the lemma is concluded by using the limits of the trace terms as derived in Appendix \ref{app:MCP_UL}, with $\lambda_j = \bal$ given by \eqref{eq:MCP_bal},
\begin{align}
&\frac{1}{N} \textrm{tr} \mathbf{D}_{k,j} \mathbf{A} \stackrel{a.s.}{\longrightarrow} \gamma \\
&\frac{1}{N} \textrm{tr} \mathbf{D}_{k,j} \mathbf{A}^2  \stackrel{a.s.}{\longrightarrow} \frac{1}{(1+\epsilon) \bal} \frac{\gamma^2}{1 -\frac{\beta \gamma^2}{(1+\gamma)^2}}
 \\
&\frac{1}{N} \textrm{tr} \mathbf{D}_{k,j} \mathbf{A} \mathbf{D}_{l,j} \mathbf{A} \stackrel{i.p.}{\longrightarrow}
\gamma^2\frac{
\frac{1+\epsilon^2}{(1+\epsilon)^2}
-  \frac{(1-\epsilon)^2}{(1+\epsilon)^2} \frac{\beta \gamma^2}{\left(1+\gamma \right)^2}
}{\left[1-
 \frac{\beta \gamma^2}{\left(1+\gamma \right)^2}\right]
\left[1- \frac{(1-\epsilon)^2}{(1+\epsilon)^2}\frac{\beta \gamma^2}{\left(1+\gamma\right)^2}\right] }
\\
&\frac{1}{N} \textrm{tr} \mathbf{D}_{k,\bar{j}} \mathbf{A} \mathbf{D}_{l,j} \mathbf{A}  \stackrel{i.p.}{\longrightarrow}
\gamma^2\frac{
\frac{2\epsilon}{(1+\epsilon)^2}}{\left[1-
 \frac{\beta \gamma^2}{\left(1+\gamma \right)^2}\right]
\left[1- \frac{(1-\epsilon)^2}{(1+\epsilon)^2}\frac{\beta \gamma^2}{\left(1+\gamma \right)^2}\right] }.
\end{align}
\end{IEEEproof}

\begin{lemma}\label{lemm:P_j_conv}
With $\mathbf{\hat{w}}_{kj} = \mathbf{A}_{k,j}  \mathbf{\tilde{h}}_{k,j}^H$, with $\bar{\lambda}$ as given by \eqref{eq:MCP_bal}, as as $K, N \rightarrow \infty$, $\frac{K}{N} \rightarrow \beta$, the following holds with probability 1, for $j = 1, 2$,
\begin{align}
\bar{P}_j \rightarrow \sum_{j'=1}^2 \frac{\beta \bar{p}_{j'}}{2},
\end{align}
where $\bar{P}_j$ is the total transmit power under the considered DL transmit power regime (see \eqref{MCP:P_j}).
\end{lemma}
\begin{IEEEproof}
Following similar steps to those in the proof of Lemma \ref{asymptotics_MCP}, we can show that the $\frac{\bal}{N}\|\mathbf{E}_j \mathbf{\hat{w}}_{kj}\|^2$ and
$\frac{\bal}{N}\|\mathbf{\hat{w}}_{kj}\|^2$ converge a.s., and uniformly in their indices, to constants such that the following holds
\begin{align}
\max_{j, i = 1,2, k \le K}&\left|\frac{\|\mathbf{E}_j \mathbf{\hat{w}}_{kj}\|^2}{\|\mathbf{\hat{w}}_{kj}\|^2}-\frac{1}{2}\right| \stackrel{a.s.}{\longrightarrow} 0.
\end{align}
As a result,
\begin{align}
\bar{P}_j = \sum_{j'=1}^2 \frac{\beta \bar{p}_{j'}}{2} + o(1).
\end{align}
This completes the proof.
\end{IEEEproof}
Thus, asymptotically, under this scheme, both BSs transmit with equal power.

\begin{lemma}\label{lemm:MCP_interf}
With $\mathbf{\hat{w}}_{kj} = \mathbf{A}_{k,j} \mathbf{\tilde{h}}_{k,j}^H$, such that $\bal$ satisfies \eqref{eq:MCP_bal}, and with $p_{kj} = \bar{p}_j$ for $k = 1, \ldots, K$, $j = 1, 2$, it follows that,
\begin{align}
&\sum_{{j}', {k}', ({k}', {j}') \neq (k,j)}
\frac{p_{k'j'}}{N} \frac{\left|\mathbf{\tilde{h}}_{k, j}
\mathbf{\hat{w}}_{k'j'}\right|^2}{\|\mathbf{\hat{w}}_{{k}'{j}'}\|^2} \nonumber \\
&\stackrel{i.p.}{\longrightarrow} \frac{\beta}{1+\gamma^2}\left[\bar{p}_j \left(
1+\epsilon - \frac{\frac{2\epsilon}{1+\epsilon}}{1-\beta \frac{(1-\epsilon)^2}{(1+\epsilon)^2} \frac{\gamma^2}{(1+\gamma)^2}}
\right) \right. \nonumber \\
&~~~~~~~~~~~~\left. + \bar{p}_{\bar{j}}
 \frac{\frac{2\epsilon}{1+\epsilon}}{1-\beta \frac{(1-\epsilon)^2}{(1+\epsilon)^2} \frac{\gamma^2}{(1+\gamma)^2}}\right]
.
\end{align}
\end{lemma}
\begin{IEEEproof}
Since DL $p_{kj}$s are all fixed to the same value, $\bar{p}_j$, we have
\begin{align}
&\sum_{{j}', {k}', ({k}', {j}') \neq (k,j)}
\frac{p_{{k}'{j}'}}{N} \frac{\left|\mathbf{\tilde{h}}_{k, j}
\mathbf{\hat{w}}_{{k}'{j}'}\right|^2}{\|\mathbf{\hat{w}}_{{k}'{j}'}\|^2} \nonumber \\
&= \frac{\bar{p}_j}{N}\sum_{k' \neq k} \frac{\left|\mathbf{\tilde{h}}_{k, j}
\mathbf{\hat{w}}_{{k}'{j}}\right|^2}{\|\mathbf{\hat{w}}_{{k}'{j}}\|^2} +
\frac{\bar{p}_{\bar{j}}}{N}\sum_{k'} \frac{\left|\mathbf{\tilde{h}}_{k, j,\bar{j}}\mathbf{\hat{w}}_{{k}'\bar{j}}\right|^2}{\|\mathbf{\hat{w}}_{{k}'\bar{j}}\|^2}
 \label{eqMCP_InterfLemmaProof}
\end{align}
Applying \eqref{eq:denom1_MCP} and \eqref{eq:denom2_MCP} of Lemma \ref{asymptotics_CBf}, \eqref{eqMCP_InterfLemmaProof} for $k = 1, \ldots, K$, $j = 1, 2$, will converge in probability to
\begin{align}
\frac{1}{1+\gamma^2}&\left[\bar{p}_j \frac{K-1}{N}\left(
1+\epsilon - \frac{\frac{2\epsilon}{1+\epsilon}}{1-\beta \frac{(1-\epsilon)^2}{(1+\epsilon)^2} \frac{\gamma^2}{(1+\gamma)^2}}
\right) \right. \nonumber \\
&~~\left. + \bar{p}_{\bar{j}}
 \frac{\frac{2\epsilon}{1+\epsilon}}{1-\beta \frac{(1-\epsilon)^2}{(1+\epsilon)^2} \frac{\gamma^2}{(1+\gamma)^2}}\right].
\end{align}
Noting that as $K, N \rightarrow \infty$, $\frac{K-1}{N} \rightarrow \beta$ completes the proof.
\end{IEEEproof}

Referring to \eqref{eq:sinr_mcp}, and applying Lemmas \ref{asymptotics_MCP} and \ref{lemm:MCP_interf}, to find the {\it minimal} pair of constants $(\bar{p}_1, \bar{p}_2)$ for users in the two cells, we therefore solve the following set of equations,
\begin{align}
&\bar{p}_1 \frac{(1+\epsilon) \left[1- \frac{ \beta \gamma^2}{\left(1+ \gamma\right)^2}\right]}{\gamma}
\nonumber \\
&=  \sigma^2 +
\frac{\beta}{1+\gamma^2}\left[\bar{p}_1 \left(
1+\epsilon - \frac{\frac{2\epsilon}{1+\epsilon}}{1-\beta \frac{(1-\epsilon)^2}{(1+\epsilon)^2} \frac{\gamma^2}{(1+\gamma)^2}}
\right) \right. \nonumber \\
&~~~~~~~~~~~~\left. + \bar{p}_{2}
 \frac{\frac{2\epsilon}{1+\epsilon}}{1-\beta \frac{(1-\epsilon)^2}{(1+\epsilon)^2} \frac{\gamma^2}{(1+\gamma)^2}}\right]
,
\end{align}
\begin{align}
&\bar{p}_2 \frac{(1+\epsilon) \left[1- \frac{ \beta \gamma^2}{\left(1+ \gamma\right)^2}\right]}{\gamma} \nonumber \\
&= \sigma^2
+ \frac{\beta}{1+\gamma^2}\left[\bar{p}_2 \left(
1+\epsilon - \frac{\frac{2\epsilon}{1+\epsilon}}{1-\beta \frac{(1-\epsilon)^2}{(1+\epsilon)^2} \frac{\gamma^2}{(1+\gamma)^2}}
\right) \right. \nonumber \\
&~~~~~~~~~~~~\left. + \bar{p}_1
 \frac{\frac{2\epsilon}{1+\epsilon}}{1-\beta \frac{(1-\epsilon)^2}{(1+\epsilon)^2} \frac{\gamma^2}{(1+\gamma)^2}}\right]
,
\end{align}
to obtain
\begin{align}
\bar{p}_1 = \bar{p}_2 = \bar{p} = \frac{\sigma^2 \gamma}{(1+\epsilon) \left[1- \frac{ \beta \gamma}{1+ \gamma}\right]}. \label{eq:Pj_converges_MCP}
\end{align}
This implies that $\bar{P}_1$ and $\bar{P}_2$ (recall Lemma \ref{lemm:P_j_conv}) both converge a.s. to $\bar{P}$, where
\begin{align}
\bar{P} := \beta \bar{p}.
\end{align}
This choice of DL transmit powers meets the SINR constraints as $K, N \rightarrow \infty$, $\frac{K}{N} = \beta$ with a probability tending to 1.

We confirm the asymptotic optimality of this deterministic power allocation, together with the beamforming directions found from analysis of the dual problem, by verifying that the duality gap is tending to zero. Indeed, the primal objective value converges to $2 \bar{P}$ ($\phi = 1$, since the power consumption in both cells is the same), whereas the dual objective value converges to $2 \beta \bar{\lambda} \sigma^2$. Recalling the definition of $\bar{\lambda}$, one can easily verify that these two quantities are the same, thereby completing the proof.

Equation \eqref{eq:Pj_converges_MCP} shows that the primal problem have a solution when effective bandwidth condition
$\displaystyle \beta
\left(\frac{\gamma }{1+\gamma} \right) < 1$ is satisfied. Conversely, if this condition does not hold, there is asymptotically no feasible solution to the primal
problem, in the limit as $N \uparrow \infty$.

\section{Proof of Theorem \ref{THEO_BETA_SCP}}
\label{appendix:proofTheoBETA_SCP}
Using \eqref{eq:GammaOpt_SCP},
$r$ simplifies to:
\begin{align}
r(\gamma^*) = \frac{1+\gamma^*}{\gamma^* \left(\frac{\sigma^2}{P} +
\epsilon\right)\left(1+\gamma^*\right)+ \gamma^*} \log\left(1 +
\gamma^* \right),  \label{eq:R_SCP}
\end{align}
which is positive at $\gamma^* = 0$, and zero at $\gamma^* =
\infty$. Defining $\eta = \frac{\sigma^2}{P} + \epsilon$ and taking
the derivative with respect to $\gamma^*$ we obtain $\displaystyle \frac{dr}{d\gamma^*}$ equals
\begin{align}
-\frac{1 + \eta
\left(1+\gamma^*\right)^2}{(\gamma^*\eta \left(1+\gamma^*\right)+
\gamma^*)^2} \log\left(1 + \gamma^* \right) +\frac{1}{\gamma^* \eta
\left(1+\gamma^*\right)+ \gamma^*}.
\end{align}
This has the same sign as
\begin{align}
h(\gamma^*) = - \log\left(1 + \gamma^* \right) +\frac{\gamma^* \eta
\left(1+\gamma^*\right)+ \gamma^*}{1 + \eta
\left(1+\gamma^*\right)^2}, \label{eq:dR}
\end{align}
which is 0 at $\gamma^* =0$, and $-\infty$ at $\gamma^* = \infty$.
Differentiating the expression in \eqref{eq:dR}, we obtain
\begin{align}
\frac{dh}{d\gamma^*} = -\frac{1}{1+\gamma^*}+\frac{\eta + 1 + \eta
\left(\eta+ 1\right)\left(1+\gamma^*\right)^2 - 2 \eta \gamma^{*2}
}{\left(1 + \eta \left(1+\gamma^*\right)^2\right)^2}.
\end{align}
This has the same sign as
\begin{align}
&-\left(1
+
\eta \left(1+\gamma^*\right)^2\right)^2 \nonumber \\
&+ \left(1+\gamma^*\right)\left(\eta + 1 +
\left(\eta+ 1\right)\eta \left(1+\gamma^*\right)^2 - 2
\eta \gamma^{*2}\right) \nonumber \\
&=-\gamma^{*4} \eta^2 -\gamma^{*3} \eta \left[1+3\eta\right]
-\gamma^{*2} \eta \left[1+3 \eta\right] -\gamma^{*}
\left[\eta^2-1\right]. \nonumber
\end{align}
If $\eta \geq 1$, then $\displaystyle \frac{dh}{d\gamma^*} < 0~~
\forall \gamma^*>0$, which implies $h(\gamma^*) < 0 ~~\forall
\gamma^*
> 0$ and hence $r(\gamma^*)$ is a decreasing function. If $\eta < 1$
then $h(\gamma^*)$ is positive for $\gamma^*$ small. This implies
that $r(\gamma^*)$ is increasing and then decreasing.

\section{Proof of Theorem \ref{THEO_BETA_COORD}}
\label{appendix:proofTheo_BETA_COORD}
The normalized achievable rate per cell is equal to
\begin{align}
&r(\gamma^*) 
= \beta \log\left(1+\gamma^*\right)
= \frac{1}{\gamma^*}\frac{\log\left(1+\gamma^*\right)}{\left[\frac{\sigma^2}{P}  + \frac{1}{1+\gamma^*} + \frac{\epsilon}{1+\epsilon \gamma^*}\right]}  \nonumber \\
&=\frac{\left(1+\gamma^*\right)\left(1+\epsilon
\gamma^*\right) \log\left(1+\gamma^*\right)}{\left[\gamma^*\left[\frac{\sigma^2}{P}\left(1+\gamma^*\right)+1\right]\left(1+\epsilon
\gamma^*\right)  + 
\epsilon \gamma^*\left(1+\gamma^*\right)\right]},  \label{eq:r_COORD}
\end{align}
which is positive at $\gamma^*=0$ and zero at $\gamma^*=\infty$. The
sign of $\displaystyle \frac{dr}{d\gamma^*}$ is the same as the sign
of the function
\begin{align}
&h(\gamma^*) \nonumber \\
&= \frac{\left(1+\epsilon
\gamma^*\right)\left[\gamma^* \left[\frac{\sigma^2}{P}\left(1+\gamma^*\right)+1\right]\left(1+\epsilon
\gamma^*\right)  + 
\epsilon \gamma^*\left(1+\gamma^*\right)\right]}{\left[(\epsilon
\gamma^{*2} +1)\left[1+\epsilon\right]
+\frac{\sigma^2}{P}\left(1+\gamma^*\right)^2\left(1+\epsilon
\gamma^*\right)^2  +4\epsilon \gamma^*\right]} \nonumber \\
&-
\log\left(1+\gamma^*\right) \label{eq:derivative}
\end{align}
which is 0 at $\gamma^* =0$, and $-\infty$ at $\gamma^* = \infty$.
The sign of $\displaystyle \frac{dh}{d\gamma^*}$ can be shown to be
{\it opposite} to the sign of a polynomial in $\gamma^*$ of degree 7, i.e. $\sum_{i=0}^7 c_i \gamma^{*i}$, such that
\begin{align}
c_7 &=  a^2 \epsilon^4, ~~c_6 = \epsilon^3\left(3 a^2 \epsilon + 4 a^2 + 2 a \epsilon\right) \nonumber
\\
c_5 &= \epsilon^2  \left(3 a^2 \epsilon^2 + 12 a^2 \epsilon + 6 a^2 + 4 a \epsilon^2 + 6 a \epsilon\right) \nonumber  \\
c_4 &= \epsilon \left(a^2 \left[\epsilon\left(\epsilon^2 + 12 \epsilon + 18\right)  + 4\right] + a \epsilon \left(3 \epsilon^2 + 8 \epsilon + 9 \right) \right. \nonumber \\&\left.- 2 \epsilon \left(\epsilon +1\right)\right) \nonumber
\end{align}
\begin{align}
c_3 &= a^2 \left[2\epsilon\left( 2 \epsilon^2 + 9 \epsilon + 6 \right) + 1\right] + a \epsilon\left(\epsilon^3 + 13 \epsilon + 6 \right) \nonumber \\&+ \epsilon^2\left(1  - \epsilon^2 - 16 \epsilon\right)\nonumber \\
c_2 &= 3 a^2 \left[2 \epsilon \left(\epsilon + 2\right) + 1\right] + a \left(3 \epsilon^2 + 10 \epsilon + 1\right)  + \epsilon \nonumber \\ &- \left(17 \epsilon^3 + 8 \epsilon^2 + 4 a \epsilon^3\right) \nonumber \\
c_1 &= a^2 \left(4 \epsilon + 3 \right)  + 6 a \epsilon+ a - \epsilon \left(2 a \epsilon^2 + 3 a \epsilon + 9 \epsilon^2 + 4\epsilon + 3\right) \nonumber \\
c_0 &= a^2  + 2 a\epsilon  - 2 a \epsilon^2 - 2\epsilon^3 - \epsilon^2 - 1 \label{eq:poly},
\end{align}
where $a = \frac{\sigma^2}{P}$. One can factor the constant term
(independent of $\gamma^*$) into
\begin{align}
(a+\epsilon+1)(a  + \epsilon - 2\epsilon^2-1).
\label{eq:constant_term}
\end{align}
We now consider two cases:

\subsection{$a  + \epsilon - 2\epsilon^2-1 \ge 0$}
Defining $\epsilon_{0,1} = \frac{1-\sqrt{8a-7}}{4}, \epsilon_{0,2} =
\frac{1+\sqrt{8a-7}}{4}$, one can show that \eqref{eq:constant_term}
will be positive if $(a \ge 1,\mbox{~and~} 0 \le \epsilon \le
\epsilon_{0,2})$, or $(\frac{7}{8} \le a \le 1, \mbox{~and~}
\epsilon_{0,1} \le \epsilon \le \epsilon_{0,2})$. One can further
establish that if this is the case, all of the other coefficients of
the polynomial \eqref{eq:poly} will be positive \footnote{This is
done by using the fact that if $a \ge \frac{7}{8}$ and
$0 \le \epsilon \le \epsilon_{0,2}$ then $\epsilon \le \sqrt{a}$ and
$\epsilon \le a$.}, and consequently $h(\gamma^*)$ will be a
decreasing function, and hence $h(\gamma^*) < 0 ~~\forall \gamma^* >
0$. In this case, $r(\gamma^*)$ is a decreasing function and the
optimum will occur at $\gamma^* = 0$, which corresponds to
$\beta^*=\infty$.

\subsection{$a  + \epsilon - 2\epsilon^2-1 < 0$}
In this case, $h(\gamma^*)$ will be positive for $\gamma^*$ small.
Thus, $r(\gamma^*)$ will be increasing initially, but it decreases
eventually (to zero), so it must attain its global maximum at a
finite value of $\gamma^* > 0$.

\bibliographystyle{IEEEbib}
\bibliography{IEEEabrv,zakhour_biblio3}

\begin{thebibliography}{10}

\bibitem{weingarten_it06}
H.~Weingarten, Y.~Steinberg, and S.~Shamai,
\newblock ``The capacity region of the {G}aussian multiple-input
  multiple-output broadcast channel,''
\newblock {\em {IEEE} Trans. Inf. Theory}, vol. 52, no. 9, Sept. 2006.

\bibitem{dahrouj_iss08}
H.~Dahrouj and W.~Yu,
\newblock ``Coordinated beamforming for the multi-cell multi-antenna wireless
  system,''
\newblock in {\em Proc. Conference on Information Sciences and Systems 2008,
  CISS 2008}, 2008.

\bibitem{dahrouj_tw10}
H.~Dahrouj and W.~Yu,
\newblock ``Coordinated beamforming for the multicell, multi-antenna wireless
  system,''
\newblock {\em {IEEE} Trans. Wireless Commun.}, May 2010.

\bibitem{karakayali_icc06}
K.~Karakayali et~al.,
\newblock ``On the maximum common rate achievable in a coordinated network,''
\newblock in {\em Proc. IEEE International Conference on Communications, ICC},
  2006.

\bibitem{jing_eurasip08}
S.~Jing et~al.,
\newblock ``Multicell downlink capacity with coordinated processing,''
\newblock {\em EURASIP Journal on Wireless Communications and Networking},
  2008.

\bibitem{somekh_it09}
O.~Somekh et~al.,
\newblock ``Cooperative multicell zero-forcing beamforming in cellular downlink
  channels,''
\newblock {\em {IEEE} Trans. Inf. Theory}, July 2009.

\bibitem{shamai_vtc2001}
S.~Shamai~(Shitz) and B.~M. Zaidel,
\newblock ``Enhancing the cellular downlink capacity via co-processing at the
  transmitting end,''
\newblock in {\em Proc. IEEE Vehicular Technology Conf.}, 2001.

\bibitem{ZH_isit11}
R.~Zakhour and S.V. Hanly,
\newblock ``Minmax coordinated beamforming via large system analysis,''
\newblock 2011,
\newblock IEEE ISIT 2011.

\bibitem{tulino04}
A.~M. Tulino and S.~Verd\'u,
\newblock ``Random matrix theory and wireless communications,''
\newblock {\em Foundations and Trends in Communications and Information
  Theory}, 2004.

\bibitem{tse_it1999}
D.~N.~C. Tse and S.~V. Hanly,
\newblock ``Linear multiuser receivers: Effective interference, effective
  bandwidth and user capacity,''
\newblock {\em {IEEE} Trans. Inf. Theory}, Mar. 1999.

\bibitem{dai_sp03}
Huaiyu Dai and H.~V. Poor,
\newblock ``Asymptotic spectral efficiency of multicell {MIMO} systems with
  frequency-flat fading,''
\newblock {\em {IEEE} Trans. Signal Processing}, Nov. 2003.

\bibitem{tulino_it05}
A.M. Tulino, A.~Lozano, and S.~Verdu,
\newblock ``Impact of antenna correlation on the capacity of multiantenna
  channels,''
\newblock {\em {IEEE} Trans. Inf. Theory}, july 2005.

\bibitem{aktas_it06}
D.~Aktas et~al.,
\newblock ``Scaling results on the sum capacity of cellular networks with
  {MIMO} links,''
\newblock {\em {IEEE} Trans. Inf. Theory}, vol. 52, no. 7, july 2006.

\bibitem{dumont_it2010}
J.~Dumont et~al.,
\newblock ``On the capacity achieving covariance matrix for {R}ician {MIMO}
  channels: An asymptotic approach,''
\newblock {\em {IEEE} Trans. Inf. Theory}, Mar. 2010.

\bibitem{nguyen_globecom08}
V.~K. Nguyen and J.~S. Evans,
\newblock ``Multiuser transmit beamforming via regularized channel inversion: A
  large system analysis,''
\newblock in {\em Proc. IEEE Global Telecomm. Conf.}, Nov.-Dec. 2008.

\bibitem{couillet_WDS08}
R.~Couillet et~al.,
\newblock ``The space frontier: Physical limits of multiple antenna information
  transfer,''
\newblock in {\em Proc. of VALUETOOLS 2008, Workshop Inter-Perf}, Oct. 2008.

\bibitem{Couillet_2010}
R.~Couillet, M~Debbah, and J.~W. Silverstein,
\newblock ``A deterministic equivalent for the capacity analysis of correlated
  multi-user {MIMO} channels,''
\newblock {\em CoRR}, vol. abs/0906.3667, 2009.

\bibitem{caire_icc10}
S.H. Moon et~al.,
\newblock ``Weighted sum rate of multi-cell {MIMO} downlink channels in the
  large system limit,''
\newblock in {\em Proc. IEEE Int. Conf. on Communication}, June 2010.

\bibitem{caire_ciss10}
H.~Huh et~al.,
\newblock ``Network {MIMO} large-system analysis and the impact of {CSIT}
  estimation,''
\newblock Mar. 2010.

\bibitem{caire_isit10}
H.~Huh et~al.,
\newblock ``Multi-cell {MIMO} downlink with fairness criteria: the large system
  limit,''
\newblock in {\em Proc. IEEE Int. Symp. on Inf. Theory}, June 2010.

\bibitem{lakshminaryana_pimrc10}
S.~Lakshminaryana, J.~Hoydis, M.~Debbah, and M.~Assaad,
\newblock ``Asymptotic analysis of distributed multi-cell beamforming,''
\newblock in {\em Proc. IEEE Int. Symp. on Personal, Indoor and Mobile Radio
  Communications}, Sept. 2010.

\bibitem{zakhour_allerton10}
R.~Zakhour and S.~V. Hanly,
\newblock ``Large system analysis of base station cooperation on the
  downlink,''
\newblock in {\em Proc. Allerton Conf. 2010}, IL, United States, Sept.-Oct.
  2010.

\bibitem{wagner_arxiv10}
S.~Wagner, R.~Couillet, M.~Debbah, and D.~T.~M. Slock,
\newblock ``Large system analysis of linear precoding in {MISO} broadcast
  channels with limited feedback,''
\newblock 2010,
\newblock arXiv:0906.3682v3.

\bibitem{hoydis_arxiv10}
J.~Hoydis, M.~Kobayashi, and M.~Debbah,
\newblock ``On the optimal number of cooperative base stations in network
  {MIMO} systems,''
\newblock 2010,
\newblock arXiv:0912.4595v3.2642.

\bibitem{telatar_99}
E.~Telatar,
\newblock ``Capacity of multi-antenna gaussian channels,''
\newblock {\em Europ. Trans. Telecomm.}, Nov. 1999.

\bibitem{farrokhi_jsac98}
F.~Rashid-Farrokhi, K.~J.~R. Liu, and L.~Tassiulas,
\newblock ``Transmit beamforming and power control for cellular wireless
  systems,''
\newblock {\em {IEEE} J. Sel. Areas Commun.}, Oct. 1998.

\bibitem{yates95}
R.D. Yates,
\newblock ``A framework for uplink power control in cellular radio networks,''
\newblock {\em {IEEE} J. Sel. Areas Commun.}, Sept. 1995.

\bibitem{visotsky_vtc99}
E.~Visotsky and U.~Madhow,
\newblock ``Optimum beamforming using transmit antenna arrays,''
\newblock in {\em Proc. IEEE Vehicular Technology Conf.}, July 1999.

\bibitem{viswanath_it03}
P.~Viswanath and D.N.C. Tse,
\newblock ``Sum capacity of the multiple antenna {G}aussian broadcast channel
  and uplink-downlink duality,''
\newblock {\em {IEEE} Trans. Inf. Theory}, Aug. 2003.

\bibitem{vishwanath_it03}
S.~Vishwanath, N.~Jindal, and A.~Goldsmith,
\newblock ``Duality, achievable rates, and sum-rate capacity of {G}aussian
  {MIMO} broadcast channels,''
\newblock {\em {IEEE} Trans. Inf. Theory}, vol. 49, no. 10, Oct. 2003.

\bibitem{yu_it04}
W.~Yu and J.~Cioffi,
\newblock ``Sum capacity of {G}aussian vector broadcast channels,''
\newblock {\em {IEEE} Trans. Inf. Theory}, Sept. 2004.

\bibitem{yu_it06}
Wei Yu,
\newblock ``Uplink-downlink duality via minimax duality,''
\newblock {\em {IEEE} Trans. Inf. Theory}, Feb. 2006.

\bibitem{tse_isit02}
D.~N.~C. Tse and P.~Viswanath,
\newblock ``Downlink-uplink duality and effective bandwidths,''
\newblock in {\em Proc. IEEE Int. Symp. on Inf. Theory}, 2002.

\bibitem{boche_vtc02}
H.~Boche and M.~Schubert,
\newblock ``A general duality theory for uplink and downlink beamforming,''
\newblock in {\em Proc. IEEE Vehicular Technology Conf.}, Sept. 2002.

\bibitem{vojcic_com98}
B.~R. Vojcic and W.~M. Jang,
\newblock ``Transmitter precoding in synchronous multiuser communications,''
\newblock {\em {IEEE} Trans. Commun.}, 1998.

\bibitem{baretto_icc01}
A.N. Baretto and G.~Fettweis,
\newblock ``Capacity increase in the downlink of spread spectrum systems
  through joint signal processing,''
\newblock in {\em Proc. IEEE Int. Conf. on Communication}, Jun. 2001, vol.~4.

\bibitem{choi_veh04}
R.~L.-U Choi and R.~D. Murch,
\newblock ``Transmit-preprocessing techniques with simplified receivers for the
  downlink of {MISO TDD-CDMA} systems,''
\newblock {\em {IEEE} Trans. Vehicular Technology}, 2004.

\bibitem{joham_sp05}
M.~Joham, W.~Utschick, and J.~A. Nossek,
\newblock ``Linear transmit processing in {MIMO} communication systems,''
\newblock {\em {IEEE} Trans. Signal Processing}, 2005.

\bibitem{wiesel_sp06}
A.~Wiesel, Y.~C. Eldar, and S.~Shamai~(Shitz),
\newblock ``Linear precoding via conic optimization for fixed {MIMO}
  receivers,''
\newblock {\em {IEEE} Trans. Signal Processing}, vol. 54, no. 1, Jan. 2006.

\bibitem{schubert_vt04}
M.~Schubert and H.~Boche,
\newblock ``Solution of the multiuser downlink beamforming problem with
  individual {SINR} constraints,''
\newblock {\em {IEEE} Trans. Vehicular Technology}, Jan. 2004.

\bibitem{boche_icc03}
H.~Boche and M.~Schubert,
\newblock ``Effective bandwidth maximization for uplink/downlink multi-antenna
  systems,''
\newblock in {\em Proc. IEEE Int. Conf. on Communication}, May 2003, vol.~5.

\bibitem{yu_sp07}
W.~Yu and T.~Lan,
\newblock ``Transmitter optimization for the multi-antenna downlink with
  per-antenna power constraints,''
\newblock {\em {IEEE} Trans. Signal Processing}, June 2007.

\bibitem{foschini_miljanic93}
Miljanic~Z. Foschini~G.J.,
\newblock ``A simple distributed autonomous power control algorithm and its
  convergence,''
\newblock {\em {IEEE} Trans. Vehicular Technology}, , no. 4, 1993.

\bibitem{peel_com05}
C.~B. Peel et~al.,
\newblock ``A vector-perturbation technique for near-capacity multiantenna
  multiuser communication-part {I}: channel inversion and regularization,''
\newblock {\em {IEEE} Trans. Commun.}, Jan. 2005.

\bibitem{A81}
A.~Abramovich,
\newblock ``A controlled method for adaptive optimization of filters using the
  criterion of maximum snr,''
\newblock {\em Radiotekh. Elektron}, vol. 26, no. 3, 1981.

\bibitem{ML06}
X.~Mestre and M.~A. Lagunas,
\newblock ``Finite sample size effect on minimum variance beamformers: optimum
  diagonal loading factor for large arrays,''
\newblock {\em {IEEE} Trans. Signal Processing}, vol. 54, no. 1, 2006.

\bibitem{karakayali_wc06}
M.~K. Karakayali, G.~J. Foschini, and R.~A. Valenzuela,
\newblock ``Network coordination for spectrally efficient communications in
  cellular systems,''
\newblock {\em IEEE Wireless Communications}, vol. 13, Aug. 2006.

\bibitem{jsac_tute10}
D.~Gesbert et~al.,
\newblock ``Multicell {MIMO} cooperative networks: a new look at
  interference,''
\newblock {\em IEEE Journal on Selected Areas}, Dec. 2010.

\bibitem{wyner_it94}
A.~Wyner,
\newblock ``Shannon-theoretic approach to a gaussian cellular multiple-access
  channel,''
\newblock {\em {IEEE} Trans. Inf. Theory}, vol. 40, no. 6, 1994.

\bibitem{hanly_it2001}
S.~V. Hanly and D.~N.~C. Tse,
\newblock ``Resource pooling and effective bandwidths in {CDMA} networks with
  multiuser receivers and spatial diversity,''
\newblock {\em {IEEE} Trans. Inf. Theory}, vol. 47, no. 4, May 2001.

\bibitem{hanly95}
S.V. Hanly,
\newblock ``An algorithm for combined cell-site selection and power control to
  maximize cellular spread spectrum capacity,''
\newblock {\em {IEEE} J. Sel. Areas Commun.}, vol. 13, no. 7, Sept. 1995.

\bibitem{hachem_08}
W.~Hachem, P.~Loubaton, and J.~Najim,
\newblock ``A clt for information-theoretic statistics of gram random matrices
  with a given variance profile,''
\newblock {\em The Annals of Applied Probability}, vol. 18, no. 6, 2008.

\bibitem{kammoun_it09}
A.~Kammoun, M.~Kharouf, W.~Hachem, and J.~Najim,
\newblock ``A central limit theorem for the {SINR} at the {LMMSE} estimator
  output for large-dimensional signals,''
\newblock {\em {IEEE} Trans. Inf. Theory}, vol. 55, no. 11, Nov. 2009.

\bibitem{hachem_07}
W.~Hachem, P.~Loubaton, and J.~Najim,
\newblock ``Deterministic equivalents for certain functionals of large random
  matrices,''
\newblock {\em The Annals of Applied Probability}, vol. 17, no. 3, 2007.

\bibitem{liang_it07}
Liang Ying-Chang, Pan Guangming, and Z.~D. Bai,
\newblock ``Asymptotic performance of {MMSE} receivers for large systems using
  random matrix theory.,''
\newblock {\em {IEEE} Trans. Inf. Theory}, vol. 53, no. 11, pp. 4173, 2007.

\bibitem{billingsley99}
P.~Billingsley,
\newblock {\em Convergence of Probability Measures, 2nd ed.},
\newblock John Wiley and Sons, 1999.

\end{thebibliography}

\end{document}